\documentclass[12pt,letterpaper]{article}
\usepackage{amsmath,amssymb}
\usepackage{graphicx,color}
\usepackage{hyperref}

\usepackage[bf]{caption}
\renewcommand{\captionfont}{\small\itshape}

\newcommand{\rr}{\mathbb{R}}

\def\theequation{\arabic{section}.\arabic{equation}}

\newcommand{\be}{\begin{equation}}
\newcommand{\ee}{\end{equation}}
\newcommand{\ba}{\begin{aligned}}
\newcommand{\ea}{\end{aligned}}
\newcommand{\ben}{\begin{displaymath}}
\newcommand{\een}{\end{displaymath}}
\newcommand{\bea}{\begin{eqnarray}}
\newcommand{\eea}{\end{eqnarray}}
\newcommand{\bean}{\begin{eqnarray*}}
\newcommand{\eean}{\end{eqnarray*}}

\newcommand{\p}{\partial}

\def\l {\lambda}
\def\th {\theta}
\def\a {\alpha}

\def\b {\beta}

\def\g {\gamma}

\def\s {\sigma}

\def\m{\mu}
\def\n{\nu}

\def\p{\partial}
\def\pb{\bar{\partial}}

\long\def\symbolfootnote[#1]#2{\begingroup
\def\thefootnote{\fnsymbol{footnote}}\footnote[#1]{#2}\endgroup}

\begin{document}

\begin{titlepage}
\vspace{10pt} \hfill {HU-EP-10/35} \vspace{20mm}
\begin{center}

{\Large \bf Vacuum type space-like string surfaces in $\mbox{AdS}_3\times\mbox{S}^3$}

\vspace{45pt}

{Harald Dorn,$^a~$
George Jorjadze,$^{a,\,b}~$
Chrysostomos Kalousios,$^a$\\[2mm] Luka Megrelidze,$^c~$
Sebastian Wuttke ${}^a$
\symbolfootnote[2]{\tt{\{dorn,jorj,ckalousi,wuttke\}@physik.hu-berlin.de}}
}
\\[15mm]

{\it\ ${}^a$Institut f\"ur Physik der
Humboldt-Universit\"at zu Berlin,}\\
{\it Newtonstra{\ss}e 15, D-12489 Berlin, Germany}\\[3mm]
{\it${}^b$Razmadze Mathematical Institute,}\\
{\it M. Aleksidze 1, 0193, Tbilisi, Georgia}\\[3mm]
{\it${}^c$Ilia State University,}\\
{\it K. Cholokashvili Ave 3/5, 0162,
Tbilisi, Georgia}

\vspace{20pt}

\end{center}

\vspace{40pt}

\centerline{{\bf{Abstract}}}
\vspace*{5mm}
\noindent We construct and classify all space-like minimal surfaces in $\mbox{AdS}_3 \times \mbox{S}^3$
which  globally  admit  coordinates with constant induced metric on both factors. Up to $\mbox{O}(2,2)\times \mbox{O}(4)$ transformations all these surfaces, except one class, are parameterized
by four real parameters. The classes of surfaces correspond to different regions
in this parameter space and show quite different boundary behavior.
Our analysis uses a direct construction of the string coordinates
via a group theoretical treatment based on the map
of  $\mbox{AdS}_3 \times \mbox{S}^3$ to $\mbox{SL}(2,\rr)\times \mbox{SU}(2)$. This is complemented by a cross check
via standard Pohlmeyer reduction.
After
embedding in $\mbox{AdS}_5\times \mbox{S}^5$ we calculate the regularized area for solutions
with a boundary spanned by a four point scattering $s$-channel momenta
configuration.

\vspace{15pt}
\end{titlepage}

\newpage

\tableofcontents

\vspace{3cm}

\section{Introduction}
In a series of papers \cite{Alday:2007hr,aldmald}  a remarkable correspondence between minimal surfaces in $\mbox{AdS}_5$ approaching a null polygonal boundary at conformal infinity
and gluon scattering amplitudes in ${\cal N}=4$ super Yang-Mills theory
has been established. For generic null polygons a lot of structural insight
concerning the dependence
of the area on the boundary data has been achieved. However,
explicit formul\ae{} for the surfaces are available in the tetragon case only
\cite{Kruczenski:2002fb, Alday:2007hr}. The tetragon case is very special,
since it turned out to be the only flat space-like minimal surface in $\mbox{AdS}_5$
\cite{Dorn:2009kq}.

The string dual to ${\cal N}=4$ SYM lives in  $\mbox{AdS}_5\times \mbox{S}^5$. There is an extensive literature on dynamical strings, i.e. time-like surfaces, in $\mbox{AdS}\times \mbox{S}$ (see for example \cite{Arutyunov:2009ga}
and references therein).
For the correspondence to scattering amplitudes one is interested in space-like surfaces extended up to infinity. Therefore, a more
complete treatment should include the study of minimal surfaces in this ten-dimensional
spacetime with the same null polygonal boundaries for their projection on $\mbox{AdS}_5$,
but with a non-trivial extension in $\mbox{S}^5$. We started a corresponding analysis
for  $\mbox{AdS}_3 \times \mbox{S}^3$ in \cite{Dorn:2009hs}. For an example in $ \mbox{AdS}_3 \times \mbox{S}^1 $ see \cite{Sakai:2009ut}. The situation now exhibits two
crucial new aspects. While at first the surface in the total product space of course
has to be minimal, its projections to the factors can be non-minimal. Secondly,
surfaces which are space-like with respect to the induced metric of the full product
space can have projections to $\mbox{AdS}$ of both Euclidean and Lorentzian signature or even with a degenerate  induced metric.
In  \cite{Dorn:2009hs}
we concentrated on the case of a space-like $\mbox{AdS}$-projection and made only
some sketchy remarks on the time-like case. Allowing non-space-like $\mbox{AdS}$-projections opens
a relatively broad set of possibilities. The present paper is devoted to a full
constructive classification within the set of surfaces which admit coordinates where
the metrics induced from the product space as well as from the individual factors are constant\footnote{In physical terms these are vacuum solutions, similar to the four cusp solution in pure $\mbox{AdS}_5$. A characterization by invariant geometrical quantities is: intrinsically flat with constant mean curvatures
.}.

The by now standard procedure for generating minimal surfaces as the solution
of string equations of motion is Pohlmeyer reduction \cite{pohlred, Dorn:2009kq}.
The reduced model inherits possible integrable structures of the original
string sigma model. After solving the reduced model a first order linear
problem has still to be solved to get the coordinates of the wanted surfaces.
Based on the bijective maps of $\mbox{AdS}_3$ and $\mbox{S}^3$ to the group manifolds $\mbox{SL}(2,\rr)$
and $\mbox{SU}(2)$, respectively, we perform an analysis which yields the
embedding coordinates directly.
The analysis of $\mbox{AdS}_3$ string equations in terms of $\mbox{SL}(2,\rr)$ group variables is usually
very helpful for the dynamics with WZ term (see \cite{Maldacena:2000hw, Jorjadze:2009tu} and references therein). It appears that the vacuum configurations of $\mbox{AdS}_3\times\mbox{S}^3$ string  also have a certain factorized structure,
which provides explicit integration of corresponding string equations.
For completeness and a cross check we also analyze
the problem in Pohlmeyer reduction.

The paper is organized as follows.
In section 2 we review the map to a group manifold and then classify and construct the solutions of vacuum type corresponding to space-like surfaces in  $\mbox{AdS}_3 \times\mbox{S}^3$.
We show that three left and three right components of Noether currents related to the isometries
of $\mbox{SL}(2,\rr)\times\mbox{SU}(2)$ are constants both in $\mbox{AdS}_3$ and $\mbox{S}^3$ sectors. That also justifies the name vacuum type solutions.
Section 3 is devoted to a parallel treatment of the problem within Pohlmeyer reduction.
In addition it contains remarks on the relation
of our solutions to the complex sin(h)-Gordon type equations, which for the light-like $\mbox{AdS}_3$ projection
degenerate to a linear equation.
Then we continue in section 4 with an elaboration of the boundary behavior of our surfaces, pointing out the qualitative
differences between the various classes and give a compact listing of their characteristic properties. In section 5 we calculate the regularized area
for the solution which is of potential use for a map to 4-point scattering
amplitudes in $s$- or $t$-channel configuration. Section 6 contains  a summary
and some conclusions. Appendices A and B contain some technical details
related to the main text.

\setcounter{equation}{0}

\section{Space-like strings in $\mbox{SL}(2,\rr)\times \mbox{SU}(2)$}

In this section we describe the $\mbox{AdS}_3 \times\mbox{S}^3$ string equations
in terms of group variables and integrate these equations in the
vacuum sector. We use conformal worldsheet coordinates and
gauge fixing conditions, based on a holomorphic structure of space-like
string surfaces.

\subsection{$\mbox{AdS}_3$ and $\mbox{S}^3$ as group manifolds}

The $\mbox{AdS}_3$ and $\mbox{S}^3$ spaces can be realized as the  group manifolds
$\mbox{SL}(2,\rr)$ and $\mbox{SU}(2)$ via
\begin{equation}\label{g=Y}
  g=\left( \begin{array}{cr}
 Y^{0'}+Y^2 &Y^1+Y^0\\Y^1-Y^0& Y^{0'}-Y^2
 \end{array}\right)~, ~~~~~ h=\left( \begin{array}{cr}
 ~~X^4+iX^3 &X^2+iX^1\\-X^2+iX^1&X^4-iX^3
 \end{array}\right)~.
\end{equation}
Here $Y^K=(Y^{0'},Y^0,Y^1,Y^2)$ are coordinates of
the embedding space $\rr^{2,2}$ and the equation for the hyperboloid
\begin{equation}\label{YY=-1}
Y\cdot Y\equiv -Y_{0'}^2-Y_0^2+Y_1^2+Y_2^2=-1~,
\end{equation}
which defines the $\mbox{AdS}_3$ space,
is equivalent to $g\in \mbox{SL}(2,\rr)$.
Similarly, the equation
for $\mbox{S}^3$ embedded in $\rr^4$
\begin{equation}\label{XX=1}
X\cdot X\equiv X_1^2+X_2^2+X_3^2+X_4^2=1
\end{equation}
is equivalent to $h\in \mbox{SU}(2)$.

Let us introduce the following basis in $\mathfrak{sl}(2,\rr)$
\begin{equation}\label{su(1,1) basis}
  {\bf{t}}_0=\left( \begin{array}{cr}
  ~0&1\\-1&0 \end{array}\right)~,~~~~
   {\bf{t}}_1=\left( \begin{array}{cr}
  0&~1\\1&~0 \end{array}\right)~,~~~~
 {\bf{t}}_2=\left( \begin{array}{cr}
  1&~0\\0&-1 \end{array}\right)~.
\end{equation}
These three matrices ${\bf t}_\mu$ $(\mu=0,1,2)$ satisfy the relations
\begin{equation}\label{tt=}
{\bf{t}}_\mu\,{\bf{t}}_\nu=\eta_{\mu\nu}\,{\bf I}+\epsilon_{\mu\nu}\,^\rho\,
{\bf{t}}_\rho~,
\end{equation}
where
$\eta_{\mu\nu}=\mbox{diag}(-1,1,1)$ and
$\epsilon_{\mu\nu\rho}$ is the Levi-Civita tensor with
$\epsilon_{012}=1$. The inner product defined by
$\langle\, {\bf t}_\mu\,{\bf t}_\nu\,\rangle\equiv\frac{1}{2}\,
\mbox{tr}({\bf t}_\mu\,{\bf t}_\nu)
=\eta_{\mu\nu}$
provides the isometry between $\mathfrak{sl}(2,\rr)$ and 3d Minkowski space.

As a basis in $\mathfrak{su}(2)$ we use
the anti-hermitian matrices ${\bf s}_n=i\boldsymbol{\sigma}_n$ $(n=1,2,3)$,
where $\boldsymbol{\sigma}_n$ are the Pauli matrices
($\boldsymbol{\sigma}_1={\bf t}_1,$
$\,\boldsymbol{\sigma}_2=-i{\bf t}_0$, $\,\boldsymbol{\sigma}_3={\bf t}_2$).
Here one has the
algebra
\begin{equation}\label{ss=}
{\bf s}_m\,{\bf s}_n=-\delta_{mn}\,{\bf I}-\epsilon_{mnl}\,{\bf s}_l~,
\end{equation}
and the inner product is introduced by a similarly normalized trace,
but with the negative sign $\langle{\bf s}_m\,{\bf s}_n\rangle
\equiv -\frac{1}{2}\,\mbox{tr}({\bf s}_m\,{\bf s}_n)=\delta_{mn}$. That
provides the isometry with $\rr^3$.

Using \eqref{su(1,1) basis}, eq. \eqref{g=Y} can be written as
$g=Y^{0'}\,{\bf I}+Y^\mu\,{\bf t}_\mu,$ $~h=X_4\,{\bf I}+X_n\,{\bf s}_n$.
The inverse group
elements respectively become
 $g^{-1}=Y^{0'}\,{\bf I}-Y^\mu\,{\bf t}_\mu,~$
$h^{-1}=X_4\,{\bf I}-X_n\,{\bf s}_n$.
With the help of \eqref{tt=} and  \eqref{ss=}  one then finds
the correspondence between the metric tensors
\begin{equation}\label{dg=dY,dh=dX}
\langle\, (g^{-1}\,dg)\,(g^{-1}\,d g)\rangle= dY\cdot dY~,~~~~~~
\langle\,(h^{-1}\,dh)\,(h^{-1}\,d h)\rangle= dX\cdot dX~.
\end{equation}
We use these relations in the next subsection to write
the string equations in terms of the group variables.

\subsection{String description in terms of group variables}

We consider space-like surfaces in $\mbox{AdS}_3 \times
\mbox{S}^3$. They can be parameterized by conformal complex
worldsheet coordinates $z=\frac{1}{2}(\s+i
\tau),~\bar{z}=\frac{1}{2}(\s-i\tau)$. In terms of group variables
we then get a pair of fields $g(z,\bar z)$ and $h(z,\bar z)$,
with $g\in \mbox{SL}(2,\rr)$ and $h\in \mbox{SU}(2)$.
Using the notation $\p\equiv\p_\s-i \p_\tau$, $\,\bar{\p}\equiv\p_\s+i \p_\tau$,
the conformal gauge conditions can be written as
\begin{equation}\label{conformal gauge}
\langle\, \left(g^{-1}\,\partial g\right)^2\,\rangle
+\langle\, \left(h^{-1}\,\partial h\,\right)^2\,\rangle=0=\langle
\left(g^{-1}\,\bar\partial g\,\right)^2\,\rangle+
\langle\,\left(h^{-1}\,\bar\partial h\right)^2\,\rangle~.
\end{equation}
The string action in this gauge corresponds to the sigma model on
$\mbox{SL}(2,\rr) \times \mbox{SU}(2)$
\begin{equation}\label{action}
S=\frac{\sqrt\lambda}{4\pi}\int \mbox{d}\sigma \mbox{d}\tau~
[\langle\, (g^{-1}\,\partial
g)\,(g^{-1}\,\bar\partial g)\rangle+
\langle\, (h^{-1}\,\partial
h)\,\,(h^{-1}\,\bar\partial h)\rangle]~,
\end{equation}
where $\lambda$ is a coupling constant.
The variation of \eqref{action} leads to the equations of motion
\begin{equation}\label{eq of motions}
\partial\left(g^{-1}\,\bar\partial g\right)+
\bar\partial\left(g^{-1}\,\partial g\right)=0~,\quad \quad \quad
\partial\left(h^{-1}\,\bar\partial h\right)+
\bar\partial\left(h^{-1}\,\partial h\right)=0~.
\end{equation}
From these equations follow the holomorphicity conditions
\begin{equation}\label{holomorphicity}
\bar\partial\langle\,\left(g^{-1}\,\partial g\right)^2\,\rangle
=0=\partial\langle \left(g^{-1}\,\bar\partial g\right)^2\,\rangle~,
\quad \bar\partial\langle\,\left( h^{-1}\,\partial h\right)^2\rangle
=0=\partial\langle \left(h^{-1}\,\bar\partial h\right)^2\,\rangle~,
\end{equation}
for the diagonal components of the induced metric tensors on $\mbox{SL}(2,\rr)$
and on $\mbox{SU}(2)$ separately.
These conditions, together with \eqref{conformal gauge},
allow to use the gauge
\begin{equation}\label{gauge fixing}
\langle\,\left(g^{-1}\,\partial g\right)^2\,\rangle
=-1=\langle\,\left(g^{-1}\,\bar\partial g \right)^2\rangle~,
\quad \quad \langle \left(h^{-1}\,\partial h\right)^2\,\rangle
=1=\langle \left(h^{-1}\,\bar\partial h\right)^2\,\rangle~.
\end{equation}
The remaining freedom of conformal transformations
$z\mapsto f(z)$ in this gauge is given by translations
$z\mapsto z+z_0$ and the reflection $z\mapsto -z$. One can
also consider  $z\mapsto -\bar z$,
which corresponds to the
reflection $\s\mapsto -\s$.

In the real worldsheet coordinates $(\sigma,\tau)$
the equations of motion \eqref{eq of motions} read
\begin{equation}\label{eq of motions 1}
\partial_\sigma\left(g^{-1}\,\partial_\sigma g\right)+
\partial_\tau\left(g^{-1}\,\partial_\tau g\right)=0~,\quad \quad\quad
\partial_\sigma\left(h^{-1}\,\partial_\sigma h\right)+
\partial_\tau\left(h^{-1}\,\partial_\tau h\right)=0~,
\end{equation}
and the gauge fixing conditions \eqref{gauge fixing} are equivalent to
\be\ba\label{gauge fixing 1}
\langle\,\left(g^{-1}\,\partial_\tau g\right)^2\,\rangle-
\langle\,\left(g^{-1}\,\partial_\sigma g\right)^2\,\rangle
=&1~,& \quad
\langle\,\left(h^{-1}\,\partial_\sigma h\right)^2\,\rangle-
\langle\,\left(h^{-1}\,\partial_\tau h\right)^2\,\rangle=&1~,\\[0.2cm]
\langle\,\left(g^{-1}\,\partial_\sigma g\right)\,
\left(g^{-1}\,\partial_\tau g\right)\,\rangle=&0~,& \quad
\langle\,\left(h^{-1}\,\partial_\sigma h\right)\,
\left(h^{-1}\,\partial_\tau h\right)\,\rangle=&0~.
\ea\ee
The isometry transformations of the group manifolds are given
by the left-right
multiplications
\begin{equation}\label{l-r}
g\mapsto g_{_L}\,g\,g_{_R}~,\quad \quad h\mapsto h_{_L}\,h\,h_{_R}~,
\end{equation}
with constant matrices $g_{_L},$ $g_{_R}$ $\in \mbox{SL}(2,\rr)$ and
$h_{_L},$ $h_{_R}$ $\in \mbox{SU}(2).$
They leave the equations of motion \eqref{eq of motions 1} and
the gauge fixing conditions \eqref{gauge fixing 1} invariant.
The system \eqref{eq of motions 1}-\eqref{gauge fixing 1}
is also invariant under the discrete
transformations $g\mapsto g^{-1}$ and $h\mapsto h^{-1}$, which correspond
to the reflections of
$Y^0$, $Y^1$, $Y^2$ and $X^1$, $X^2,$ $X^3$, respectively. Hence,
the composition of $h\mapsto h_{_L}\,h\,h_{_R}$ and $h\mapsto h^{-1}$ form
the complete group of isometry transformations of  $\mbox{SU}(2).$
The complete isometry group of $\mbox{SL}(2,\rr)$ is obtained as a composition
of $g\mapsto g_{_L}\,g\,g_{_R}$, $g\mapsto g^{-1}$ and $g\mapsto {\bf t}_1\,g\,{\bf t}_1$,
where the latter corresponds to the reflection of $Y^0$ and $Y^2$.

Using the covariant notation:
$(\sigma,\tau)=(\xi^1,\xi^2)$, $\p_a=\p_{\xi^a}$ $(a=1,2)$,
the induced metric tensors on the $\mbox{SL}(2,\rr)$ and
$\mbox{SU}(2)$ projections can be written as\footnote{In this paper
(as in \cite{Dorn:2009hs}) the
index $s$ is used for some variables of the spherical part to distinguish
them from similar variables of the AdS part.}
\be\label{induced metric A-S}
f_{ab}=\langle\,\left(g^{-1}\,\p_a g\right)
\left(g^{-1}\,\p_b g\right)\,\rangle~,\quad (f_s)_{ab}=\langle\,\left(h^{-1}\,\p_a h\right)
\left(h^{-1}\,\p_b h\right)\,\rangle~.
\ee
In the next two subsections we construct solutions corresponding to constant $f_{ab}$ and $(f_s)_{ab}$.
First we consider the $\mbox{SU}(2)$ projection, since it is easier to treat. The same method then we
apply to the $\mbox{SL}(2,\rr)$ part.

\subsection{Vacuum solutions in $\mbox{SU}(2)$}

Let us introduce $\mathfrak{su}(2)$ valued fields related
to the right derivatives of $h$
\begin{equation}\label{R}
R_\sigma=h^{-1}\,\partial_\sigma h~,\quad \quad \quad
R_\tau=h^{-1}\,\partial_\tau h~.
\end{equation}
These fields obey the zero curvature condition
\begin{equation}\label{dR=}
\p_\tau R_\sigma-\p_\sigma R_\tau=[R_\sigma,\,R_\tau]~,
\end{equation}
and the second equation in  \eqref{eq of motions 1} is equivalent to
\begin{equation}\label{dR=1}
\p_\sigma R_\sigma+\p_\tau R_\tau=0~.
\end{equation}
The norms and the scalar product of $R_\sigma$ and $R_\tau$
define the induced metric tensor on the $\mbox{SU}(2)$ projection.
We denote the norm of $R_\tau$ by $\rho_s$, and write the $\mbox{SU}(2)$
part of the gauge fixing conditions \eqref{gauge fixing 1} in the form
\begin{equation}\label{R-norms}
 \langle\,R_\sigma\,R_\sigma\,\rangle=1+\rho_s^2~,\quad \quad
 \langle\,R_\tau\,R_\tau\,\rangle=\rho_s^2~, \quad \quad
 \langle\,R_\sigma\, R_\tau\,\rangle=0~.
\end{equation}
Our aim is to describe solutions of \eqref{R}-\eqref{R-norms}
for constant $\rho_s$.
The vectors $R_\sigma$, $\,R_\tau$ and $[R_\sigma,R_\tau]$ form
an orthogonal basis in $\mathfrak{su}(2)$.
Expanding the first derivatives of $R_\sigma$ and $R_\tau$
in this basis,
from \eqref{dR=}-\eqref{R-norms} one finds that they have vanishing
projections on $R_\sigma$ and $\,R_\tau$. Therefore, the first
derivatives can be written as\footnote{The coefficients on the r.h.s
are interpreted as the matrix elements of the second fundamental form.}
\be\ba
\label{partial R}
\p_\sigma R_\sigma&=a\,[R_\sigma,R_\tau]~,& \p_\tau R_\sigma &=b\,[R_\sigma,R_\tau]~, \\[0.2cm]
\p_\sigma R_\tau &=(b-1)\,[R_\sigma,R_\tau]~,&\p_\tau R_\tau&=-a\,[R_\sigma,R_\tau]~.
\ea\ee
The consistency conditions of this system
are given by the equations
\begin{equation}\label{consist-cond}
\p_\sigma b=\p_\tau a~, \quad \quad\p_\tau b=-\p_\sigma a~,\quad \quad
a^2+b^2-b=0~.
\end{equation}
From the first two equations follows that the coefficients
$a$ and $b$ are
harmonic functions $a=f_1(z)+\bar f_1(\bar z)$, $\,\,b=f_2(z)+\bar f_2(\bar z)$,
with $f_2'(z)=if_1'(z)$.
Then, the third equation of \eqref{consist-cond} leads to
$f_1'(z)=0$, which means that $a$ and $b$
are $(\sigma,\tau)$-independent.
Due to the algebraic relation \eqref{consist-cond},
the coefficients $a$ and $b$
are parameterized by one angle variable
\be\label{a,b=}
a=\sin\phi_s\,\cos\phi_s~, \quad\quad b=\cos^2\phi_s~,\quad \quad
\phi_s\in [0,\pi)~.
\ee
As a result, the following linear combination
\begin{equation}\label{R=}
 R_0=\sin\phi_s\,R_\sigma +\cos\phi_s\,R_\tau
\end{equation}
is constant.

Now we introduce $\mathfrak{su}(2)$ valued fields with the left derivatives
\begin{equation}\label{L}
L_\sigma=\partial_\sigma h\,h^{-1}~, \quad \quad
L_\tau=\partial_\tau h\,h^{-1}~,
\end{equation}
which are related to the right fields in a standard way
\begin{equation}\label{L=R}
L_\sigma=hR_\sigma h^{-1}~,\quad \quad L_\tau=hR_\tau h^{-1}~.
\end{equation}

The differentiations of \eqref{L=R} lead to a system
similar to  \eqref{partial R}
\be\ba\label{partial L}
\p_\sigma L_\sigma=&a\,[L_\sigma, L_\tau]~,& \p_\tau L_\sigma &=(b-1)\,[L_\sigma,L_\tau]~,&\\[0.2cm]
\p_\sigma L_\tau =&b\,[L_\sigma,L_\tau]~,& \p_\tau L_\tau&=-a\,[L_\sigma,L_\tau]~,
\ea\ee
and one finds that the linear combination
\begin{equation}\label{L=}
 L_0=\cos\phi_s\,L_\sigma-\sin\phi_s\,L_\tau
\end{equation}
is also $(\sigma,\tau)$-independent.

From \eqref{R=} and \eqref{L=} follows that the field $h$
satisfies the equations
\begin{equation}\label{dh=}
\frac{\p h}{\p l_s}= L_0\,h~,\quad \quad \quad \quad \frac{\p h}{\p r_s}=
h\,R_0~,
\end{equation}
where the new coordinates  $(l_s, r_s)$  are related to
$(\sigma,\tau)$ by the rotation
\be\label{l,r=s,t}
l_s=\cos\phi_s\,\,\s-\sin\phi_s\,\,\tau~,
\quad
r_s=\sin\phi_s\,\,\s+
\cos\phi_s\,\,\tau~.
\ee

The integration of the system \eqref{dh=} is straightforward and yields
\begin{equation}\label{h0}
h=e^{l_s\,L_0}\,h_0\,e^{r_s\,R_0}~,
\end{equation}
with an integration constant $h_0\in\mbox{SU}(2)$.
The isometry transformation $h\mapsto h\,h_0^{-1}$ brings the solution
to the form $h=e^{l_s\,L_0}\,e^{r_s\,R_1},\,$
with $R_1=h_0\,R_s\,h_0^{-1}$.
Denoting $L_0$ by $L_s$ and $ R_1$  by $R_s$, we obtain
\begin{equation}\label{solution h}
  h=e^{l_s\,L_s}\,e^{r_s\, R_s}~.
\end{equation}
This field indeed solves equation \eqref{dR=1}
for any pair of constant vectors $L_s$ and $R_s$.

To verify the orthonormality conditions \eqref{R-norms}
we calculate the induced metric tensor in the coordinates $\,(l_s,r_s)\,$.
From \eqref{solution h} follows its matrix form
\begin{equation}\label{S^3 metric}
(f_s)_{a b}(l_s, r_s)=\left( \begin{array}{cr}
  \langle\,L_s\,L_s\,\rangle&\langle L_s\, R_s\,\rangle\\[0.2cm]
\langle\, L_s\, R_s\,\rangle&\langle R_s\, R_s\,\rangle \end{array}\right)~.
\end{equation}
On the other hand, the map \eqref{l,r=s,t} defines the same tensor as
\begin{equation}\label{S^3 metric 1}
M\,\left( \begin{array}{cr}
1+ \rho_s^2&0~\\[0.1cm]
0& \rho_s^2\end{array}\right)\,M^T =\left( \begin{array}{cr}
 \rho_s^2+\cos^2\phi_s&\sin\phi_s\cos\phi_s\\[0.2cm]
\sin\phi_s\cos\phi_s& \rho_s^2+\sin^2\phi_s\end{array}\right)~,
\end{equation}
where $M$ is the rotation matrix in \eqref{l,r=s,t}.
Comparing \eqref{S^3 metric} and \eqref{S^3 metric 1}, one finds
\be\label{Ls,Rs}
\langle\,L_s\,L_s\,\rangle=\rho_s^2+\cos^2\phi_s~, \quad
\langle\,R_s\,R_s\,\rangle=\rho_s^2+\sin^2\phi_s~, \quad
\langle\,L_s\,R_s\,\rangle=\sin\phi_s\,\cos\phi_s~.
\ee
The obtained norms of $L_s$ and $R_s$ are consistent
with \eqref{L=} and \eqref{R=}. A new result from this calculation is
the scalar product $\langle\,L_s\,R_s\,\rangle$, which defines the
angle between these vectors.

\begin{figure}
\centering{
\begin{tabular}{ccc}
\includegraphics[height=5cm]{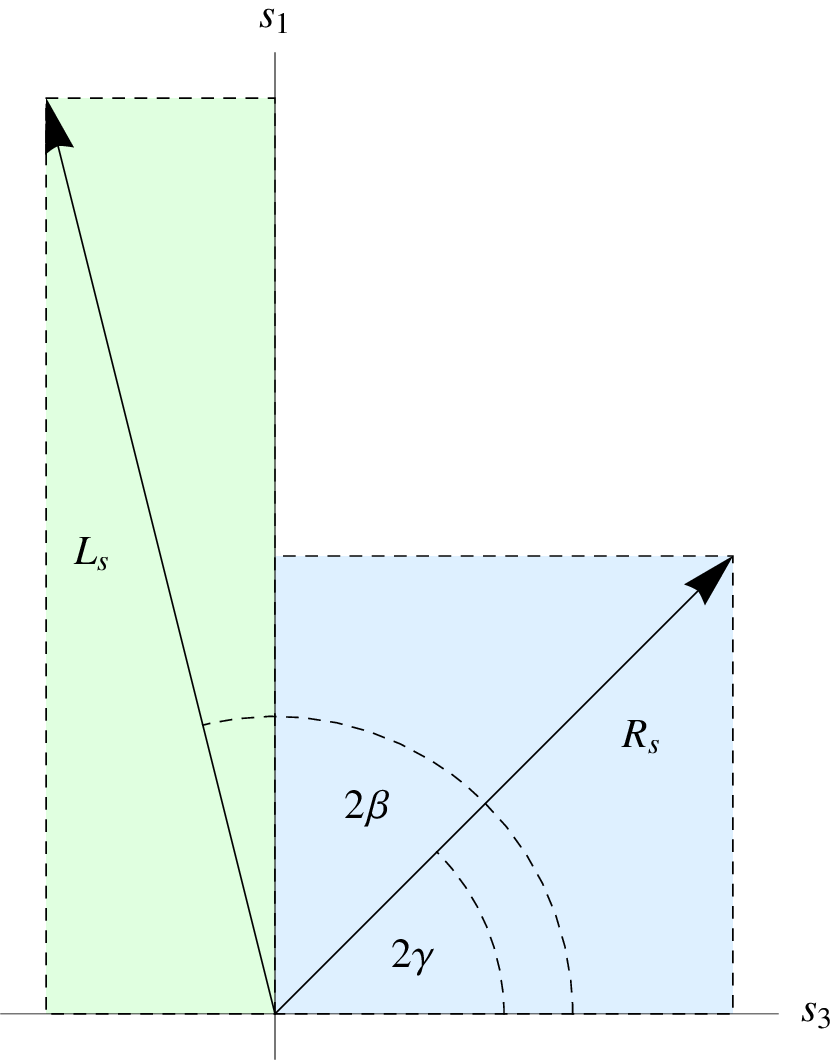}&\hspace{2cm}&\includegraphics[height=5cm]{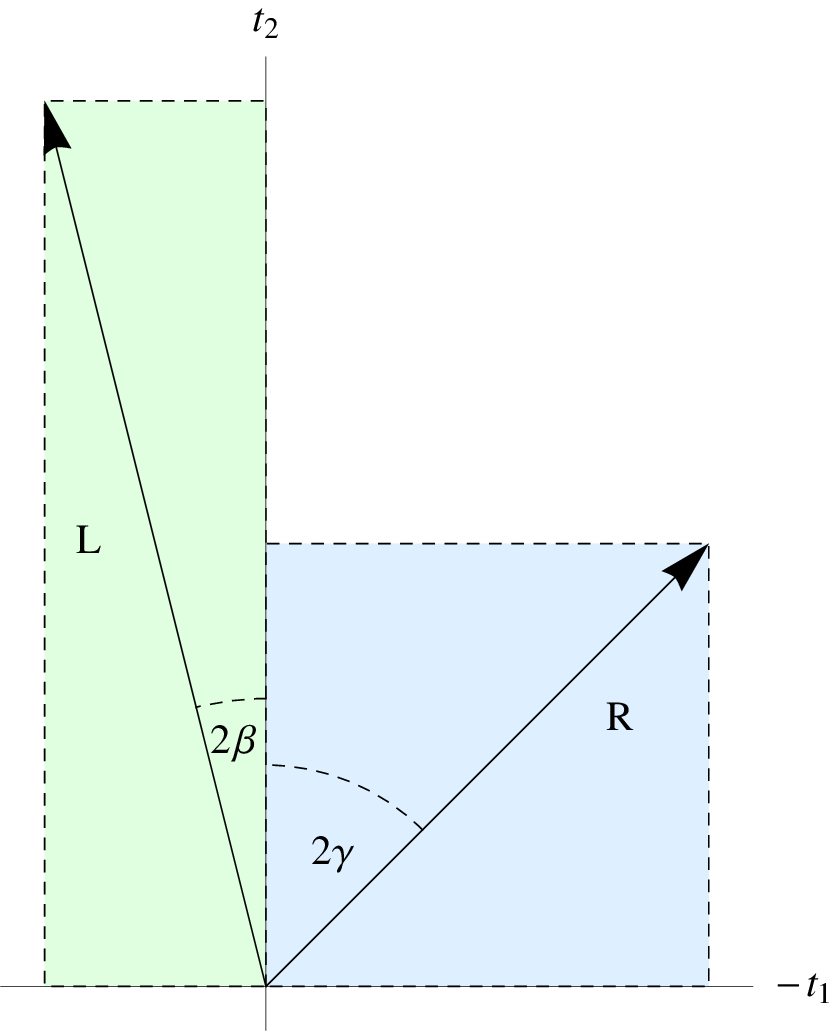}\\
(a)&&(b)
\end{tabular}}
\caption{The pair $(L_s,\, R_s)$ in the first plot is identified with
two vectors in $\rr^3$ located in the plane $(3,1)$. Similarly, $L$ and $R$ in the second plot
are 3d Minkowski vectors with vanishing time components.}
\label{s1s3}
\end{figure}

Now we describe how to factorize the left-right symmetry \eqref{l-r}
and parameterize the $\mbox{SU}_L(2)\otimes\mbox{SU}_R(2)$ orbits
of solutions by the pairs $(\rho_s, \phi_s)$.

Since the general case \eqref{h0} reduces to \eqref{solution h}
by  right (or left) multiplications, it is enough to classify
the fields \eqref{solution h}.
This factorized form of solutions is invariant under
the similarity transformations $h\mapsto h_{_L}\,h\,h_{_L}^{-1}$, which
rotate the vectors $L_s$ and $R_s$ by the adjoint representation:
$L_s \mapsto h_{_L}\, L_s\,h_{_L}^{-1},$
$\,R_s \mapsto h_{_L}\,R_s\,h_{_L}^{-1}$.
The adjoint representation is equivalent to $\mbox{SO}(3)$
and it is convenient to interpret $L_s$ and $R_s$ as vectors in $\rr^3$.
Using this picture, one can easily bring any pair $(L_s,\, R_s)$
to some canonical form defined only by the scalar invariants
\eqref{Ls,Rs}. Note that the commutator $[L_s,\, R_s]$ transforms also
by the adjoint representation,
its norm is given by $\left\langle [L_s,\, R_s]^2\right\rangle=4\rho_s^2(1+\rho_s^2)$
and one can use this vector to specify a canonical pair $(L_s,\, R_s)$.
We choose the pair
\begin{equation}\label{Ls,Rs new}
L_s =-\rho_s\,\sin\phi_s\,{\bf s}_3+
\sqrt{1+\rho_s^2}\,\cos\phi_s\,{\bf s}_1~,
\quad \quad
R_s =\rho_s\,\cos\phi_s\,{\bf s}_3
+\sqrt{1+\rho_s^2}\,\sin\phi_s\,{\bf s}_1~.
\end{equation}
It corresponds to
\begin{equation}\label{commutator}
 [L_s,\, R_s]=2\rho_s\,\sqrt{1+\rho_s^2}\,{\bf s}_2~,
\end{equation}
and, in addition, the vectors $L_s$ and $R_s$ are oriented in some symmetric
way in the $({\bf s}_3,\,{\bf s}_1)$ plane. Namely,
the two rectangles formed by the components of the vectors $L_s$ and $R_s$
have the same area (see fig.~\ref{s1s3}(a)).
The choice \eqref{Ls,Rs new} is motivated by its relevance for the generalization to the $\mbox{SL}(2,\rr)$ case.
It is also helpful to establish contact with Pohlmeyer reduction.

The calculation of the exponents $e^{l_s L_s} $ and $e^{r_s R_s}$ corresponding
to \eqref{Ls,Rs new} is straightforward. However, in general,
the solution \eqref{solution h} has a rather complicated matrix form. One can simplify it
by an isometry transformation. For this purpose we rewrite \eqref{Ls,Rs new} as
\begin{equation}\label{Ls,Rs=}
 L_s=|L_s|\,e^{-i\beta\,\boldsymbol{\s}_2}\,{\bf s}_3\,
e^{i\beta\,\boldsymbol{\s}_2}~,\quad\quad
 R_s=|R_s|\,e^{-i\gamma\,\boldsymbol{\s}_2}\,{\bf s}_3\,
e^{\gamma\,\boldsymbol{\s}_2}~,
\end{equation}
where $(|L_s|,\,2\beta)$ and $(|R_s|,\,2\g)$ are the polar coordinates
in the  $({\bf s}_3,\,{\bf s}_1)$ plane for the vectors $L_s$ and $R_s$, respectively
(see fig.~\ref{s1s3}(a)). Using then \eqref{e^ue^v} and multiplying the solution \eqref{solution h} by $e^{i\b\,\boldsymbol{\s}_2}$
from the l.h.s and by $e^{-i\g\,\boldsymbol{\s}_2}$ from the r.h.s.,
we obtain the new solution
\begin{equation}\label{solution h 1}
h=e^{i\tilde l_s\,\boldsymbol{\s}_3}\,e^{i\theta_s\,\boldsymbol{\s}_2}\,
e^{i\tilde r_s\,\boldsymbol{\s}_3}=
\left( \begin{array}{cr}
  \cos\theta_s\,e^{i\xi_s}&\sin\theta_s\,e^{i\eta_s}~\\[0.2cm]
-\sin\theta_s\,e^{-i\eta_s}
& ~\cos\theta_s\,e^{-i\xi_s}~\end{array}\right)~.
\end{equation}
Here $\tilde l_s$ and $\tilde r_s$ are the rescaled worldsheet coordinates
\be\label{tilde l,r}
\tilde l_s=|L_s|\,\,l_s~,\quad\quad
\tilde r_s=|R_s|\,r_s~,
\ee
$\xi_s=\tilde l_s+\tilde r_s,$ $\eta_s=\tilde l_s-\tilde r_s$
and $2\theta_s=2\beta-2\gamma$ corresponds to the angle between
$L_s$ and $R_s$
\be\label{theta-s}
\cos2\theta_s=\frac{\left\langle L_s\,R_s\right\rangle}
{|L_s|\,|R_s|}~,
\qquad \qquad \qquad 2\theta_s\in(0,\pi)~.
\ee
By \eqref{Ls,Rs} one has
\be\label{cot th}
\cot 2\theta_s=\frac{\sin\phi_s\,\cos\phi_s}{\rho_s\sqrt{1+\rho_s^2}}~.
\ee

The embedding coordinates for the solution \eqref{solution h 1}
are given by
\begin{eqnarray}\label{solution X}
X_1=\sin\theta_s\,\sin\eta_s~,\quad
X_2=\sin\theta_s\,\cos\eta_s~,\quad
X_3=\cos\theta_s\,\sin\xi_s~,\quad
X_4=\cos\theta_s\,\cos\xi_s~.
\end{eqnarray}
The pairs ($X_1$, $X_2$) and
($X_3$, $X_4$) here are on the circles
\begin{equation}\label{quadratic}
X_1^2+X_2^2=\sin^2\theta_s~, \quad \quad  X_3^2+X_4^2=\cos^2\theta_s~.
\end{equation}
Therefore, eq. \eqref{solution X} describes a torus in $S^3\subset \rr^4$
with the radii
$\sin\theta_s$ and $\cos\theta_s$.

The solution \eqref{solution X} was obtained in
\cite{Dorn:2009hs}
via Pohlmeyer reduction. It was shown
that the mean curvature of the embedding in $\mbox{S}^3$ is equal to
\begin{equation}
H_s = \cot2\theta_s~.
\end{equation}
From \eqref{solution h} follows the equation
\begin{equation}\label{quadratic relation}
\langle\, L_s\,h\,R_s\,h^{-1}\,\rangle=\langle\, L_s\, R_s\,\rangle~,
\end{equation}
which defines a surface in $\mbox{SU}(2)$ in a form
independent of worldsheet coordinates.
Since $h$ and $h^{-1}$ are linear functions of the
embedding coordinates,
the l.h.s. of \eqref{quadratic relation} provides a quadric in these variables.
The corresponding surface expressed in $\rr^4$ embedding coordinates is given as an intersection of the $S^3\subset\rr^4$ and
\be
X_4^2\langle L_sR_s\rangle +X_4X_n\langle L_s[s_n,R_s]\rangle -X_nX_m\langle L_ss_m
R_ss_n\rangle ~=~\langle L_sR_s\rangle ~.
\ee
Taking $L_s$ and $R_s$ from \eqref{Ls,Rs=},
after simple transformations one can rewrite \eqref{quadratic relation}
in the form
\begin{equation}\label{quadratic relation 1}
\langle\,{\bf s}_3\,h\,{\bf s}_3\,h^{-1}\,\rangle
=\cos2\theta_s~,
\end{equation}
which reproduces \eqref{quadratic}.
Though this surface in $\mbox{SU(2)}$ is described only by $\theta_s$, this parameter
is not a complete characteristic of the $\mbox{SU(2)}$ part for our system in
$\mbox{SL}(2,\rr)\times\mbox{SU(2)}$. Since we are in a fixed gauge,
$\rho_s$ also has a certain gauge invariant meaning.
Namely, $\rho_s\sqrt{1+\rho_s^2}\,d\sigma\,d\tau$
defines the area measure on the surface induced from $\mbox{S}^3$.

Finally, note that $L_s$ and $R_s$ are related to the Noether integrals
of the isometry group and therefore they have  also a gauge invariant meaning.

\subsection{Vacuum solutions in  $\mbox{SL}(2,\rr)$}

Based on the isometry between the  $\mathfrak{sl}(2,\rr)$ algebra and
3d Minkowski space,
we call a vector $\mathfrak{a}\in \mathfrak{sl}(2,\rr)$ space-like
if $\langle \mathfrak{a}\, \mathfrak{a}\rangle>0$,
time-like if $\langle\mathfrak{a}\, \mathfrak{a}\rangle<0$ and
light-like if $\langle \mathfrak{a}\, \mathfrak{a}\rangle=0$.

Note that the vector  $g^{-1}\p_\tau g$ is space-like in
the gauge \eqref{gauge fixing 1}.
Similarly to the $\mbox{SU}(2)$ case, we introduce the notation
$\rho^2\equiv\langle(g^{-1}\p_\tau g)^2\rangle$
and write the induced metric tensor on the $\mbox{SL}(2,\rr)$ projection
in the form\footnote{\label{foot4}This form differs from the one used in
\cite{Dorn:2009hs}. There the $(\tau\tau)$-component of the induced
metric tensor was denoted by $1+\rho^2$ for the space-like surfaces
and by $1-\rho^2$ for the time-like ones. Note also, that in our convention $\sigma$ is time-like in the time-like case.}
\begin{equation}\label{AdS3 metric}
f_{a b}(\s, \tau)=\left( \begin{array}{cr}
  \rho^2-1&0~\\
0& \rho^2\end{array}\right)~.
\end{equation}
The sum with the $\mbox{SU}(2)$ metric (\ref{R-norms}) gives the total metric induced from $\mbox{SL}(2,\rr)\times \mbox{SU}(2)$ and is by construction a multiple of the identity matrix. The metric (\ref{AdS3 metric}) is space-like if $\rho^2>1,\,$ it is time-like if $0 < \rho^2<1\,$
and it becomes light-like if $\rho^2=1$.
The last case corresponds
to a degenerate metric with a light-like vector $g^{-1}\p_\sigma g$.

We consider constant $\rho^2$ and apply the scheme of
the previous subsection. The case of light-like $R_\s$ is special and we consider it separately.
If $R_\s$ is space-like or time-like,  the vectors $R_\s,$  $\,R_\tau\,$ and $\,[R_\s,\,R_\tau]$ form
an orthonormal basis in $\mathfrak{sl}(2,\rr)$. Repeating then the same steps as in $\mbox{SU}(2)$
we come to the factorized type
solution \eqref{solution h}
\begin{equation}\label{solution g}
  g=e^{l\,L}\,e^{r\,R}~.
\end{equation}
Here $(L,\,R)$ is a pair of constant elements of the $\mathfrak{sl}(2,\rr)$ algebra
and ($l, r)$ are  worldsheet coordinates obtained from
$(\sigma,\tau)$ by a rotation on an angle $\phi$ which parameterizes the coefficients
of the linear system  (see \eqref{partial R})
\be\label{SL-a,b}
a=\frac{\langle\,\p_\s R_\s\,[R_\s, R_\tau]\,\rangle}
{\langle\,[R_\s,R_\tau]^2\,\rangle}~, \qquad
b=\frac{\langle\,\p_\tau R_\s\,[R_\s, R_\tau]\,\rangle}
{\langle\,[R_\s,R_\tau]^2\,\rangle}~.
\ee
Now $R_\s=g^{-1}\p_\s g,$ $R_\tau=g^{-1}\p_\tau g\,$ and $\,a,$ $b$ satisfy again \eqref{consist-cond}.
For further convenience, we use here a different parameterization
\be\label{SL-a,b=}
a=-\sin\phi\,\cos\phi~, \quad\quad b=\sin^2\phi~,\quad\quad \phi\in (-\pi/2,\,\pi/2],
\ee
which is obtained from \eqref{a,b=} by the replacements $\cos\phi_s\mapsto \sin\phi,\,$
$\,\sin\phi_s \mapsto -\cos\phi$.
This provides the following $(l,\, r)$ coordinates (compare with \eqref{l,r=s,t})
\be\label{l,r=s,t a}
l=\cos\phi\,\,\tau+\sin\phi\,\,\s~,
\quad
r=\sin\phi\,\,\tau-
\cos\phi\,\,\s~.
\ee

If $R_\s$ is light-like, one gets the commutator\footnote{Note that \eqref{comm, rho=1} is equivalent to $[(\bf{t}_1-\bf{t}_0),\,\bf{t}_2]=2(\bf{t}_1-\bf{t}_0) $.}
\be\label{comm, rho=1}
[R_\s,\,R_\tau]= 2\,R_\s~.
\ee
Hence, in this case, $R_\s,$  $\,R_\tau\,$ and $\,[R_\s,\,R_\tau]$ do not form a basis in $\mathfrak{sl}(2,\rr)$.
Completing $R_{\sigma}$ and $R_{\tau}$ in a suitable way to a basis, one can
show that in the end the third independent direction is not needed to express
the derivatives of $R_{\sigma}$ and $R_{\tau}$. For details see the parallel
discussion of this issue in the framework of Pohlmeyer reduction in section 3.3 below.
Altogether the system  (\ref{partial R}) is
modified and replaced by the linear system
\be\ba
\label{partial R AdS}
\p_\sigma R_\sigma&=2a\,R_\sigma~,& \p_\tau R_\sigma &=2b\,R_\sigma~, \\[0.2cm]
\p_\sigma R_\tau &=2(b- 1)\,R_\sigma~,&\p_\tau R_\tau&=-2a\,R_\sigma~.
\ea\ee
The consistency conditions for this system lead to the equations
\begin{equation}\label{consist-cond AdS}
\p_\tau a-\p_\sigma b=0~, \quad \quad \p_\sigma a+\p_\tau b+2(a^2+b^2-b)=0~,
\end{equation}
which, in contrast to other cases, do not necessarily require constant $a$ and $b$.
In section 3.3  we describe also the general solution of the consistency conditions \eqref{consist-cond AdS} 
and indicate how to integrate the linear system \eqref{partial R AdS}. 
Here we identify  vacuum configurations with the solutions for
constant $a$ and $b$. In this case \eqref{consist-cond AdS} reduces again to $a^2+b^2-b=0$ and one can use the parameterization \eqref{SL-a,b=}. Constant currents $(L,\,R)$ are then constructed in a same way and one again
obtains solutions in the factorized form \eqref{solution g}.
Thus, \eqref{solution g} represents solutions in the vacuum sector for all three cases: $\rho^2>1,\,$ $\rho^2<1$ and
$\rho^2=1$ as well.

The induced metric tensor in $(l,r)$-coordinates, calculated
by \eqref{AdS3 metric} and \eqref{l,r=s,t a}, becomes
\begin{equation}\label{AdS metric 1}
f_{ab}(l,r)=\left( \begin{array}{cr}
 \rho^2-\sin^2\phi&\sin\phi\cos\phi\\[0.1cm]
\sin\phi\cos\phi& \rho^2-\cos^2\phi\end{array}\right),
\end{equation}
and comparing it with the calculation from \eqref{solution g},
one finds
\be\label{La,Ra}
\langle\,L\,L\,\rangle=\rho^2-\sin^2\phi~, \quad
\langle\,R\,R\,\rangle=\rho^2-\cos^2\phi~, \quad
\langle\,L\,R\,\rangle=\sin\phi\,\cos\phi~.
\ee
These equations fix the norm of the commutator $[L,\,R]$ to
\begin{equation}\label{|[L,R]|}
 \left\langle\, [L,\,R]^2\,\right\rangle =4\rho^2(1-\rho^2)~.
\end{equation}
Thus, $[L,\,R]$ is time-like for space-like surfaces $(\rho^2>1)$,
$[L,\,R]$ is light-like for light-like surfaces
($\rho^2=1$) and  $[L,\,R]$ is space-like for time-like surfaces $(\rho^2<0)$.

We use these metric characteristics
of the commutator $[L,R]$ and also the scalars \eqref{La,Ra} to classify the fields $g$
by the adjoint orbits similarly to the previous subsection.
Since the adjoint representation of $\mbox{SL}(2,\rr)$ is given by the
group of proper Lorentz transformations $\mbox{SO}_\uparrow(1,2)$,
the orbits can be identified with
the hyperbolas and cones in 3d Minkowski space.
Choosing canonical pairs $(L, R)$ from the orbits, one can construct
solutions by \eqref{solution g} and then simplify them as in \eqref{solution h 1}.
This procedure also simplifies the form of the quadratic relation between the $\rr^{2,2}$
embedding coordinates, which is now given by
\begin{equation}\label{quadratic AdS}
\langle\, L\,g\,R\,g^{-1}\,\rangle~=(Y^{0'})^2\langle LR\rangle+Y^{\mu}Y^{0'}\langle L[t_{\mu},R]\rangle-Y^{\mu}Y^{\nu}\langle Lt_{\mu}Rt_{\nu}\rangle~=\langle\, L\, R\,\rangle~.
\end{equation}
It is natural to divide the solutions in three classes according to the signature
of the induced metric tensor and then continue classification inside the classes.

Before starting classification it is useful to note that
the reflection freedom  $\s\mapsto -\s$ allows to set $a\leq0$ (see eqs. \eqref{SL-a,b}-\eqref{SL-a,b=} and \eqref{partial R AdS}).
This corresponds to $\phi\in [0,\pi/2]$. Below we assume this condition.
\vspace{6mm}

\noindent
{\bf 1}. {\bf Space-like surfaces $\,\,(\rho^2>1)$}.
\vspace{2mm}\\*
This case corresponds to space-like $L$, $R$ and time-like $[L,\,R]$. One can take the commutator $[L,\,R]$ proportional to ${\bf t}_0$.
However, in contrast to the $\mbox{SU}(2)$ case there are two possibilities
\begin{equation}\label{[L,R] 1}
[L,\,R]=\pm 2\rho\sqrt{\rho^2-1}\,{\bf t}_0~,
\end{equation}
which are not on the same adjoint orbit. These two cases are related by the
discrete isometry transformation $g\mapsto {\bf t}_1\,g\,{\bf t}_1$.

Let us consider the negative sign in \eqref{[L,R] 1} and choose the pair
\begin{equation}\label{L,R new}
L=\rho \cos\phi \,{\bf t}_2+\sqrt{\rho^2-1}\,\sin\phi\, {\bf t}_1~,
\quad \quad
R=\rho\sin\phi\,{\bf t}_2-\sqrt{\rho^2-1}\,\cos\phi\, {\bf t}_1~.
\end{equation}
Using the same trick as in \eqref{Ls,Rs=}, based now on
\eqref{e^t1 t2 e^t1}, we first rewrite \eqref{L,R new} in the form
\begin{equation}\label{L,R new 1}
L=|L|\,e^{-\b\,{\bf t}_0}\,{\bf t}_2\,e^{\b\,{\bf t}_0}~,
\quad \quad
R=|R|\, e^{-\g\,{\bf t}_0}\,{\bf t}_2\,e^{\g\,{\bf t}_0}~,
\end{equation}
where $|L|=\sqrt{\rho^2-\sin^2\phi},$ $\,|R|=\sqrt{\rho^2-\cos^2\phi},$
and  ($2\b,\,$ $2\g$) are the polar angles in the $({\bf t}_2,{\bf t}_1)$ plane,
(see fig.~\ref{s1s3}(b)). Then, as in \eqref{solution h 1}, the left-right multiplications reduce
the solution \eqref{solution g} to
\begin{equation}\label{solution g 1}
  g=e^{\tilde l\,{\bf t}_2}\,e^{\theta\,{\bf t}_0}\,
  e^{\tilde r\,{\bf t}_2}=\left( \begin{array}{cr}
  \cos\theta\,e^{\xi}&\sin\theta\,e^{\eta}~\\[0.1cm]
-\sin\theta\,e^{-\eta}
& ~\cos\theta\,e^{-\xi}\end{array}\right)~.
\end{equation}
Here  $\tilde l=|L|\,l$ and
$\tilde r=|R|\,r$
are the rescaled worldsheet coordinates,
$\,\xi=\tilde l+\tilde r,$  $\eta=\tilde l-\tilde r$ and
$2\theta=2\b-2\g$ is the angle between the vectors \eqref{L,R new},
with
\be\label{theta 1}
\cot2\theta=\frac{\sin\phi\,\cos\phi}
{\rho\sqrt{\rho^2-1}}~,\qquad\qquad \th\in(0,\pi/4]~.
\ee

The embedding coordinates for the solution \eqref{solution g 1} are given by
\begin{eqnarray}\label{solution Y 1}
Y^{0'}=\cos\theta\,\cosh\xi~,\quad
Y^0=\sin\theta\,\cosh\eta~,\quad
Y^1=\sin\theta\,\sinh\eta~,\quad
Y^2=\cos\theta\,\sinh\xi~.
\end{eqnarray}
This surface was also obtained in \cite{Dorn:2009hs}
by  Pohlmeyer reduction. Its mean curvature is
\begin{equation}
H= \cot 2 \theta ~.
\end{equation}
Similarly to \eqref{quadratic relation 1}, the quadric \eqref{quadratic AdS} reduces to
\begin{equation}\label{quadratic AdS 1}
(Y^{0})^2-(Y^{1})^2=\sin^2\theta~,
\end{equation}
and for $\theta=\pi/4$ it reproduces
the four cusp surface of \cite{Kruczenski:2002fb, Alday:2007hr}.

\begin{figure}
\centering
\begin{tabular}{ccc}
\includegraphics[scale=1]{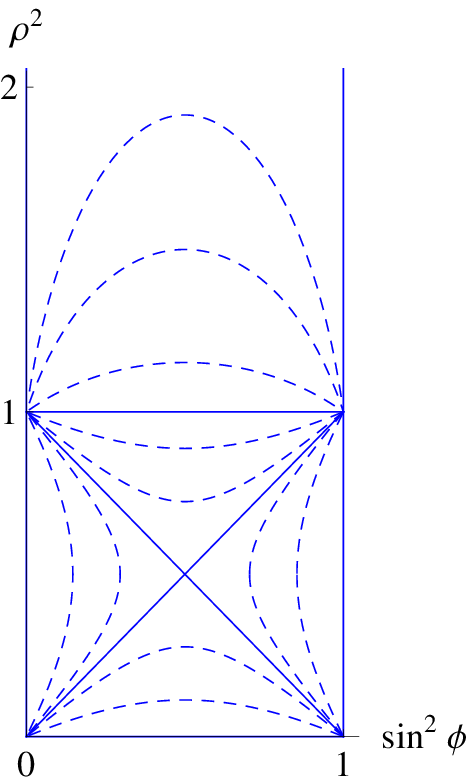}
&\hspace{50pt}&
\includegraphics[scale=1]{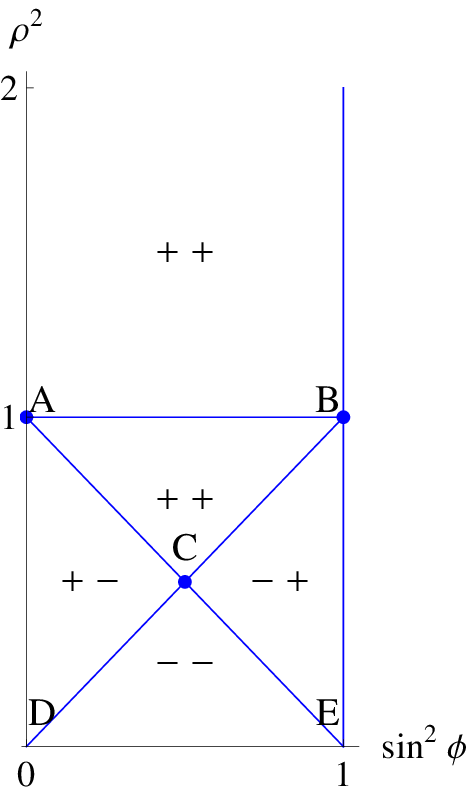}\\
(a)~~~~~~&\hspace{50pt}&(b)~~~~~~
\end{tabular}
\caption{The dashed lines in the first plot correspond to constant $\theta$.
The left and right signs in the second plot correspond to the signs of $\langle L^2\rangle$
and $\langle R^2\rangle$, respectively. $\langle L^2\rangle$ vanishes at the diagonal $BD$ and $\langle R^2\rangle$
at $AE$.}
\label{Fig:RhoPhi}
\end{figure}

Like in \eqref{solution X}, the shape of the surface \eqref{solution Y 1} depends only on
the parameter $\theta$, and not on $\rho$ and $\phi$ separately.
The first plot in fig.~\ref{Fig:RhoPhi}  shows the constant $\theta$ lines on the strip with the coordinates
$(\sin^2\phi,\,\rho^2)$. From \eqref{theta 1} follows that these lines are arcs of ellipses
for the space-like surfaces ($\rho^2>1$).
One can check (see below the analog of \eqref{theta 1} ) that the lines become arcs of
hyperbolas for $\rho^2<1.$
The second plot in fig.~\ref{Fig:RhoPhi} is used for a classification of the surfaces which we consider below.
Note again that different points on the constant $\theta$ lines describe different
solutions of the system in $\mbox{AdS}_3\times \mbox{S}^3$, though they have the same $\mbox{AdS}_3$ projection.

\begin{figure}
\begin{center}
\begin{tabular}{ccccc}
\includegraphics[scale=0.23]{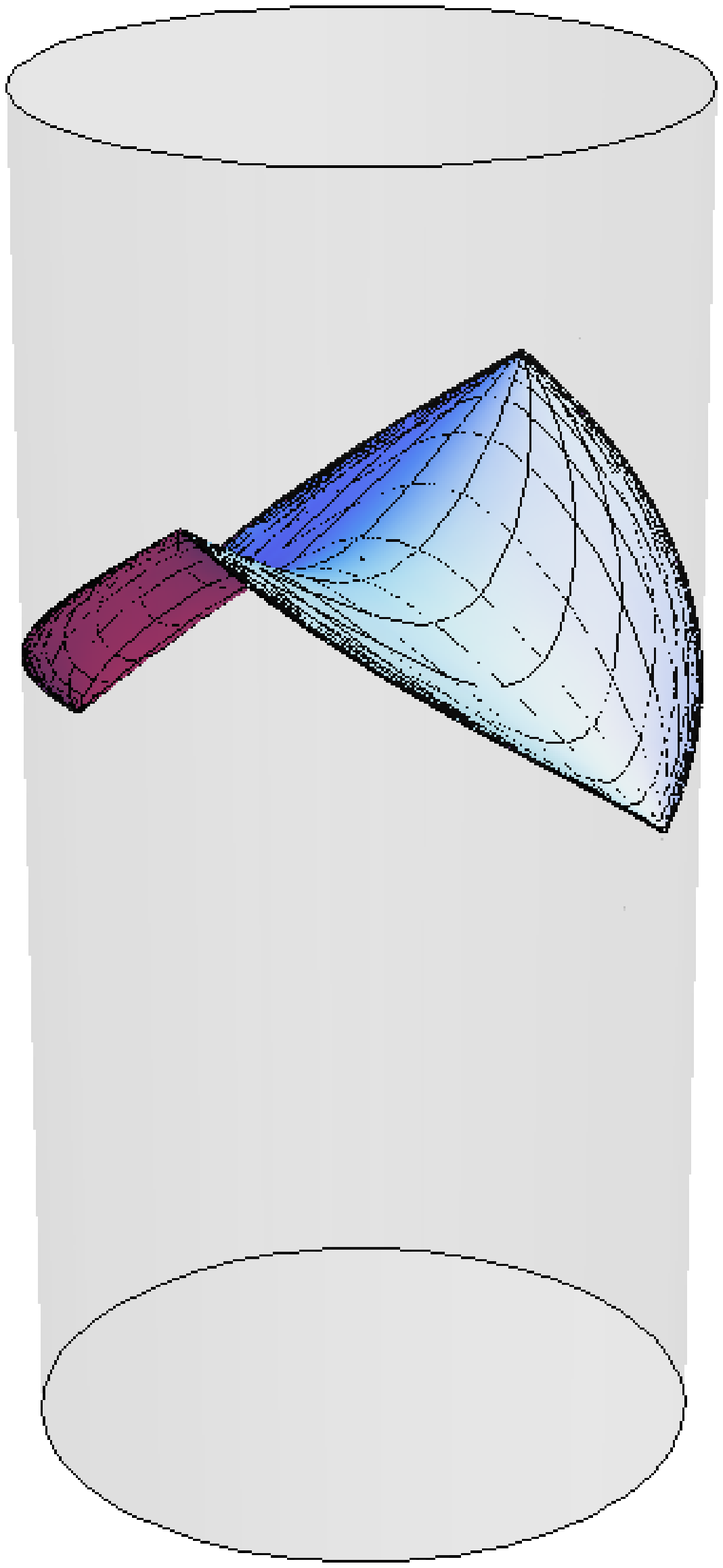}&\hspace{50pt}&\includegraphics[scale=0.23]{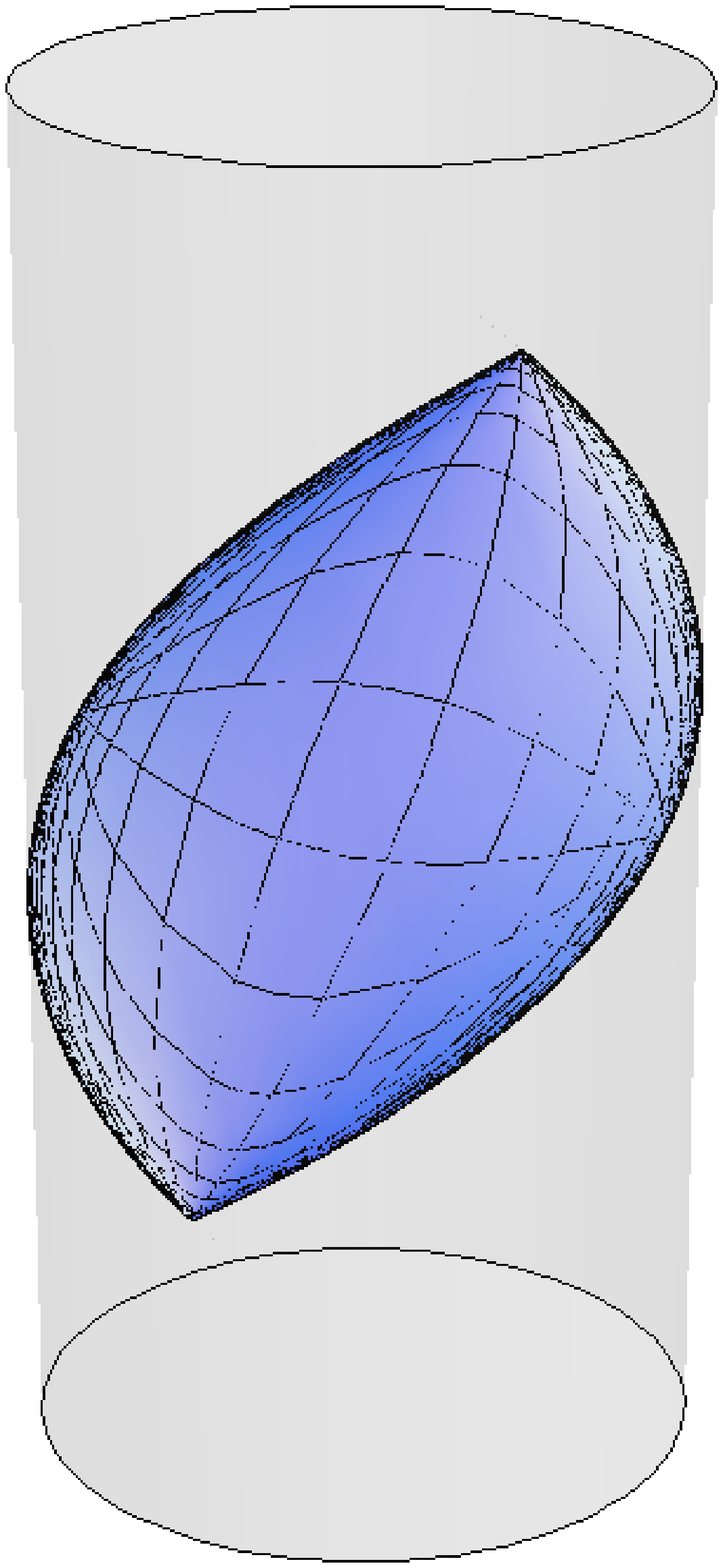}&\hspace{50pt}&\includegraphics[scale=0.23]{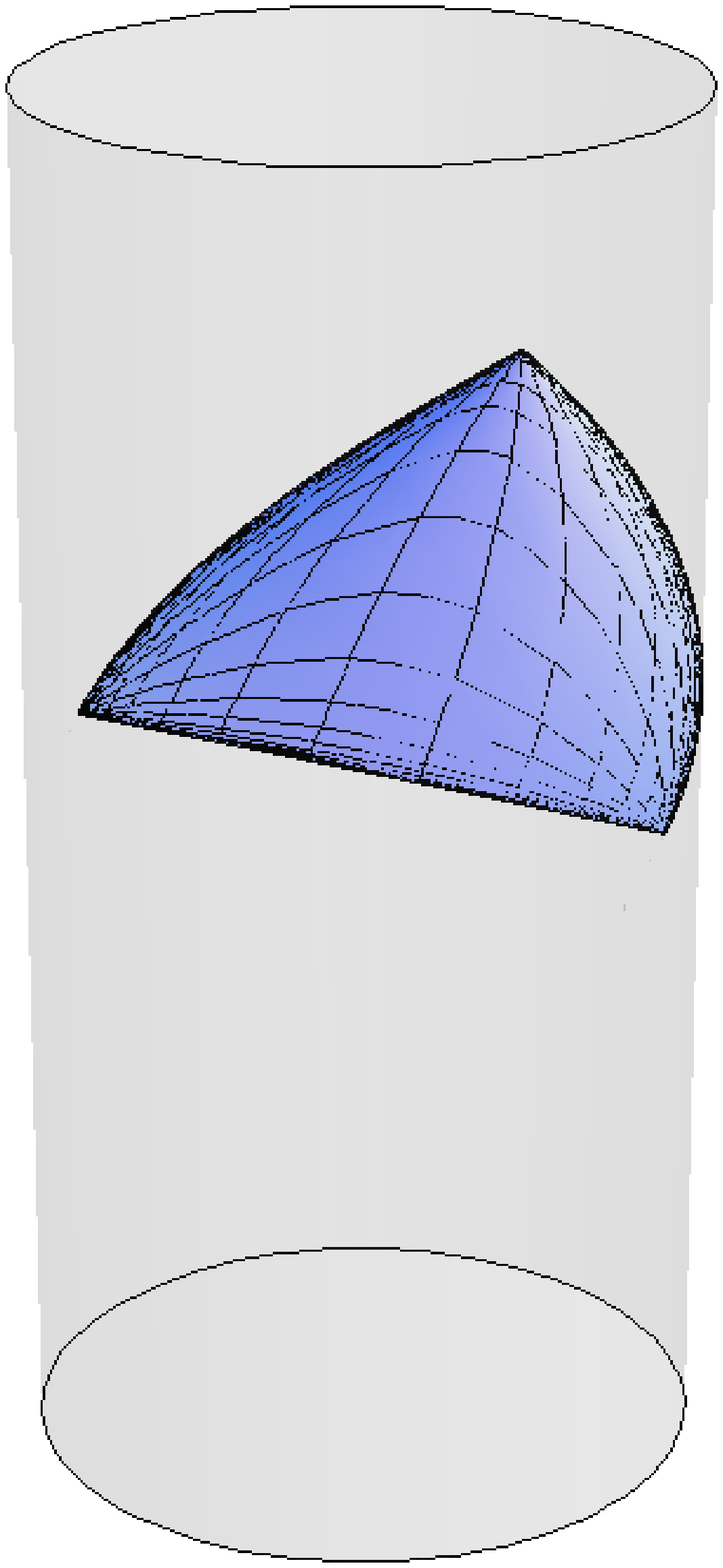}\\
(a)&\hspace{50pt}&(b)&\hspace{50pt}&(c)\\ \\
\includegraphics[scale=0.23]{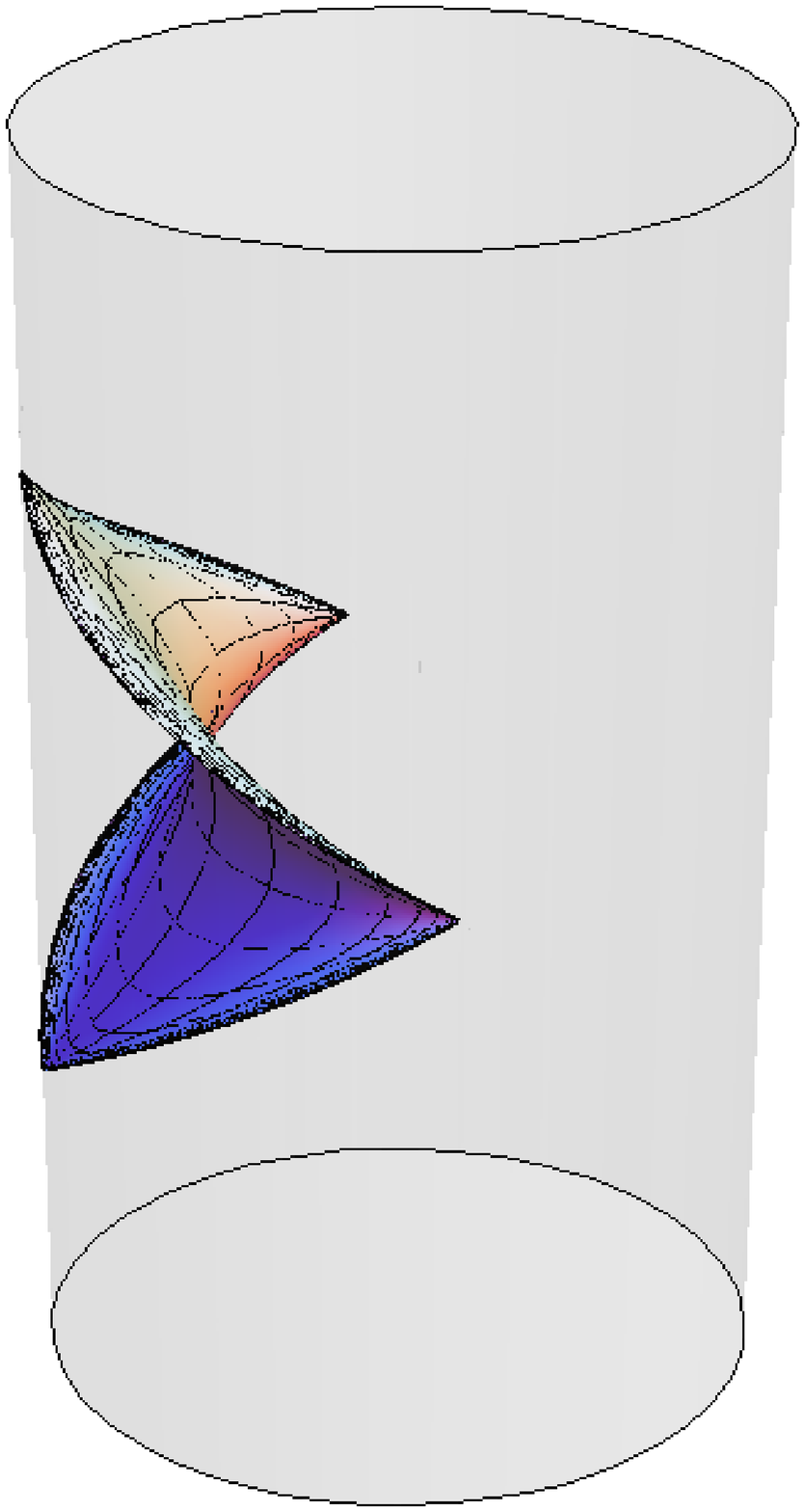}&\hspace{50pt}&\includegraphics[scale=0.23]{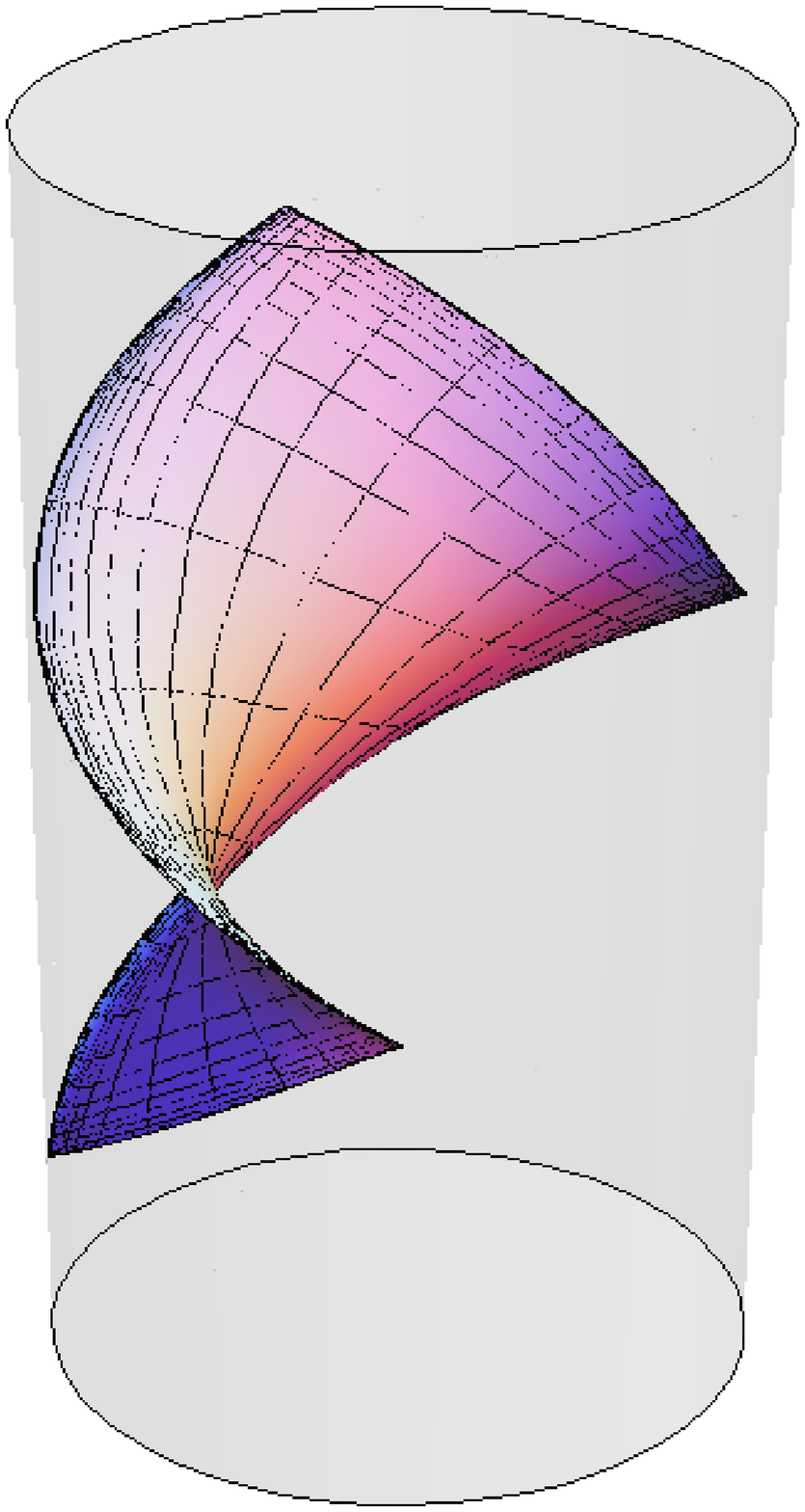}&\hspace{50pt}&\includegraphics[scale=0.23]{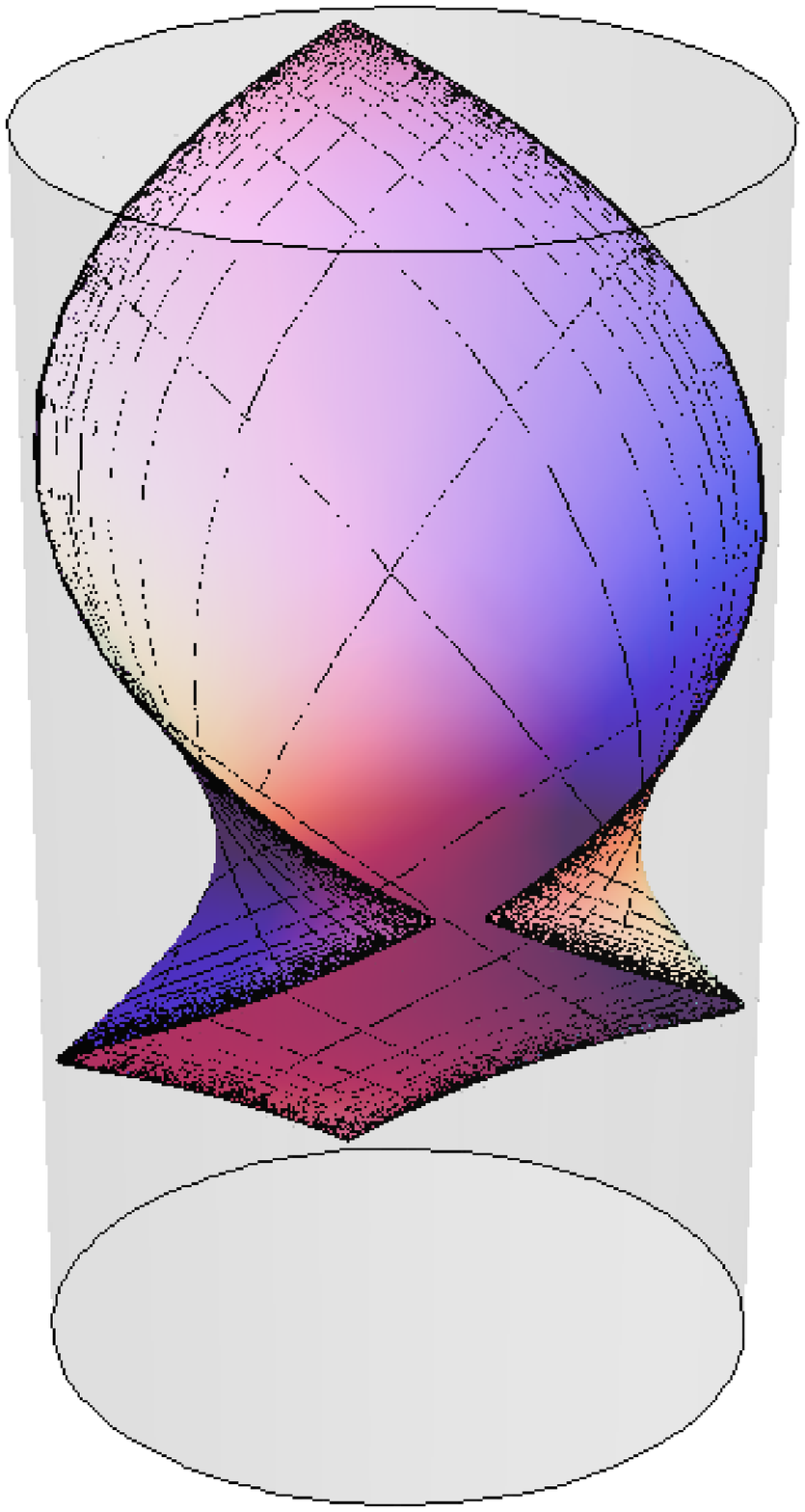}\\
(d)&\hspace{50pt}&(e)&\hspace{50pt}&(f)\\ \\
\includegraphics[scale=0.23]{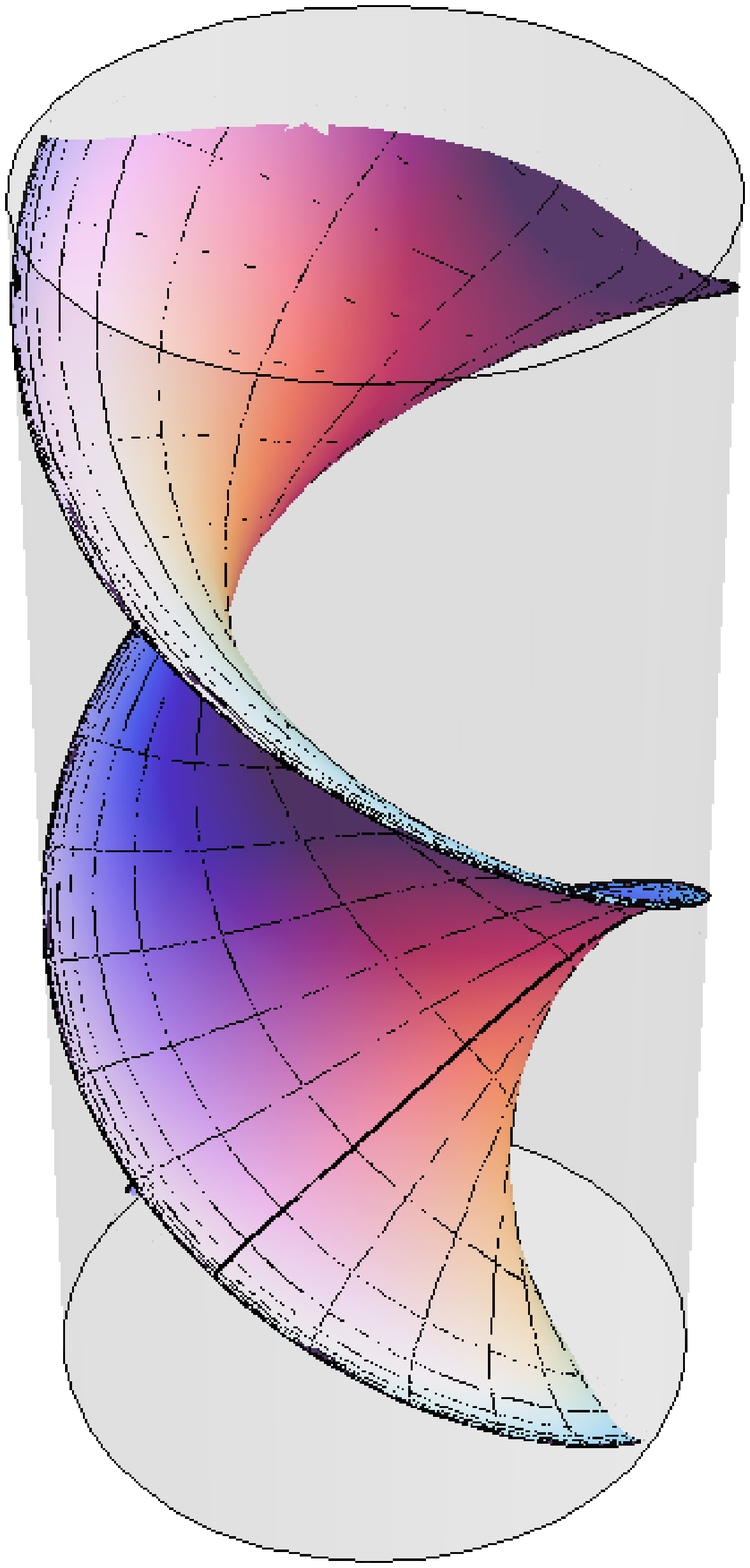}&\hspace{50pt}&\includegraphics[scale=0.23]{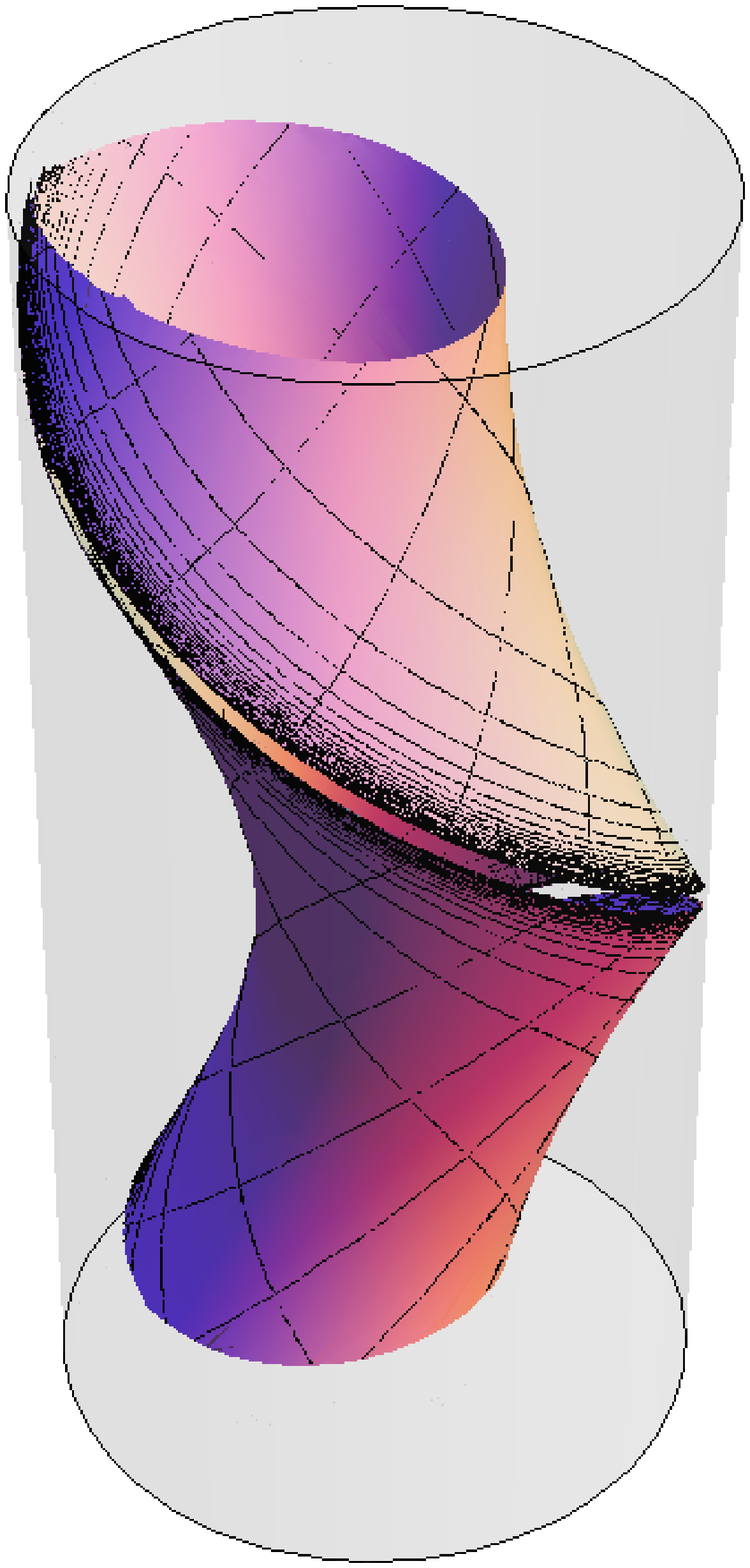}&\hspace{50pt}&\includegraphics[scale=0.23]{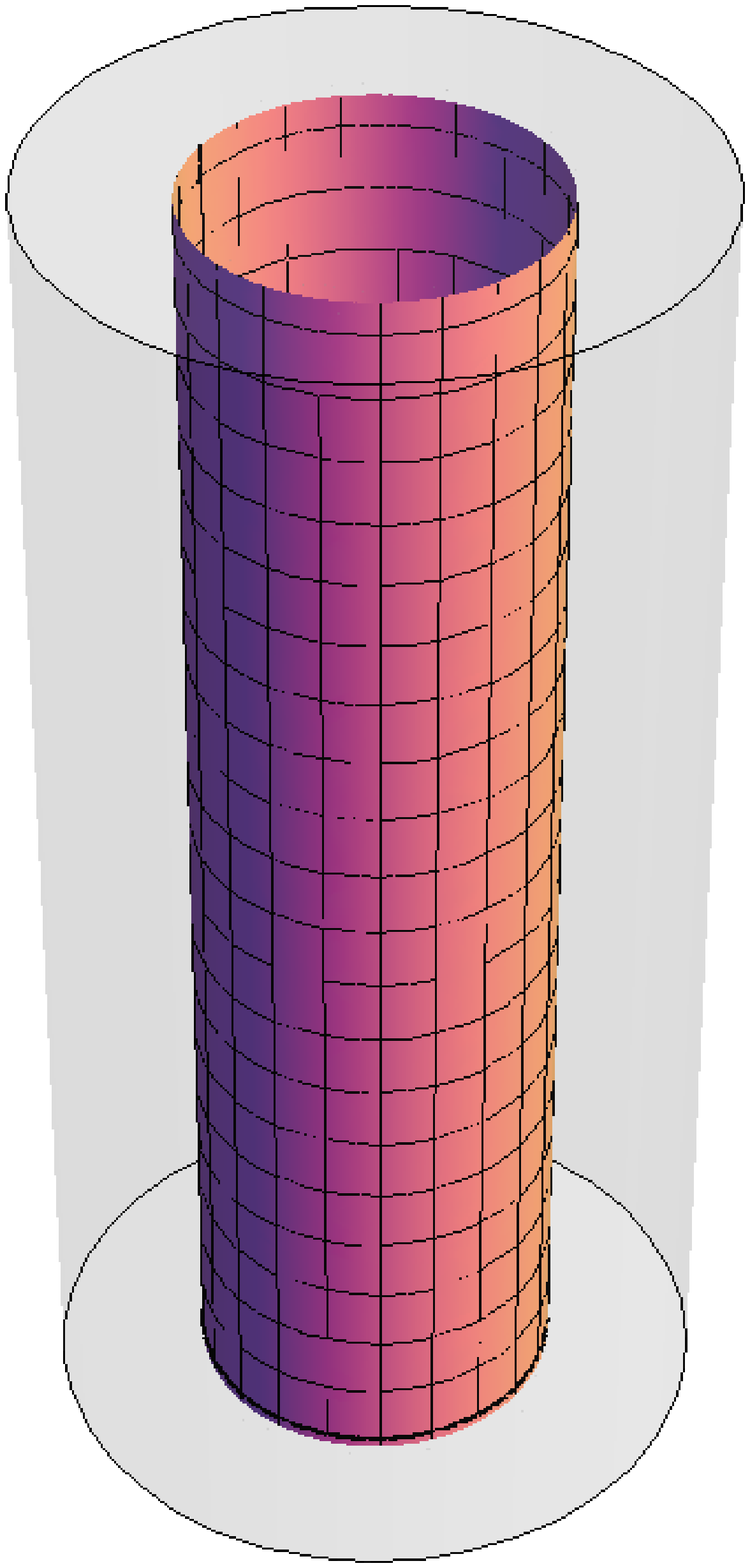}\\
(g)&\hspace{50pt}&(h)&\hspace{50pt}&(i)
\end{tabular}
\caption{Different vacuum string surfaces in $\mbox{AdS}_3$ described in section 2.4. The correspondence to formulas is: (a) $~$-$~$\eqref{solution Y 1}, (b)$~$-$~$\eqref{solution Y 2.1}, (c)$~$-$~$\eqref{solution Y 2.3},
(d)$~$-$~$\eqref{solution Y 3.1}, (e)$~$-$~$\eqref{solution Y 3.2}, (f)$~$-$~$\eqref{solution Y 3.5}, (g)$~$-$~$\eqref{solution Y 3.3},
(h)$~$-$~$\eqref{solution Y 3.6}, (i)$~$-$~$\eqref{solution Y 3.9}.}
\label{plot}
\end{center}
\end{figure}

In fig.~\ref{plot} we present the surfaces in $\mbox{AdS}_3$ associated with different solutions of our system.
The surface \eqref{solution Y 1} is shown in the plot \ref{plot}(a).

\vspace{6mm}
\noindent {\bf 2}. {\bf Light-like surfaces $\,\,(\rho^2=1).$}
\vspace{2mm}\\*
The commutator $[L,\,R]$ is light-like for $\rho^2=1$ and one has again
two options
\begin{equation}\label{[L,R] 2.1}
[L,\,R]= \pm\, 2{\bf t}_+\equiv \pm({\bf t}_0+{\bf t}_1)~,
\end{equation}
which are not on the same adjoint orbit. The option $[L,\,R]=-2{\bf t}_+$
can be realized by
\begin{equation}\label{L,R new 2}
L=\cos\phi \,{\bf t}_2+\sin\phi\, {\bf t}_+~,
\quad \quad
R=\sin\phi\,{\bf t}_2-\cos\phi\, {\bf t}_+~.
\end{equation}
Note that the pair $(-L,\,-R)$ has the same commutator and the scalar invariants \eqref{La,Ra}
as $(L,R)$. However, these two pairs are on different adjoint orbits.
We have the same situation for $[L,\,R]=2{\bf t}_+$. Hence, for $\rho^2=1$, there are four sets
of adjoint orbits of $(L,R)$ pairs. They are related to each other by discrete isometry transformations.
Therefore, it is sufficient to consider only \eqref{L,R new 2}.
It splits into three subcases:

\vspace{2mm}

2.1) $\phi=0;~$ 2.2) $\phi=\pi/2;~$ 2.3) $\phi\in (0,\pi/2)$.

\vspace{3mm}

\noindent
{\bf Case 2.1}.  $\phi=0\,$ corresponds to the point $A$ in fig.~\ref{Fig:RhoPhi}(b).
\vspace{2mm}\\*
Now $\langle\,L\,L\,\rangle=1,~$ $\langle\,R\,R\,\rangle=0,~$
$\langle\,L\,R\,\rangle=0;~$
$l=\tau,~$ $r=-\s.$
Then, eq.  \eqref{solution g} yields the solution
\begin{equation}\label{solution g 2.1}
g=e^{\tau\,{\bf t}_2}\,e^{\s\,{\bf t}_+}=
\left( \begin{array}{cr}
  e^{\tau}&\s\,e^{\tau}\\[0.1cm]
0 & e^{-\tau}\end{array}\right)~,
\end{equation}
with the embedding coordinates
\begin{eqnarray}\label{solution Y 2.1}
Y^{0'}=\cosh\tau~,\quad
Y^0=\frac{\sigma}{2}\,\,e^{\tau}~,\quad Y^1=\frac{\sigma}{2}\,\,e^{\tau}~,\quad
Y^2=\sinh\tau~.
\end{eqnarray}
The quadric \eqref{quadratic AdS} here reduces to
\begin{equation}\label{quadratic 2.1}
 (Y^0-Y^1)(Y^{0'}+Y^2)=0~,
\end{equation}
which is realized by \eqref{solution Y 2.1} in the linear form $Y^0-Y^1=0.$
The surface \eqref{solution Y 2.1} is given by the plot \ref{plot}(b). It has two cusps at the boundary.

\vspace{2mm}

\noindent
{\bf Case 2.2}. $\phi=\pi/2\,$ corresponds to the point $B$ in fig.~\ref{Fig:RhoPhi}(b).
\vspace{2mm}\\*
Here $\langle\,L\,L\,\rangle=0,~$ $\langle\,R\,R\,\rangle=1,~$
$\langle\,L\,R\,\rangle=0;~$ $l=\s$, $~r=\tau.$
The solution \eqref{solution g} now becomes
\begin{equation}\label{solution g 2.2,}
   g=e^{\s\,{\bf t}_+}\,e^{\tau\,{\bf t}_2}=
\left( \begin{array}{cr}
  e^{\tau}&\s\,e^{-\tau}\\[0.1cm]
0 & e^{-\tau}\end{array}\right)~,
\end{equation}
with the embedding coordinates
\begin{eqnarray}\label{solution Y 2.2}
Y^{0'}=\cosh\tau~,\quad
Y^0=\frac{\sigma}{2}\,\,e^{-\tau}~,\quad Y^1=\frac{\sigma}{2}\,\,e^{-\tau}~,\quad
Y^2=\sinh\tau~,
\end{eqnarray}
and the quadric $(Y^0-Y^1)(Y^{0'}-Y^2)=0.$
This surface is obtained from \eqref{quadratic 2.1} by
the reflection $Y^2\mapsto -Y^2$. Hence, the plot \ref{plot}(b) represents \eqref{solution Y 2.2} as well.

\vspace{2mm}

\noindent
{\bf Case 2.3}. $\phi\in(0,\pi/2)\,$ corresponds to the segment between the points $A$ and $B$ in fig.~\ref{Fig:RhoPhi}(b).
\\*[2mm]In this case $\langle\,L\,L\,\rangle=\cos^2\phi>0,\,$ $\langle\,R\,R\,\rangle=\sin^2\phi>0.$
Here we simplify the solution \eqref{solution g} by the same trick as in \eqref{Ls,Rs=}-\eqref{solution h 1}.
Using \eqref{e^t t2 e^t}, we write \eqref{L,R new 2} as
\begin{equation}\label{L,R new 2.3}
L=\cos\phi \,e^{-\beta\,{\bf t}_+}\,{\bf t}_2\,e^{\beta\,{\bf t}_+}~,
\quad \quad
R=\sin\phi\,\,e^{-\g\,{\bf t}_+}\,{\bf t}_2\,e^{\g\,{\bf t}_+}~,
\end{equation}
and bring the solution \eqref{solution g} to the form  $e^{l\,\cos\phi\,{\bf t}_2}\,e^{\theta{\bf t}_+}\,
e^{r\,\sin\phi\,{\bf t}_2}$, with $\theta=\beta-\g=1/\sin2\phi$.
By the rescaling property $e^{\l{\bf t}_2}\,{\bf t}_+\,e^{-\l{\bf t}_2}=e^{2\l}t_+$,
we then set $\theta=1$ and get
\begin{equation}\label{solution g 2.3}
   g=e^{l\,\cos\phi\,{\bf t}_2}\,e^{{\bf t}_+}\,
e^{r\sin\phi\,{\bf t}_2}=\left( \begin{array}{cr}
e^{\tau}&e^{\tilde\tau}\\[0.1cm]
0 & e^{-\tau}\end{array}\right)~,\qquad  \qquad
\tilde\tau=\tau\,\cos2\phi +\s\,\sin2\phi~.
\end{equation}
This solution satisfies the relation $\left\langle {\bf t}_2\,g\,{\bf t}_2\,
g^{-1}\right\rangle=1$, which reduces to
\begin{equation}\label{quadratic AdS 2.3}
(Y^{0})^2-(Y^{1})^2=0~,
\end{equation}
and it can be treated as a limiting case of \eqref{quadratic AdS 1} at $\theta\rightarrow 0$. The embedding coordinates
\begin{eqnarray}\label{solution Y 2.3}
Y^{0'}=\cosh\tau~,\quad
Y^0=\,\frac{e^{\tilde\tau}}{2}~,\quad Y^1=\frac{e^{\tilde\tau}}{2}~,\quad
Y^2=\sinh\tau
\end{eqnarray}
realize \eqref{quadratic AdS 2.3} by $Y^0-Y^1=0$ as in \eqref{solution Y 2.1}.
The difference with the case 2.1
is that now $Y^0$ and $Y^1$ are positive on the worldsheet.
This surface is given by the plot \ref{plot}(c) and has one cusp at the boundary.
The line crossing the interior of $\mbox{AdS}_3$ corresponds to $\bar\tau\rightarrow -\infty $. Although now the projection to $\mbox{AdS}_3$ has a boundary inside
$\mbox{AdS}_3$, the surface itself has there no boundary inside $\mbox{AdS}_3\times \mbox{S}^3$, since
$\bar\tau\rightarrow -\infty$ is correlated with infinitely wrapping the torus
in $\mbox{S}^3$.

Concluding this part we note that the calculation of
\eqref{solution g} for $L$ and $R$ given by \eqref{L,R new 2} leads to the answer
\begin{equation}\label{solution g 2 gen}
   g=e^{l\,L}\,e^{r\,R}=\left( \begin{array}{cr}
  e^{\tau}&\mu \\[0.1cm]
0& e^{-\tau}\end{array}\right)~,\quad\quad
\mu=\frac{e^{\tilde\tau}-\cosh\tau-\cos2\phi\sinh\tau}{\sin2\phi}~.
\end{equation}
The isometry transformations described above simplify its $g_{12}$ component to \eqref{solution g 2.3}.

The solution \eqref{solution g 2 gen} provides the following embedding coordinates
\begin{equation}\label{gen 2}
\begin{aligned}
Y^{0'}= \cosh \tau~, \quad Y^{0}=
\frac{e^{\tilde\tau}-\cosh \tau -\cos2\phi \sinh \tau  }{2\sin 2\phi }=Y^1~,\quad
 Y^{2}= \sinh \tau~,
\end{aligned}
\end{equation}
which in the limit $\phi\rightarrow 0$ and $\phi\rightarrow \pi/2$ reproduces \eqref{solution Y 2.1} and \eqref{solution Y 2.2},
respectively.

\vspace{6mm}

\noindent
{\bf 3}. {\bf Time-like surfaces $~(\rho^2<1).$}
\vspace{2mm}\\*
Here $[L,\,R]$ is space-like and we can fix it by
\begin{equation}\label{[L,R] 3}
[L,\,R]=-2\rho\sqrt{1-\rho^2}\,{\bf t}_1~.
\end{equation}
$L$ and $R$ are then in the plane $({\bf t}_2, {\bf t}_0)$ and
the pairs $(L,R)$ and $(-L,-R)$ belong to different adjoint orbits,
generated by ${\bf t}_1$.
Hence, there are two options for $(L,\,R)$, which are related to each other through the discrete
isometry transformation $g\mapsto {\bf t}_1\,g\,{\bf t}_1$.

Similarly to \eqref{L,R new}, we choose the pair
\begin{equation}\label{L,R new 3}
L =\rho\, \cos\phi \,{\bf t}_2+\sqrt{1-\rho^2}\,\sin\phi\, {\bf t}_0~,
\quad \quad
R =\rho\,\sin\phi\,{\bf t}_2-\sqrt{1-\rho^2}\,\cos\phi\, {\bf t}_0~,
\end{equation}
which now allows all three possibilities (space-like, light-like, time-like)
for both $L$ and $R$. The corresponding nine cases
are represented in fig.~\ref{Fig:RhoPhi}(b) inside the square $\rho^2<1$
and we enumerate them in the following order:

\vspace{2mm}

\noindent
{\bf Case 3.1} $\langle\,L\,L\,\rangle>0$, $\langle\,R\,R\,\rangle>0$.
\vspace{2mm}\\*
This corresponds to the area inside the triangle $ABC$ in fig.~\ref{Fig:RhoPhi}(b).
\\*
Here $\rho^2>\mbox{max}[\sin^2\phi,\,\cos^2\phi],$
which implies $\rho\,\sqrt{1-\rho^2}<\sin\phi\,\cos\phi$ and $\rho^2>1/2$.

The space-like vectors \eqref{L,R new 3} can be written in the form \eqref{Ls,Rs=}
\begin{equation}\label{L-R 3.1}
 L=\sqrt{\rho^2-\sin^2\phi}\,\,e^{-\beta\,{\bf t}_1}\,{\bf t}_2\,
 e^{\beta\,{\bf t}_1}~,\quad\quad
 R=\sqrt{\rho^2-\cos^2\phi}\,\,e^{-\g\,{\bf t}_1}\,{\bf t}_2\,
 e^{\g\,{\bf t}_1}~,
\end{equation}
using the boost parameters $\beta$ and $\g$.
Applying then the same trick as above,  we find the solution
\begin{equation}\label{solution g 212}
 g=e^{\tilde l\,{\bf t}_2}\,e^{\theta\,{\bf t}_1}\,
  e^{\tilde r\,{\bf t}_2}=\left( \begin{array}{cr}
\cosh\theta\,e^{\xi}&\sinh\theta\,e^{\eta}\\[0.2cm]
\sinh\theta\,e^{-\eta}& \cosh\theta\,e^{-\xi} \end{array}\right)~,
\end{equation}
where
\be\label{rescal}
\tilde l=\sqrt{\rho^2-\sin^2\phi}\,\,l~, \quad \quad \tilde r=\sqrt{\rho^2-\cos^2\phi}\,\,r
\ee
are the rescaled coordinates,
$\xi=\tilde l+\tilde r,$ $\eta=\tilde l-\tilde r$, and the new boost parameter $\theta=\b-\g >0$ is given by
\be\label{thetadef}
\tanh 2\theta=\frac{\rho\sqrt{1-\rho^2}}{\sin\phi\cos\phi}~.
\ee
The embedding coordinates
\begin{equation}\label{solution Y 3.1}
Y^{0'}=\cosh\theta\,\cosh\xi~,\quad~
Y^0=\sinh\theta\,\sinh\eta~,\quad~
Y^1=\sinh\theta\,\cosh\eta~,\quad
Y^2=\cosh\theta\,\sinh\xi~
\end{equation}
satisfy the quadric
\be\label{quadratic 3.1}
(Y^1)^2-(Y^0)^2=\sinh^2\theta~.
\ee
The solution \eqref{solution Y 3.1} was obtained in \cite{Dorn:2009hs} by an analytical continuation
of the space-like case \eqref{solution Y 1}. The corresponding surface is shown in the plot \ref{plot}(d). It also has four cusps, however, now only two of them are separated by a space-like interval and the other two by a time-like one.

\vspace{6mm}

\noindent
{\bf Case 3.2} $\langle\,L\,L\,\rangle>0$, $\langle\,R\,R\,\rangle=0.$
\vspace{2mm}\\*
This corresponds to the open line segment between the points $A$ and $C$ in  fig.~\ref{Fig:RhoPhi}(b).
\\*
Here $\rho^2=\cos^2\phi>\sin^2\phi$, i.e. $\sin\phi=\sqrt{1-\rho^2},\,$
$\phi\in (0,\pi/4)$ and $\,\rho^2>1/2$.

The pair \eqref{L,R new 3} now takes the form
\begin{equation}\label{L,R 3.2}
 L=\rho^2\,{\bf t}_2+(1-\rho^2)\,{\bf t}_0~,\quad\quad
 R= \rho\sqrt{1-\rho^2}\,({\bf t}_2-{\bf t}_0)~,
\end{equation}
and one gets
 $L=\sqrt{2\rho^2-1}\,e^{-\beta\,{\bf t}_1}\,{\bf t}_2\,e^{\beta\,{\bf t}_1}$,
 with $e^{2\beta}=(2\rho^2-1)^{-\frac{1}{2}}.$ Using then the identity
$e^{\beta\,{\bf t}_1}\,({\bf t}_2-{\bf t}_0)\,e^{-\beta\,{\bf t}_1}=
e^{2\beta}\,({\bf t}_2-{\bf t}_0),\,$ and the multiplication trick,
we obtain the solution
\begin{equation}\label{solution g 3.2}
g=e^{\tilde{l}\, {\bf t}_2}\, e^{\tilde{r}\,({\bf t}_2-{\bf t}_0)}=
\begin{pmatrix}
 e^{\tilde{l}}\,(1+\tilde r) &  -e^{\tilde{l}}\,\,\tilde r \\[0.2cm]
e^{-\tilde{l}}\,\, \tilde r & e^{-\tilde{l}}\,(1-\tilde r)
\end{pmatrix}~,
\end{equation}
where $\tilde l$ and $\tilde r$ are the rescaled worldsheet coordinates
\be
\tilde{l}=\sqrt{2\rho^2-1}\,\,l, \qquad \tilde{r}=
\frac{\rho\,\sqrt{1-\rho^2}}{\sqrt{2\rho^2-1}} \, r~.
\ee
The embedding coordinates for \eqref{solution g 3.2}
\begin{equation}\label{solution Y 3.2}
Y^{0'}=\cosh\tilde{l}+\tilde r\,\sinh\tilde{l}~,\quad~
Y^0=-\tilde r\,\cosh\tilde{l}~,\quad~
Y^1=-\tilde r\,\sinh\tilde{l}~,\quad
Y^2=\tilde r\,\cosh\tilde{l}+\sinh\tilde{l}~,
\end{equation}
satisfy the quadric
\be\label{quadratic 3.2}
(Y^{0'}+Y^1)^2-(Y^0+Y^2)^2=1~.
\ee
This surface is depicted in the plot \ref{plot}(e). It has again four cusps at the boundary,
but the separation of the opposite points at the diagonals are different than the one
in the previous cases. We analyze the boundary behavior of the surfaces in section 4.

\vspace{6mm}

\noindent
{\bf Case 3.3} $\langle\,L\,L\,\rangle>0$, $\langle\,R\,R\,\rangle<0.$
\vspace{2mm}\\*
This corresponds to the area inside the triangle $ACD$ in fig.~\ref{Fig:RhoPhi}(b).

Here $\sin^2\phi<\rho^2<\cos^2\phi$, i.e.
$\sin\phi\,\cos\phi<\rho\,\sqrt{1-\rho^2}$, and one writes the pair \eqref{L,R new 3} as
\begin{equation}\label{L-R 3.3}
 L=\sqrt{\rho^2-\sin^2\phi}\,e^{-\beta\,{\bf t}_1}\,{\bf t}_2\,
 e^{\beta\,{\bf t}_1}~,\quad\quad
 R=-\sqrt{\cos^2\phi-\rho^2}\,e^{-\g\,{\bf t}_1}\,{\bf t}_0\,
 e^{\g\,{\bf t}_1}~,
\end{equation}
with the boost parameters $\beta$ and $\gamma$.
Then, similarly to \eqref{solution g 212}, we find the solution
\begin{equation}
 g=e^{\tilde l\,{\bf t}_2}\,e^{\theta\,{\bf t}_1}\,
  e^{-\tilde r\,{\bf t}_0}=\left( \begin{array}{cr}
\cosh\th\, e^{\tilde l}\,\cos\tilde r+\sinh\th\, e^{\tilde l}\,\sin\tilde r &
\sinh\th\, e^{\tilde l}\,\cos\tilde r-\cosh\th\, e^{\tilde l}\sin\tilde r\\[0.2cm]
\sinh\th e^{-\tilde l}\cos\tilde r+\cosh\th e^{-\tilde l}\sin\tilde r &
\cosh\th e^{-\tilde l}\cos\tilde r-\sinh\th e^{-\tilde l}\sin\tilde r
\end{array}\right),
\end{equation}
where the rescaled coordinates are $\tilde l=\sqrt{\rho^2-\sin^2 \phi}\,\,\,l$, $~\tilde r=\sqrt{\cos^2\phi-\rho^2}\,\,\,r$,
and the new boost parameter $\theta=\beta-\gamma>0$ is given by
\be\label{theta 3.3}
\tanh 2\theta=\frac{\sin\phi\cos\phi}{\rho\sqrt{1-\rho^2}}~.
\ee
The embedding coordinates in this case
\begin{equation}
\label{solution Y 3.3} \begin{aligned}
Y^{0'} &=\cosh \theta\cosh \tilde l \cos \tilde r +\sinh \theta\sinh \tilde l\sin \tilde r ~,~
Y^{0}  =\sinh \theta\sinh \tilde l\cos \tilde r-\cosh \theta\cosh \tilde l\sin \tilde r ~,\\
Y^{1}  &= \sinh \theta \cosh \tilde l\cos \tilde r-\cosh \theta \sinh \tilde l \sin \tilde r ~,~
Y^{2}  =\sinh \theta \cosh \tilde l \sin \tilde r +\cosh \theta \sinh \tilde l\,\cos \tilde r
\end{aligned} \end{equation}
satisfy the quadric
\begin{equation}
Y^{0'}Y^{1} - Y^{0}Y^{2} = \sinh \theta\,\cosh\theta~.
\end{equation}
This surface is shown in the plot \ref{plot}(g). It describes an infinite open string with two ends on the conformal boundary.

\vspace{6mm}

\noindent
{\bf Case 3.4} $\langle\,L\,L\,\rangle=0$, $\langle\,R\,R\,\rangle>0.$
\vspace{2mm}\\*
This case corresponds to the open line segment between the points $B$ and $C$ in fig.~\ref{Fig:RhoPhi}(b).
\\*
Here $\rho^2=\sin^2\phi$ and $~\phi\in (\pi/4,\,\pi/2).$  One obviously gets the same picture as in 3.2,
since these two cases are related to each other by exchange of the left and right elements.

\vspace{6mm}

\noindent
{\bf Case 3.5} $\langle\,L\,L\,\rangle=0$, $\langle\,R\,R\,\rangle=0.$
\vspace{2mm}\\*
This corresponds to the point $C$ in fig.~\ref{Fig:RhoPhi}(b), with $\rho^2=1/2$ and $\phi=\pi/4$.
From \eqref{l,r=s,t a} then follows: $l=\frac{1}{\sqrt{2}}\,(\tau+\s)$ and $~r=\frac{1}{\sqrt{2}}(\tau-\s)$.

The pair \eqref{L,R new 3} here reduces to
\be\label{L,R 3.5}
L=\frac{1}{2}({\bf t}_2+{\bf t}_0)~,
\qquad R=\frac{1}{2}({\bf t}_2-{\bf t}_0)~,
\ee
and by \eqref{solution g} we obtain the solution
\be\label{g 3.5}
g=e^{\frac{\sqrt{2}}{4}\,(\tau+\s)({\bf t}_2+{\bf t}_0)}\,
e^{\frac{\sqrt{2}}{4}\,(\tau-\s)({\bf t}_2-{\bf t}_0)}=
\frac{1}{4}\begin{pmatrix}
 4-\s^2+\tau^2+2\sqrt{2}\,\tau & \s^2-\tau^2+2\sqrt{2}\,\s \\[0.2cm]
 \s^2-\tau^2-2\sqrt{2}\,\s&  4-\s^2+\tau^2-2\sqrt{2}\,\tau
\end{pmatrix}.
\ee
The quadric \eqref{quadratic AdS} now yields
\begin{equation}\label{quadratic 3.5}
 (Y^{0'}+Y^{1})^2=1~,
\end{equation}
and the embedding coordinates
\begin{equation}
\label{solution Y 3.5} \begin{aligned}
Y^{0'}=1-\frac{\s^2-\tau^2}{4}~,\quad
Y^{0} =\frac{\sqrt{2}}{2}\,\s~,\quad
Y^{1}= \frac{\s^2-\tau^2}{4}~,\quad
Y^{2}  =\frac{\sqrt{2}}{2}\,\tau~
\end{aligned} \end{equation}
realize \eqref{quadratic 3.5} in the linear form $Y^{0'}+Y^{1}=1$.
This surface is given by the plot \ref{plot}(f).

\vspace{6mm}

\noindent
{\bf Case 3.6} $\langle\,L\,L\,\rangle=0$, $\langle\,R\,R\,\rangle<0.$
\vspace{2mm}\\*
This corresponds to the open segment between the points  $C$ and $D$ in fig.~\ref{Fig:RhoPhi}(b).
\\*
Here $\rho^2=\sin^2\phi<\cos^2\phi,\,$
 i.e $~\cos\phi=\sqrt{1-\rho^2},$ $~\phi\in (0,\pi/4)$ and $\rho^2<1/2.$

We follow the scheme of the case 3.2. The pair \eqref{L,R new 3} now becomes
\begin{equation}\label{L,R 3.6}
 L=\rho\sqrt{1-\rho^2}\,({\bf t}_2+{\bf t}_0)~,\quad\quad
 R=\rho^2\,{\bf t}_2-(1-\rho^2)\,{\bf t}_0 ~,
\end{equation}
and one has the representation
 $R=-\sqrt{1-2\rho^2}\,e^{-\gamma\,{\bf t}_1}\,{\bf t}_0\,e^{\gamma\,{\bf t}_1}$,
 with $e^{-2\gamma}=(1-2\rho^2)^{-\frac{1}{2}}.$  This leads to the solution
\begin{equation}\label{solution g 3.6}
g=e^{\tilde{l}\,({\bf t}_2+{\bf t}_0)}\,e^{-\tilde{r}\,{\bf t}_0}=
\begin{pmatrix}
 (1+\tilde{l})\,\cos\tilde r+\tilde{l}\,\sin\tilde r & \tilde{l}\,\cos\tilde r-(1+\tilde{l})\,\sin\tilde r \\[0.2cm]
 (1-\tilde{l})\,\sin\tilde r-\tilde{l}\,\cos\tilde r & \tilde{l}\,\sin\tilde r+(1-\tilde{l})\,\cos\tilde r
\end{pmatrix}~,
\end{equation}
where $\tilde l$ and $\tilde r$ are the rescaled coordinates
\be
\tilde{l}=\frac{\rho\,\sqrt{1-\rho^2}}{\sqrt{1-2\rho^2}}\,l, \qquad
\tilde{r}=\sqrt{1-2\rho^2}\,\, r~.
\ee
The embedding coordinates for \eqref{solution g 3.6} are
\begin{equation}\label{solution Y 3.6}
Y^{0'}=\tilde{l}\,\sin\tilde r+\cos\tilde{r}~,\quad~
Y^0=\tilde{l}\,\cos\tilde r-\sin\tilde r~,\quad~
Y^1=-\tilde{l}\,\sin\tilde r~,\quad
Y^2=\tilde{l}\,\cos\tilde r~,
\end{equation}
and they satisfy the quadric
\be\label{quadratic 3.6}
(Y^{0'}+Y^1)^2+(Y^0-Y^2)^2=1~.
\ee
This surface is given by the plot \ref{plot}(h). It also describes an infinite open string, but in contrast to the case 3.3,
now the end points coincide at the boundary.

\vspace{6mm}

\noindent
{\bf Case 3.7} $\langle\,L\,L\,\rangle<0$, $\langle\,R\,R\,\rangle>0.$
\vspace{2mm}\\*
This corresponds to the area inside the triangle $BCE$ in fig.~\ref{Fig:RhoPhi}(b).
\\*
Here $\sin^2\phi\cos^2\phi<\rho^2(1-\rho^2),~$
$\phi\in (\pi/4,\pi/2),~$ and one has the same picture as in 3.3.

\vspace{6mm}

\noindent
{\bf Case 3.8} $\langle\,L\,L\,\rangle<0$, $\langle\,R\,R\,\rangle=0.$
\vspace{2mm}\\*
This corresponds to the open segment between the points $C$ and $E$ in fig.~\ref{Fig:RhoPhi}(b).
\\*
Here $\sin^2\phi>\rho^2=\cos^2\phi$,
$~\phi\in (\pi/4,\pi/2),$ and the picture is the same as in 3.6.

\vspace{6mm}

\noindent
{\bf Case 3.9} $\langle\,L\,L\,\rangle<0$, $\langle\,R\,R\,\rangle<0.$
\vspace{2mm}\\*
This corresponds to the area inside the triangle $DCE$ in fig.~\ref{Fig:RhoPhi}(b).
\\*
Here $\rho^2<\mbox{min}\,[\sin\phi , \,
\cos\phi]$ and it implies $\rho^2(1-\rho^2)<\sin^2\phi\,
\cos^2\phi.$

Writing the pair \eqref{L,R new 3} as in \eqref{L-R 3.3}
\begin{equation}\label{L-R 3.9}
 L=\sqrt{\sin^2\phi-\rho^2}\,e^{-\beta\,{\bf t}_1}\,{\bf t}_0\,
 e^{\beta\,{\bf t}_1}~,\quad\quad
 R=-\sqrt{\cos^2\phi-\rho^2}\,e^{-\g\,{\bf t}_1}\,{\bf t}_0\,
 e^{\g\,{\bf t}_1}~,
\end{equation}
one obtains the solution
\be\label{g 3.9}
g=e^{\tilde l\,{\bf t}_0}\,e^{\theta\,{\bf t}_1}\,
e^{-\tilde r\,{\bf t}_0}=\begin{pmatrix}
  \sinh\th\sin\xi+\cosh\th\cos\eta &
\cosh\th\sin\eta+\sinh\th\cos\xi\\[0.2cm]
 \sinh\th\cos\xi-\cosh\th\sin\eta&
\cosh\th\cos\eta-\sinh\th\sin\xi
  \end{pmatrix},
\ee
where $\tilde l = \sqrt{\sin^2 \phi-\rho^2}~l,\,$  and $\tilde r=\sqrt{\cos^2 \phi-\rho^2}~ r$
are the rescaled coordinates, $\xi=\tilde l+\tilde r,~\eta=
\tilde l-\tilde r$, and $\theta=\beta-\g$ is defined by
\be
\tanh2\th=\frac{\rho \sqrt{1-\rho^2}}{\sin\phi\cos\phi}~.
\ee
The embedding coordinates
\begin{equation}\label{solution Y 3.9}
Y^{0'}=\cosh\th\cos\eta~,\quad~
Y^0=\cosh\th\sin\eta~,\quad~
Y^1=\sinh\th\cos\xi~,\quad
Y^2=\sinh\th\sin\xi~
\end{equation}
satisfy the quadric
\be\label{quadratic 3.9}
(Y^1)^2+(Y^2)^2=\sinh^2\th~.
\ee
This surface is shown in the plot \ref{plot}(i). Only this case
does not extend to the boundary of $\mbox{AdS}_3$.

At the end of this section we comment on the solution \eqref{solution g} in the general time-like
case with the pair $(L,\,R)$ given by \eqref{L,R new 3}. The simple rules for the
exponentiation of the $\mathfrak{sl}(2,\rr)$ elements (see appendix A) enable us to calculate the exponents
in \eqref{solution g} and then find the embedding coordinates. In the domain bounded by
the triangle $ABC$ in fig.~\ref{Fig:RhoPhi}(a), this calculation leads to the answer
\begin{equation}\begin{aligned}\label{gen 3}
Y^{0'} &=\cosh(\lambda\,l) \cosh(\mu\, r)- \sin\phi\,\cos\phi~ \frac{\sinh(\lambda\,l)}
{\lambda} \frac{\sinh(\mu\,r)}{\mu}  ~,\\
Y^{0}  &=\sqrt{1-\rho^2}\,\,\sin \phi\, \frac{\sinh(\lambda\,l)}{\lambda}\,\cosh(\mu\,r)
+\sqrt{1-\rho^2}~\cos \phi \cosh (\lambda\,l)\,\,\frac{\sinh(\mu\,r)}{\mu}  ~,\\
Y^{1}  &=  \rho \sqrt{1-\rho^2}~\,\frac{\sinh(\lambda\, l)}{\lambda}~\frac{\sinh(\mu\,r)}{\mu} ~,\\
Y^{2}  &= \rho\,\cos \phi\,\, \frac{\sinh(\lambda\,l)}{\lambda}~\cosh(\mu\,r)-
\rho\,\sin \phi \,\cosh(\lambda\,l)~\frac{\sinh(\mu\,r)}{\mu}~,
\end{aligned} \end{equation}
where $\lambda=\sqrt{\rho^2-\sin^2\phi}$ and $\mu=\sqrt{\rho^2-\cos^2\phi}$.

Note that $\lambda$ and $\mu$ vanish at the edges $BC$ and $AC$, respectively.
However, the solution \eqref{gen 3} remains regular there, and has a smooth continuation
to other domains of the square $\rho^2<1$. When crossing the diagonals of the square one has to replace
the hyperbolic functions with the corresponding trigonometric ones. This means that in any
compact domain of $\mbox{AdS}_3$ the different solutions are smoothly related to each other via variations of the parameters $\rho$ and $\phi$. However, the solutions differ essentially in their global properties which are reflected at the boundary of $\mbox{AdS}_3$. We analyze the boundary behavior
of the solutions in section 4.

In the next section we consider the scheme of Pohlmeyer reduction for the time-like and light-like surfaces
and establish a connection with the group theoretical treatment of this section.
Namely we show how the general cases described by eqs. \eqref{gen 2} and \eqref{gen 3}
are reproduced by the Pohlmeyer reduction.

\setcounter{equation}{0}

\section{Analysis via Pohlmeyer reduction}

In this section we treat minimal  space-like surfaces in $\mbox{AdS}_3 \times \mbox{S}^3$
within the framework of the Pohlmeyer reduction.
As it was mentioned above, the $\mbox{AdS}_3$ projection of a space-like surface in
$\mbox{AdS}_3 \times \mbox{S}^3$ may be space-like, time-like, or even light-like.
The $\mbox{S}^3$ part and the case of space-like $\mbox{AdS}_3$ was studied in \cite{Dorn:2009hs}.
Here we consider the time-like and light-like cases.

\subsection{Time-like  $\mbox{AdS}_3$ projection}
We realize the $\mbox{AdS}_3$ space as the hyperboloid
\eqref{YY=-1} embedded  in $\rr^{2,2}$.
Using conformal worldsheet coordinates on $\mbox{AdS}_3 \times \mbox{S}^3$ one gets
the following equation for the $\mbox{AdS}_3$ projection
\begin{equation}\label{eq. AdS}
\bar\p\partial Y=(\p Y\cdot\bar\p Y)\,Y~.
\end{equation}
The gauge fixing conditions \eqref{gauge fixing}
can be written as
\begin{equation}\label{gauge fixing AdS}
\partial Y\cdot\partial Y=-1=\bar\partial Y\cdot\bar\partial Y~,
\end{equation}
and in the real worldsheet coordinates $(\sigma,\tau)$ this means
\begin{equation}\label{gauge fixing AdS 1}
(\partial_\tau Y)^2-(\partial_\sigma Y)^2=1~,\quad \quad \quad
\partial_\sigma Y \cdot\partial_\tau Y=0~.
\end{equation}
We assume that $\p_\sigma Y$ is time-like
$\partial_\sigma Y\cdot\partial_\sigma Y<0$ and $\p_\tau Y$ is space-like
 $\partial_\tau Y\cdot\partial_\tau Y>0$.
 These conditions  yield  $-1<\partial Y\cdot\bar\partial Y<1$
and one can use the parameterization
\begin{equation}\label{ind-metric AdS}
\partial Y\cdot\bar\partial Y=\cos\alpha~.
\end{equation}
We introduce
the basis $\mathcal B = \{ Y,\p Y,\bar\p Y, N\}$ in $\rr^{2,2}$, where $N$
is the normalized orthogonal vector to the surface
\begin{equation}\label{N dot N}
N\cdot N=1~,~~~~~~Y\cdot N=\p Y\cdot N=\bar\p Y\cdot N=0~.
\end{equation}
The real vector $N$ has to be space-like, since the worldsheet
in $\mbox{AdS}_3$ is time-like.

Differentiating with respect to $z$ on gets the linear equations
\be\ba\label{system1}
\p^2 Y &= -Y+\frac{\p\a}{\sin \a}(\cos\a\,\, \p Y+
\bar{\p}Y)+u N~,\\
\p\bar{\p} Y &= \cos\a\,\, Y~,\\
\p N&=\frac{u} {\sin^2\a}\, (\p Y+\cos\a\,\, \bar{\p}Y)~,
\ea\ee
where
\begin{equation}\label{u=}
u=\p^2 Y\cdot N~
\end{equation}
is the $zz$-component of the second fundamental form. Introducing also
the equations complex conjugated to \eqref{system1},
one finds the consistency conditions for the linear system
\be\ba\label{consistency conditions}
\p\bar{\p} \a &=\sin \a -\frac{u\bar{u}}{\sin\a}~,\\
\bar{\p}u &=-\frac{\p\a}{\sin\a}\, \bar{u}~,
\qquad \p\bar{u}=-\frac{\bar{\p}\a}{\sin\a}~u~.
\ea\ee
In the next subsection we describe constant solutions
of these equations and integrate the corresponding linear system
\eqref{system1}.

\subsection{Integration of the linear system}
Due to the consistency conditions \eqref{consistency conditions} constant $\alpha$ implies constant $u$ and we parameterize
\begin{equation}\label{sol}
u=u_0~,~~~~~~~~~\sin^2 \a =|u_0|^2 ~,
\end{equation}
\begin{equation}\label{u_0=}
u_0=\frac{2}{i}\,\rho\,\sqrt{1-\rho^2}\,\,e^{2i\phi}~,~~~~~~~~0<\rho <1~.
\end{equation}
The induced metric tensor $f_{ab}=\p_a Y \cdot \p_b Y$
and the second fundamental
form $U_{ab}=\p^2_{ab}Y\cdot N$ corresponding to the solution
\eqref{sol}-\eqref{u_0=} become
\begin{equation}\label{f=, s=}
f_{ab}(\sigma,\tau)=\left(
  \begin{array}{cc}
    \rho^2-1 & 0 \\
    0 & \rho^2\\
  \end{array}
\right),\quad U_{ab}=
\rho\,\sqrt{1-\rho^2}\,\left(\begin{array}{cc}
    \sin 2\phi & ~~\cos 2\phi \\
    \cos 2\phi & -\sin 2\phi \\
  \end{array}\right)~.
\end{equation}
This form of the metric tensor (compare with \eqref{AdS3 metric}) justifies  the parameterization of
the  norm of $u_0$ by $\rho$. One can show
that the angle variable
in \eqref{u_0=} also coincides with $\phi$ used in the previous section.
This implies for the mean curvature of the $\mbox{AdS}_3$ projection ($\theta $ defined as in \eqref{thetadef}) \footnote{Taking the conformal metric of the surface instead of $f_{ab}$ will of course provide a vanishing mean curvature due to minimality of the surface as a whole.}
\be
H~=~\frac{1}{2}(f^{-1})_{ab}U_{ab}~=~-\frac{\sin 2\phi}{2\rho\sqrt{1-\rho ^2}}~=~-\coth 2\theta~.
\ee
Let us introduce the following real orthonormal basis vectors
\begin{equation}\label{o-n basis}
\mathcal E_{0'}=Y~,\quad \mathcal E_0=\frac{\partial_1 Y}{\sqrt{1-\rho^2}}~,
\quad \mathcal E_1=N~,\quad \mathcal E_2=\frac{\p_2 Y}{\rho}~.
\end{equation}
They satisfy the conditions $\mathcal E_J\cdot \mathcal E_K=G_{JK},$ where
$G_{JK}$ is the metric tensor of $\rr^{2,2}$. Hence, the matrix $\mathcal E_J\,^K$ built from
vector components is in  $\mbox{O}(2,2)$.
The system \eqref{system1} is then equivalent to the matrix equations
\begin{equation}\label{e'=}
\p_a \mathcal E_J\,^K =(\mathcal A_a)_J\,^{J'} \,\mathcal E_{J'}\,^K~~~~~~~~~~~~~~~(a=1,2)~.
\end{equation}
The matrices $\mathcal A_a$ belong to the $\mathfrak{so}(2,2)$ algebra and
they have the following block structure
\begin{equation}\label{matrices}
\mathcal A_1 = \left(
       \begin{array}{cc}
         C_1 & B_1 \\
         B_1^T &D_1  \\
       \end{array}
     \right),
~~~~~~~~~
  \mathcal A_2 = \left(
       \begin{array}{cc}
         0 & B_2 \\
         B_2^T &D_2  \\
       \end{array}
     \right),
\end{equation}
with $2\times 2$ matrices
\begin{equation}\label{B_sigma=}
B_1=\left(
       \begin{array}{cc}
        0  & 0 \\[0.1cm]
         \rho\,\sin2\phi &  0\\
       \end{array}
     \right),\quad
B_2=\left(
       \begin{array}{cc}
         0& \rho  \\[0.1cm]
          \rho\,\cos2\phi &0\\
         \end{array}
     \right)~,
\end{equation}
\begin{equation}
C_1=\sqrt{1-\rho^2}\,{\bf t}_0~,\quad
D_1=-\sqrt{1-\rho^2}\,\cos2\phi\,{\bf t}_0~,
\quad D_2=\sqrt{1-\rho^2}\,\sin2\phi\,{\bf t}_0~.
\end{equation}
The integrability of the system \eqref{e'=} is provided by
$[\mathcal{A}_1,\,\mathcal{A}_2]=0$, and one obtains
\begin{equation}\label{e^A}
\mathcal E=\exp (\xi^a\mathcal A_a)\,\,C~,
\end{equation}
where $C$ is a constant $\mbox{O}(2,2)$ matrix.

To perform the exponentiation we use
the decomposition $\mathfrak{so}(2,2)=
\mathfrak{sl}(2,\mathbb R) \oplus \mathfrak{sl}(2,\mathbb R)$,
described in appendix B. The basis of this decomposition can
be chosen as
\be \label{L,R basis} \ba
(L_0)_J\,^K &=\left(
       \begin{array}{cc}
         {\bf t}_0 & 0 \\
         0 & {\bf t}_0 \\
       \end{array}
     \right),
& (L_1)_J\,^K &=\left(
            \begin{array}{cc}
              0 & {\bf t}_2 \\
             {\bf t}_2 & 0 \\
            \end{array}
          \right),
&(L_2)_J\,^K &=\left(
            \begin{array}{cc}
              0 & {\bf t}_1 \\
              {\bf t}_1 & 0 \\
            \end{array}
          \right), \\[0.2cm]
(R_0)_J\,^K &=\left(
       \begin{array}{cc}
         -{\bf t}_0 & 0 \\
         0 & {\bf t}_0 \\
       \end{array}
     \right),
&(R_1)_J\,^K &=\left(
            \begin{array}{cc}
              0 & -{\bf I} \\
              -{\bf I} & 0 \\
            \end{array}
          \right),
&(R_2)_J\,^K &=\left(
            \begin{array}{cc}
              0 & -{\bf t}_0 \\
              {\bf t}_0 & 0 \\
            \end{array}
          \right).
\ea\ee
The matrices $\,L_\mu$ $(\mu=0,1,2)\,$ and $\,R_\nu$ $(\nu=0,1,2)\,$  commute:  $[L_\mu,\,R_\nu]=0,$ and
in addition they satisfy the algebraic relations
similar to \eqref{tt=}
\begin{equation}\label{LmuLnu}
L_\mu\,L_\nu=\eta_{\mu\nu}\,{I}-\epsilon_{\mu\nu}\,^\rho\,
L_\rho~,\quad \quad
R_\mu\,R_\nu=\eta_{\mu\nu}\,{I}-\epsilon_{\mu\nu}\,^\rho\,
R_\rho~.
\end{equation}
The expansion of the matrices $\mathcal A_a$ in
the basis \eqref{L,R basis} takes the form
\begin{equation}\begin{aligned}
\label{A_1,2}
\mathcal A_1 &=\sin\phi\,(\rho\,\cos\phi\,L_2+\sqrt{1-\rho^2}\,\sin\phi\,L_0)+
\cos\phi\,(\rho\,\sin\phi\,R_2-\sqrt{1-\rho^2}\,\cos\phi\,R_0)~,\\   \\
\mathcal A_2 &=\cos\phi\,(\rho\,\cos\phi\,L_2+\sqrt{1-\rho^2}\,\sin\phi\,L_0)-
\sin\phi\,(\rho\,\sin\phi\,R_2-\sqrt{1-\rho^2}\,\cos\phi\,R_0)~,
\end{aligned}
\end{equation}
and the exponential in \eqref{e^A} becomes
\begin{equation}
\label{exp A_1,2}
\exp[l(\rho\,\cos\phi\,L_2+\sqrt{1-\rho^2}\,\sin\phi\,L_0)+r
(\rho\,\sin\phi\,R_2-\sqrt{1-\rho^2}\,\cos\phi\,R_0)]~,
\end{equation}
where $l$ and $r$ are given by \eqref{l,r=s,t a}.
The calculation of \eqref{exp A_1,2} can be done similarly to the $\mathfrak{sl}(2,\rr)$ case.
The solution $Y(\s,\,\tau)$ is defined by the first row of \eqref{e^A}.
Taking $C=I$, one obtains just \eqref{gen 3}.
In appendix B we give an alternative proof of the same statement, which establishes direct relation between the two integration methods.

\subsection{Light-like $\mbox{AdS}_3$ projection}

The induced metric on a light-like $\mbox{AdS}_3$ projection
\begin{equation}
f_{ab} = \begin{pmatrix} 0 & 0 \\ 0 & 1  \end{pmatrix}~
\end{equation}
is obtained
from \eqref{f=, s=} in the limit $\rho\rightarrow 1,$ and it
corresponds to a light-like tangent vector $\p_\s Y$.
Due to this light-like tangent vector the usual definition of the normal direction via orthogonality conditions or constructions with $\epsilon$-tensor break down. Instead one can fix unambiguously a normal vector in a covariant way by the conditions
\be\label{lightlike N}
N\cdot Y=N\cdot\partial_{\tau}Y=N\cdot N=0~,\quad\quad N\cdot \partial _{\sigma} Y=\frac 1 2~.
\ee
Then we introduce the orthonormal basis
\begin{equation}\label{light-like basis}
\mathcal E_{0'}=Y~, \quad \mathcal E_0=\p_\sigma Y - N~,
\quad \mathcal E_1= \p_\sigma Y + N~,\quad
\mathcal E_2=\p_\tau Y~.
\end{equation}
The linear system for this basis takes the standard form \eqref{e'=}
\begin{equation}\label{linear sys 1}
\p_\s \mathcal E = \mathcal A_\s\, \mathcal E~, \quad \quad \quad
\p_\tau \mathcal E = \mathcal A_{\tau}\, \mathcal E~,
\end{equation}
with the $\mathfrak{so}(2,2)$ matrices
\begin{equation}\label{A_s,A_t}
\mathcal A_\s=
\begin{pmatrix}
 ~~0 & 1/2& 1/2 & 0\\
-1/ 2 & 0 &  2u_{\s\s} & u_{\tau\sigma}\\
~~1/2& 2u_{\s\s} & 0 & -u_{\tau\s}\\
~~0 & u_{\tau\sigma} &  u_{\tau\s}  & 0
\end{pmatrix}~, \quad
\mathcal A_\tau=
\begin{pmatrix}
0 & 0 & 0 & 1\\
0 & 0 & 2u_{\tau\s} & -u_{\s\s} \\
0 &  2u_{\tau\s} & 0 & u_{\s\s}\\
1 & -u_{\s\s} &  -u_{\s\s} & 0
\end{pmatrix}~,
\end{equation}
where $u_{\s\s} = N \cdot \p_\sigma \p_\sigma Y = - N \cdot \p_\tau \p_\tau
Y$ and $u_{\tau\s} = N \cdot \p_\tau \p_\sigma Y$ are the coefficients of
the second fundamental form. Note that this analysis implies that all second derivatives of $Y$ are linear
combinations only of $Y$ and $\partial _{\sigma}Y$ and hence as vectors from
the AdS point of view parallel to $\partial _{\sigma}Y$ . This in agreement with
the analogous situation in the group theoretical treatment above. The consistency conditions for \eqref{linear sys 1} yield the equations
\begin{equation}\label{consist-cond Pohlmeyer}
\p_\tau u_{\s\s}=\p_\s u_{\tau\s}~, \quad \quad \quad
\p_\s u_{\s\s}+\p_\tau u_{\tau\s}+2u_{\s\s}^2+2u_{\tau\s}^2-\frac{1}{2}=0~,
\end{equation}
which are equivalent to \eqref{consist-cond AdS} with $u_{\s\s}=a,$ $u_{\tau\s}=b-1/2$, and
they allow non constant solutions as well.

The vacuum surfaces are associated with constant $u_{\s\sigma}$ and $u_{\tau\s}$. In this case
\eqref{consist-cond Pohlmeyer} reduces to $u_{\s\sigma}^2 + u_{\tau\s}^2 =1/4$ and one can use
the parameterization $u_{\sigma\s}=\frac 1 2\,\sin2\phi$ and $u_{\tau\s}=\frac 1 2\,\cos2\phi$.

Since the metric induced from $\mbox{AdS}_3$ is degenerate, the mean curvature as a standard invariant geometric quantity is ill defined. To relate the information carried by $\phi$ to an invariant,  one could replace $(f^{-1})_{ab} $ in the definition of the mean curvature by $(f_{s}^{-1})_{a\,b}$. In any case, $\phi$ has meaning only for the mutual
relation of $\mbox{AdS}_3$ and $\mbox{S}^3$ projection.
The expansion of the matrices \eqref{A_s,A_t} in the basis \eqref{L,R basis} now becomes
\begin{equation}\begin{aligned}
\label{A_1,2-0}
\mathcal A_\s &=\sin\phi\,(\cos\phi\,L_2+\sin\phi L_+)+
\cos\phi\,(\sin\phi\,R_2-\cos\phi\,R_+)~,\\[0.2cm]
\mathcal A_\tau &=\cos\phi\,(\cos\phi\,L_2+\sin\phi\,L_+)-
\sin\phi\,(\sin\phi\,R_2-\cos\phi\,R_+)~,
\end{aligned}
\end{equation}
where $L_+=\frac{1}{2}(L_0+L_1)$ and $R_+=\frac{1}{2}(R_0+R_1)$.
The left-right decomposition simplifies the calculation of the exponent
$\text{exp}(\sigma\,\mathcal A_\sigma  + \tau\,\mathcal A_\tau )$, and one finds that
its first row is given by \eqref{gen 2}, like for the time-like surfaces.
The cases $\phi=0$ and $\phi=\pi/2$
have to be treated separately and they reproduce \eqref{solution Y 2.1} and \eqref{solution Y 2.2},
respectively.

The derivation of the equivalence between the Pohlmeyer reduction and group variable construction,
presented in appendix B, is valid for this case as well.

Now we describe light-like $\mbox{AdS}_3$ string surfaces in the general case (with non-constant $u_{\s\s}$ and $u_{\tau\s}$).
For this purpose let us consider the analog of the linear system \eqref{system1}, which now takes the form
\be\ba\label{system2}
\p^2 Y &= -Y+u(\p Y+\bar{\p}Y)~,\\
\p\bar{\p} Y &=  Y~,\\
\p N&=\frac{Y}{2}+\frac{u} {4}\, (\p Y-\bar{\p}Y)-u N~,
\ea\ee
where $u=\p^2 Y\cdot N$ and $N$ is defined by \eqref{lightlike N}. 
We will use only the first two equations. From them follows that $\p(\p Y+\bar\p Y)=u(\p Y+\bar\p Y)$,
and, therefore, the light-like vector $\p_\s Y=\frac 1 2(\p Y+\bar\p Y)$
can be writen as $\p_\s Y=\psi e_+$, with a constant $e_+$ and an arbitrary scalar function $\psi$.
For $e_+=(0,1,1,0)$, which one can choose using the $\mbox{O}(2,2)$ transformations, the vector $Y$
becomes $Y=(\cosh\gamma,\, F,\,F,\,\sinh\g)$,
where $\gamma$ depends only on $\tau$ and $\p_\s F=\psi.$ Then, from the second equation of \eqref{system2} follows $\gamma=\tau$
and $F$ satisfies the free field equation
\be\label{ff-eq}
\pb\p F=F~.
\ee
Thus, the general light-like $\mbox{AdS}_3$ surface, up to $\mbox{O}(2,2)$ transformations, is given by
\begin{eqnarray}\label{solution Y 2 genral}
Y^{0'}=\cosh\tau~,\quad
Y^0=F(\s,\tau)~,\quad Y^1=F(\s,\tau)~,\quad
Y^2=\sinh\tau~.
\end{eqnarray}
Being interested in the AdS projection only, this surface satisfies the same quadric as the constant $u$ light-like surfaces.

Concluding this section note that the consistency conditions for the system \eqref{system2}
reads
\be\label{l-l consist cond}
\bar\p u+u\bar u=1~, \quad \quad \p\bar u+u\bar u=1~,
\ee
which is equivalent to \eqref{consist-cond Pohlmeyer} with $u=2(u_{\s\s}-iu_{\tau\s})$. According to the construction above,
these consistency conditions are solved by  $u=\p\psi/\psi$.

\subsection{Relation to complex sin(h)-Gordon models}

We will now discuss the Lagrangean structure of the consistency conditions. It is well known that Pohlmeyer reduction of sigma models on $\mbox{S}^3$ and $\mbox{AdS}_3$ leads to
complex sine-Gordon and complex sinh-Gordon models, respectively \cite{pohlred}.
Considering space-like strings on $\mbox{AdS}_3\times\mbox{S}^3$ with different signatures of the induced
metric tensor on the $\mbox{AdS}$ projection, one obtains various modifications of these models
on Euclidean worldsheets. All these models are Lagrangean. Below we present the Lagrange functions and the corresponding simple solutions related to our string configurations.

Let us consider the case of a time-like $\mbox{AdS}$ projection.
The consistency conditions \eqref{consistency conditions} allow a parameterization of $u$ and $\bar u$ by one real field $\varphi$ (and the field $\alpha$) in the following form
\be\label{u=varphi}
u=\p\varphi\, \tan(\a/2), \qquad \bar{u}=\pb\varphi\, \tan(\a/2)~.
\ee
The obtained two second order differential equations for $\alpha$ and $\varphi$ are Lagrangean with
\be\label{Lagrangian AdS-T}
{\cal L}=\frac{1}{2}\,\p\a\,\bar\p\a-\frac{1}{2}\,\tan^2({\a}/{2})~\p\varphi\,\pb\varphi-\cos\a~.
\ee
In terms of the new fields
\be\label{eq psi}
\psi_\pm=e^{\pm\frac \varphi 2}\,\,\sin(\a/2)~,
\ee
the Lagrangian \eqref{Lagrangian AdS-T} becomes algebraic
\be\label{Lagrangian Ads-T psi}
{\cal L}=\frac{\p\psi_+\,\pb\psi_-+\pb{\psi}_+\,\p\psi_-}{1-\psi_+\,\psi_-}-2\psi_+\psi_-~,
\ee
and the constant solution \eqref{u_0=} corresponds to
\be\label{solution psi}
\psi_\pm=\sqrt{1-\rho^2}~e^{\pm\rho^2[\cos(2\phi) \, \tau+\sin(2\phi) \, \s ]}~.
\ee

In the case of light-like $\mbox{AdS}_3$ projection (see \eqref{l-l consist cond}) one
can introduce the parameterization $u=\p\psi/\psi,\,$ $\bar u=\pb\psi/\psi$,
with real $\psi,$ which satisfies the linear equation $\pb\p\psi=\psi$. Thus, the Lagrangian here
is quadratic
\be\label{Lagrangian L-L}
{\cal L}=\p\psi\pb\psi+\psi^2~,
\ee
and our constant solution $u=-ie^{2i\phi}$ corresponds to $\psi=e^{\cos(2\phi) \, \tau+\sin(2\phi) \, \s }.$

One can similarly derive a Lagrangian system for
the space-like \mbox{AdS} projection. Instead of  \eqref{Lagrangian AdS-T}
one now gets the complex sinh-Gordon model with
\be\label{Lagrangian Ads-S psi}
{\cal L}=\frac{\p\psi_+\,\pb\psi_-+\pb{\psi}_+\,\p\psi_-}{1+\psi_+\,\psi_-}+2\psi_+\psi_-~.
\ee
Here $\psi_\pm$  parameterize the first and the second fundamental forms as in
\eqref{u=varphi} and \eqref{eq psi},
replacing there the trigonometric functions by hyperbolic ones. The solution for $\psi_\pm$
is also obtained from \eqref{solution psi} by changing the sign under the square root, that
keeps \eqref{solution psi} real.

Finally, for the spherical part one has the consistency conditions (see for example in \cite{Dorn:2009hs})
\be\label{S^3 consist-cond}
\p \pb \b+\sinh\b-\frac{v \bar{v}}{\sinh\b}=0~,\quad \pb v-\frac{\p \beta}{\sinh\beta}\bar{v}=0~,
\ee
where $v=M\cdot \p^2X$ (with normal vector $M$) is the coefficient of the second fundamental form
and $\beta$ parameterizes the induced metric on the $\mbox{S}^3$ projection like $\a$ on $\mbox{AdS}_3$
\be\label{S^3-metric}
(f_s)_{ab}= \begin{pmatrix}\cosh^2(\b/2) & 0\\0&\sinh^2(\b/2) \end{pmatrix}~.
\ee
Then, in a similar way (with $v = i \p \varphi_s \tanh (\beta / 2)$) one finds the Lagrangian
\be\label{Lagrangian Sp psi}
{\cal L}=\frac{\p\psi\,\pb\bar\psi+\pb{\psi}\,\p\bar\psi}{1+\psi\,\bar\psi}-2\psi\bar\psi~,
\ee
with a complex field
\be\label{S^3 psi}
\psi=e^{\frac{i\varphi_s}{2}}~\sinh({\beta}/{2})~.
\ee
Constant solutions of \eqref{S^3 consist-cond} leading to \eqref{solution X} correspond to
$\psi=\rho_s\,e^{i\sqrt{1+\rho_s^2}\,[\sin(2\phi_s) \, \tau -\cos(2\phi_s) \, \s]}$.

\setcounter{equation}{0}

\section{Analysis of the boundaries}
For the investigation of the boundary behavior of our surfaces it is very convenient to use their description as (part of the) intersections of quadrics with the $\mbox{AdS}_3$ hyperboloid in $\rr ^{(2,2)}$. Global $\mbox{AdS}_3$ coordinates just implementing the conformal map to one half of the Einstein static universe (ESU) are

\be
Y^{0'}=\frac{1}{\cos\vartheta}~\sin t~,~~~Y^0=\frac{1}{\cos\vartheta}~\cos t~,~~~Y^1=
\tan\vartheta ~\cos\gamma ~,~~~Y^2=\tan\vartheta ~\sin\gamma~,
\ee
with $0\leq\vartheta<\frac{\pi}{2},~~0\leq\gamma<2\pi,~~-\pi\leq t<\pi$. This map is the basis for the fig.~\ref{plot}, $\vartheta=\frac{\pi}{2}$ corresponds to the boundary of $\mbox{AdS}_3$. We demonstrate the procedure with a combined discussion of the space-like and time-like tetragon solution, fig.~\ref{plot}(a) and \ref{plot}(d).

Here the quadrics (\ref{quadratic AdS 1}) and (\ref{quadratic 3.1}) imply
\be
\label{esu}
\cos t ~=~\pm~\sqrt{ \sin^2\vartheta\cos^2\gamma +\kappa \cos^2\vartheta}~,
\ee
with $\kappa=\sin^2\theta\in (0,\frac{1}{2}]$ in the space-like case and $\kappa=-\sinh^2\theta <0$ in the time-like case. Obviously, for the space-like case all values of $\vartheta$ inside the ESU cylinder are allowed, and for a given  $\vartheta$ the angle variable $\gamma$ can
go around a full circle. On the other side, in the time-like case reality
of the square root requires $\vert\kappa\vert\leq\tan^2\vartheta$ and $\cos^2\gamma\geq\vert\kappa\vert\cot^2\vartheta$. Now there is an inner region of the ESU cylinder
not reached by the surface, and beyond this region the allowed values for $\gamma$
fall into two disjoint arcs of a circle.

In both cases, due to the symmetry and periodicity properties of $\cos t$, we see that to each allowed pair $(\vartheta,\gamma)$ belong just four different values in $t\in[-\pi,\pi)$, as long as $\vartheta<\frac{\pi}{2}$. Therefore, the intersections under discussion both
consist out of four disconnected pieces touching each other only at the boundary.
\begin{figure}
\begin{center}
\begin{tabular}{ccccc}
\includegraphics[width=3.9cm,bb=0 0 240 244]{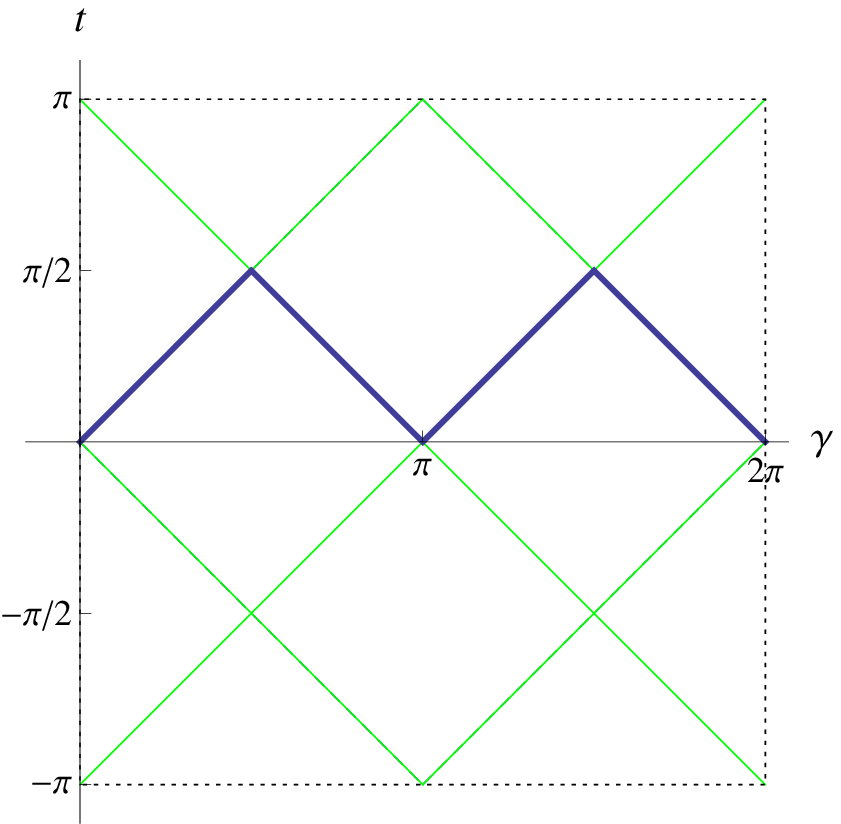}&\hspace{200pt}&\includegraphics[width=3.9cm,bb=0 0 240 244]{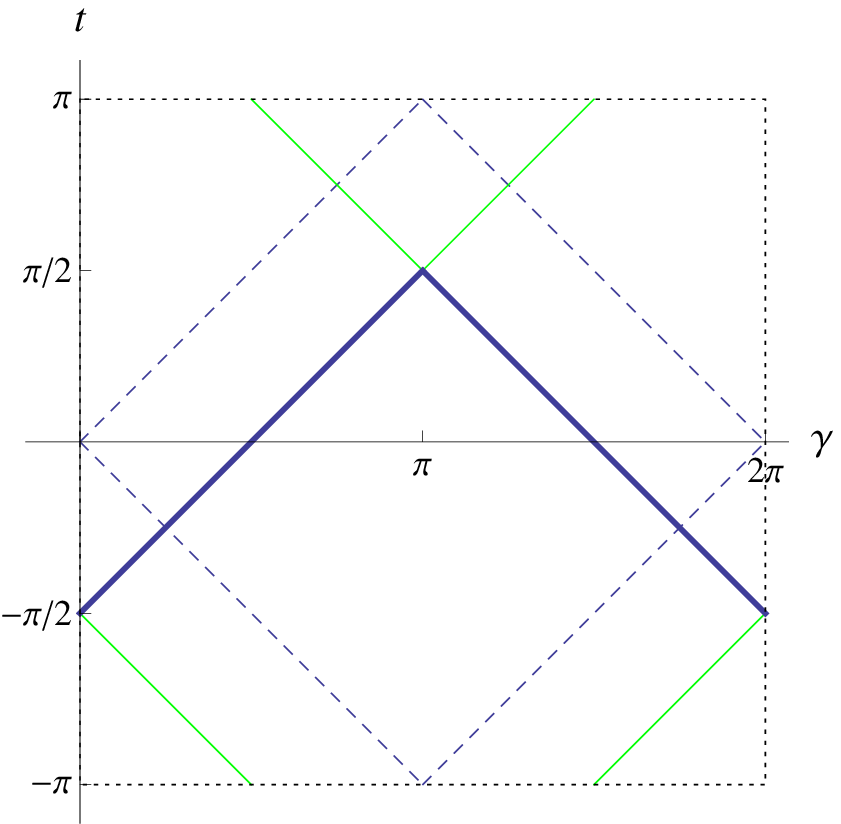} &\hspace{200pt}& \includegraphics[width=3.9cm,bb=0 0 240 244]{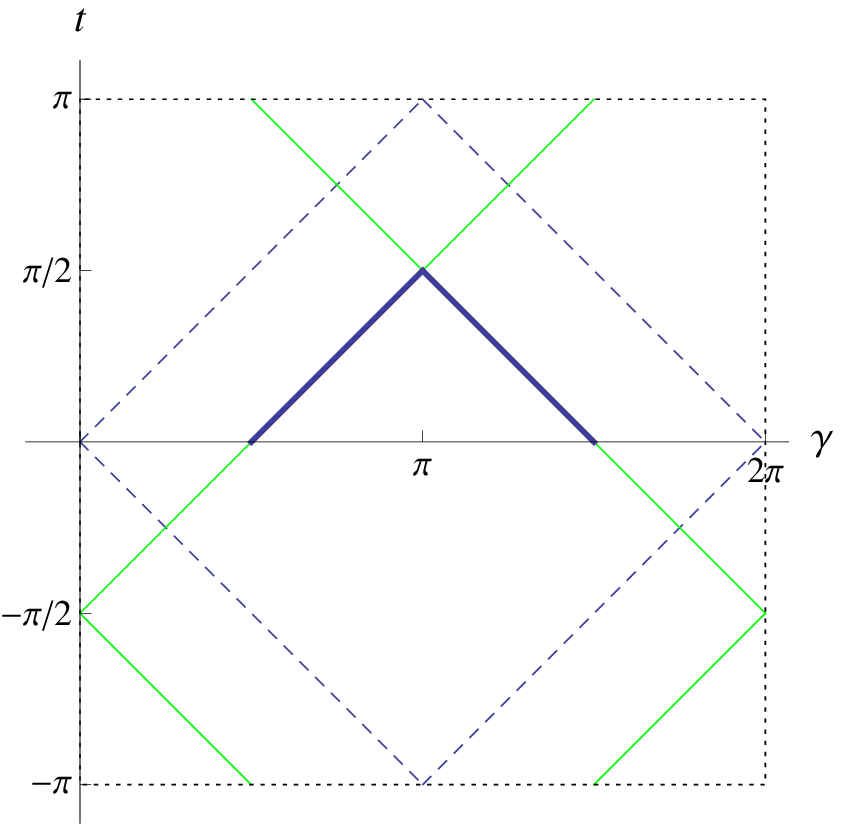}
\\(a)&&(b)&&(c)\\
\includegraphics[width=3.9cm,bb=0 0 240 244]{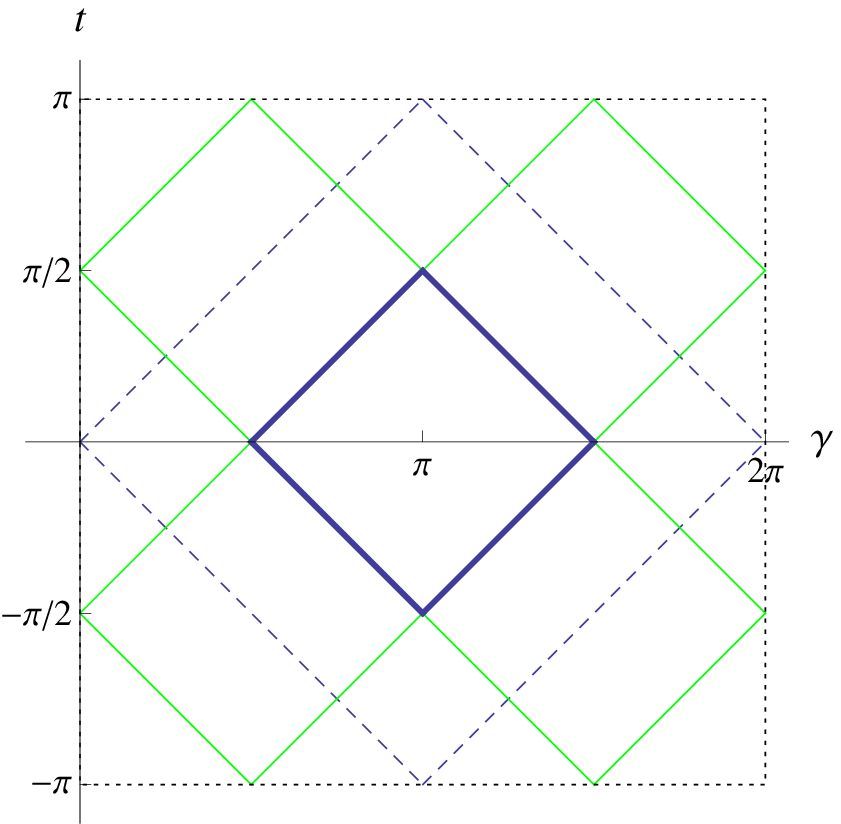} &\hspace{200pt}&\includegraphics[width=3.9cm,bb=0 0 240 244]{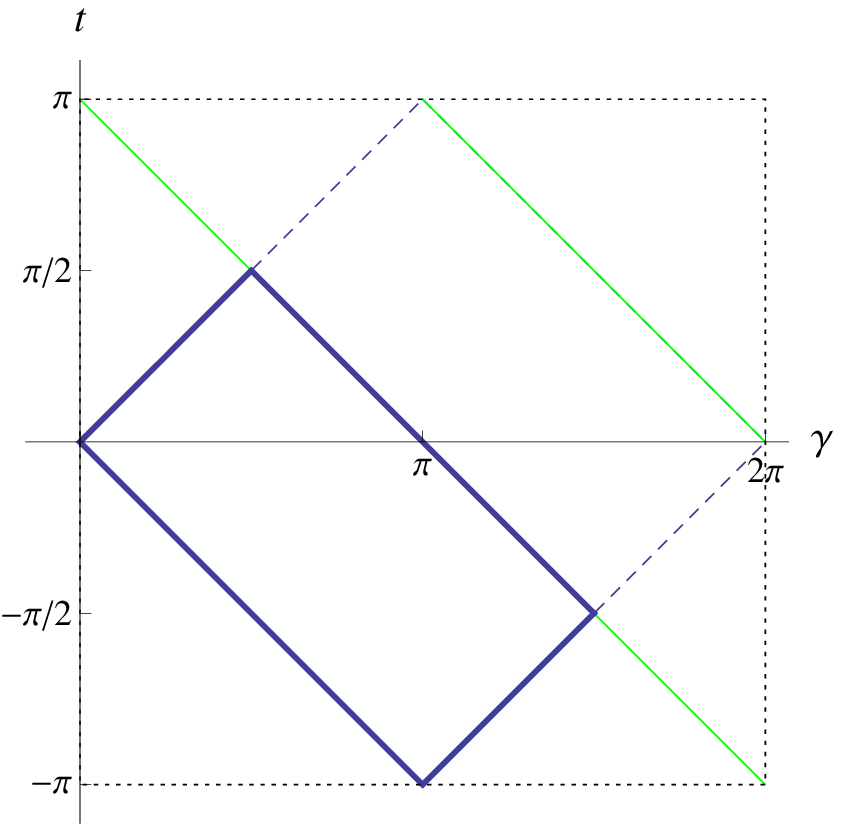}&\hspace{200pt}& \includegraphics[width=3.9cm,bb=0 0 240 244]{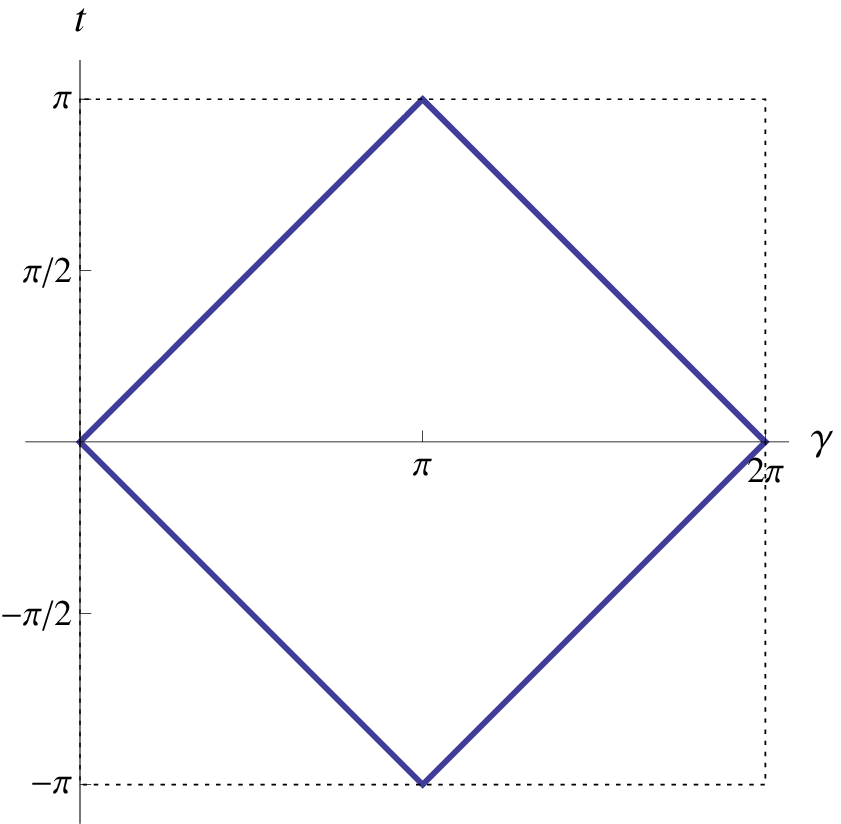}
\\(d)&&(e)&&(f)
\end{tabular}
\begin{tabular}{ccc}
\includegraphics[width=3.9cm,bb=0 0 240 244]{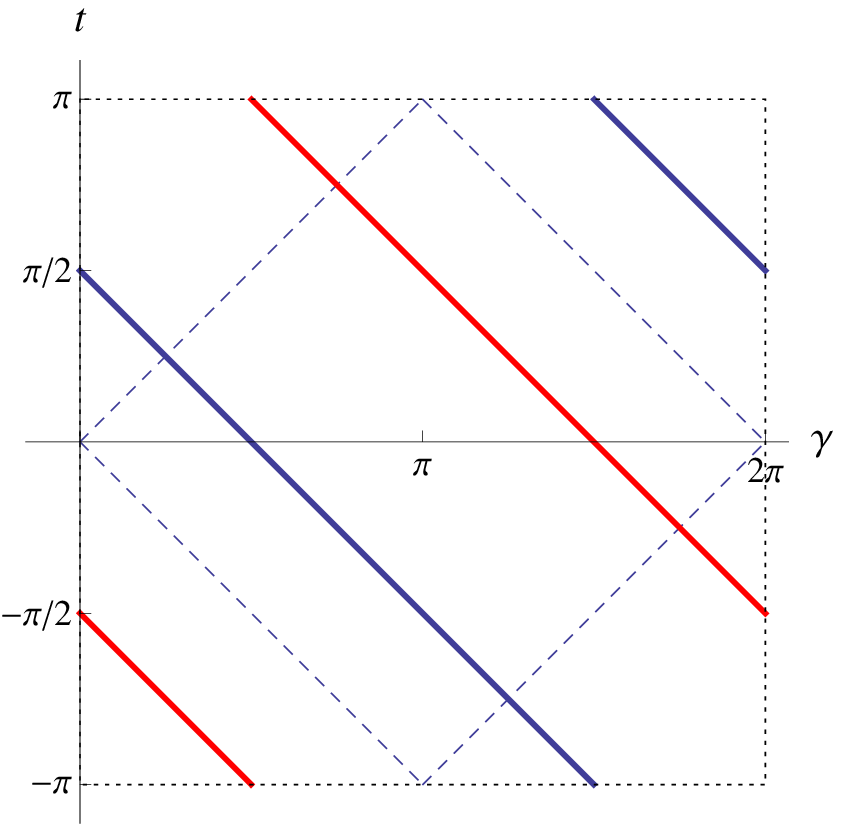}&\hspace{200pt}& \includegraphics[width=3.9cm,bb=0 0 240 244]{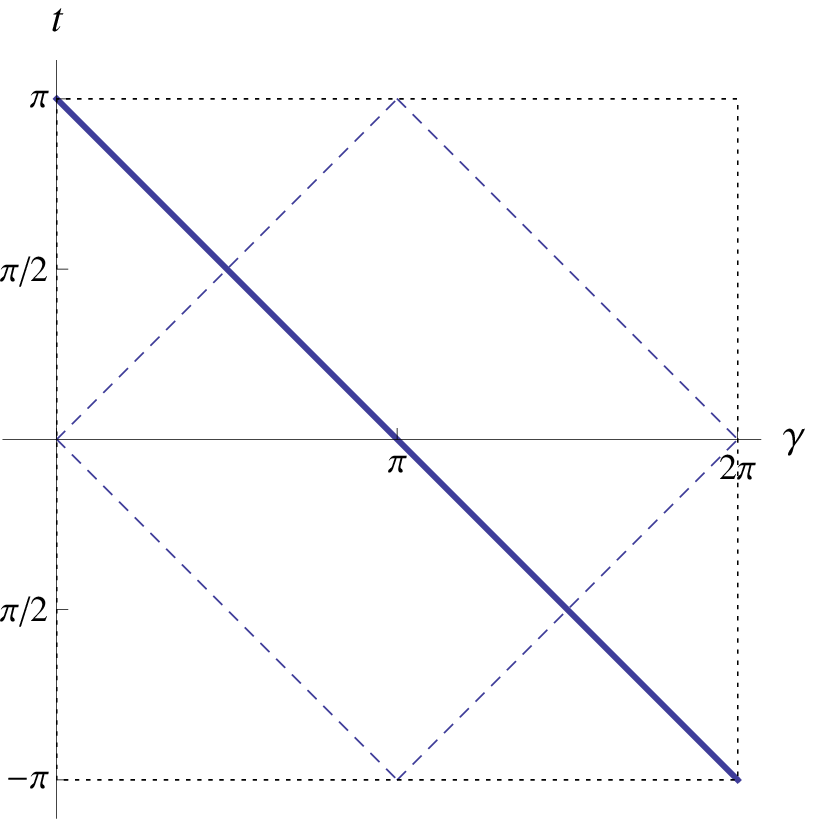}
\\(g)&&(h)
\end{tabular}
\end{center}
\caption{These are the conformal boundaries of all classes from fig.~\ref{plot} that touch the boundary. Additionally, the dashed line is the boundary of a single Poincar\'{e} patch.}
\label{fig:Bound}
\end{figure}
Each of these four pieces is a surface of the type depicted in fig.~\ref{plot}(a) or fig.~\ref{plot}(d), respectively.

Independent of the value of $\kappa$, \eqref{esu} at the boundary of $\mbox{AdS}_3$ becomes
$\cos t=\pm \vert\cos\gamma\vert$ which in $0\leq\gamma<2\pi,~~-\pi\leq t<\pi$
has the null line solutions
\be
\label{esu-bound}
t=\pm \gamma ~,~~~t=\pi-\gamma ~,~~~t=-\pi+\gamma~,~~~t=2\pi-\gamma~,~~~t=-2\pi+\g~.
\ee
Out of this net of null lines the boundary of the four parts of the intersection
of (\ref{quadratic AdS 1}) (space-like case) with the hyperboloid is formed from four zigzag null lines going around
the cylinder and each staying within a time slice of width $\frac{\pi}{2}$.
For the intersection of (\ref{quadratic 3.1}) (time-like case) the boundaries
are formed by four squares organized in two pairs with partners just being mirror
images under reflection at the cylinder axis.

\begin{figure}[ht]
\begin{center}
\begin{tabular}{cccccccccc}
\includegraphics[height=4cm,bb=0 0 266 432]{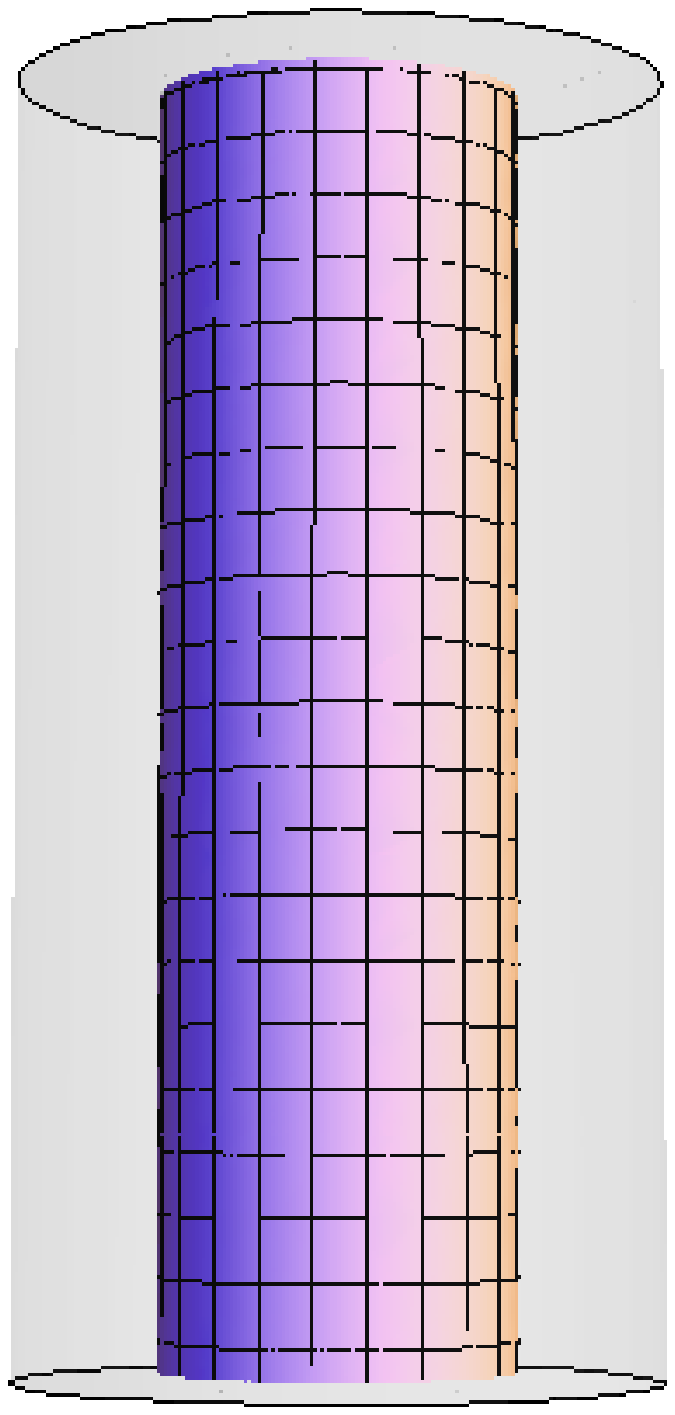} &&&\includegraphics[height=4cm,bb=0 0 266 432]{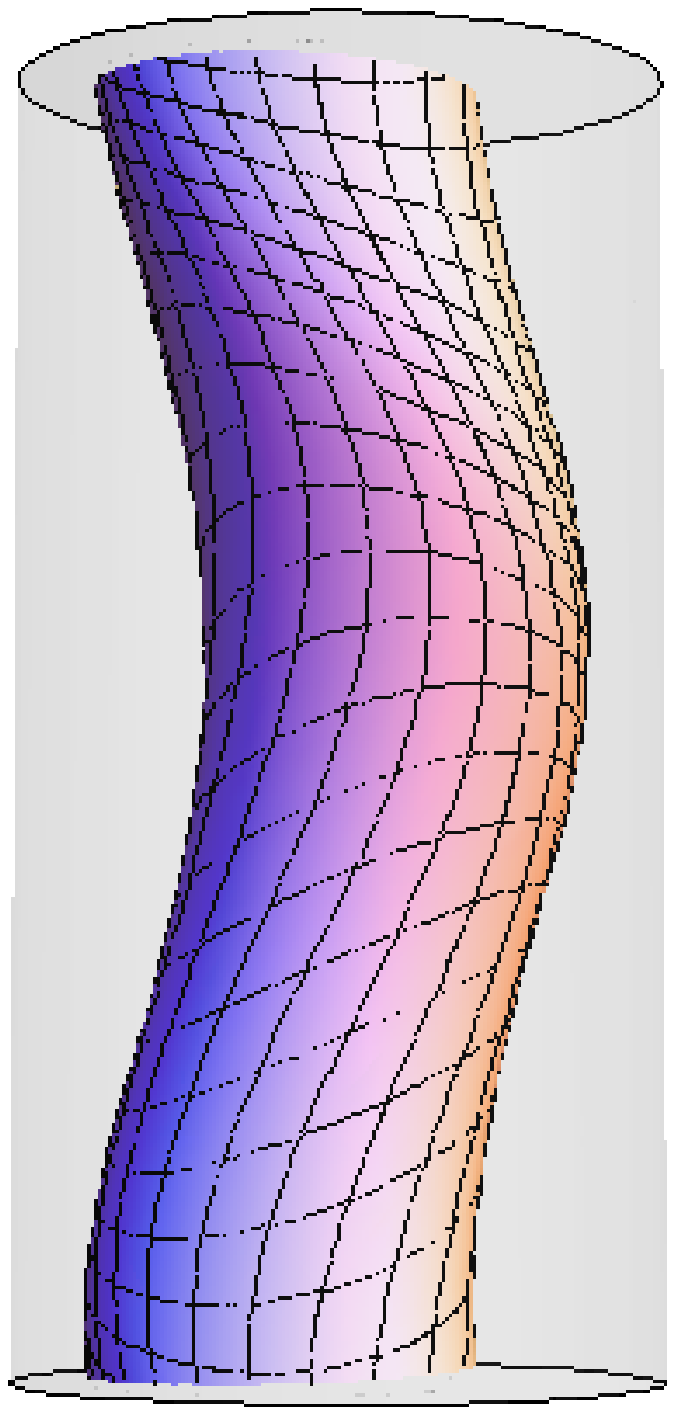} &&&\includegraphics[height=4cm,bb=0 0 266 432]{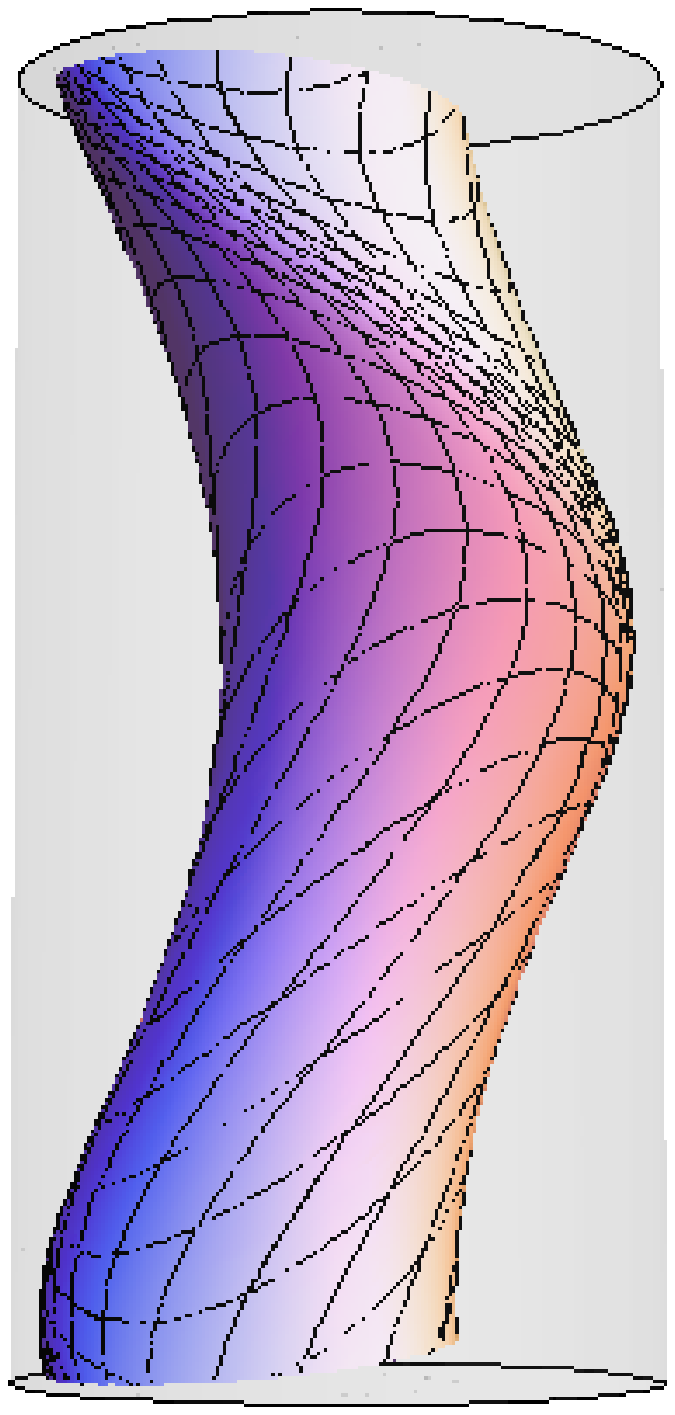}&&&\includegraphics[height=4cm,bb=0 0 266 432]{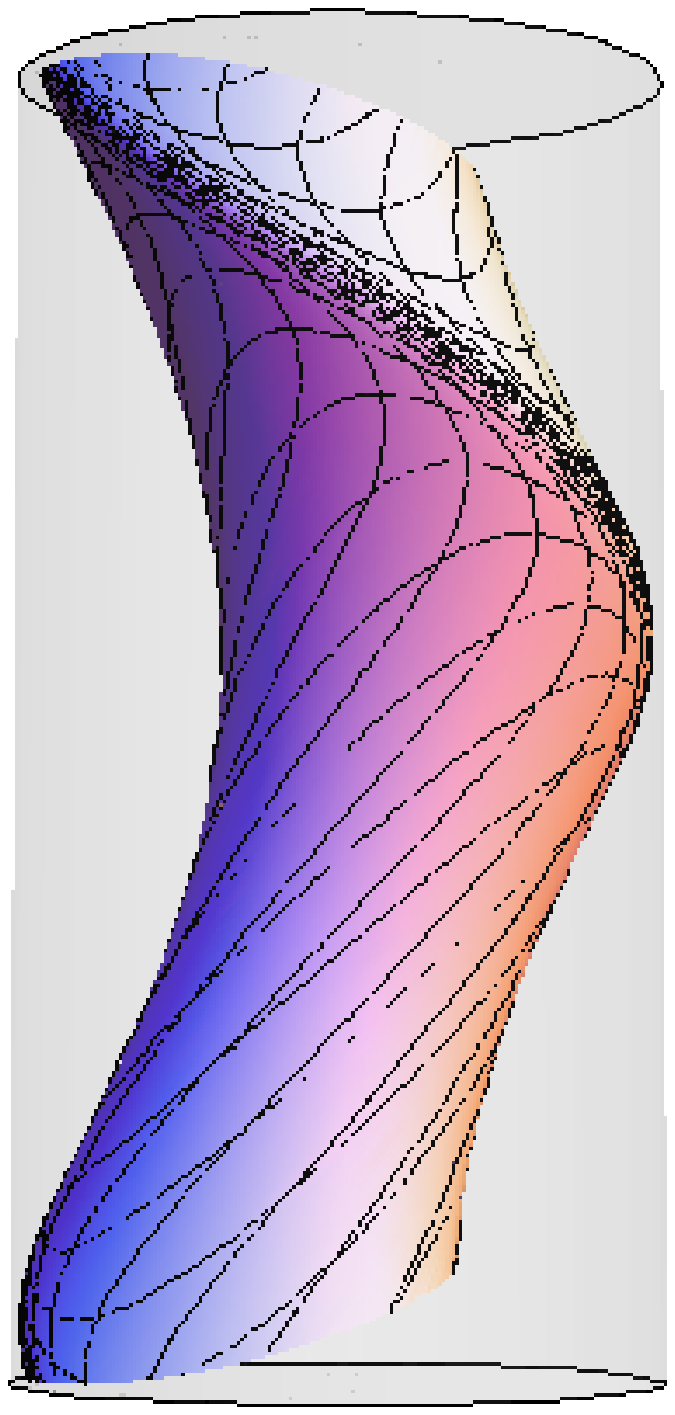} \\
\includegraphics[height=4cm,bb=0 0 266 432]{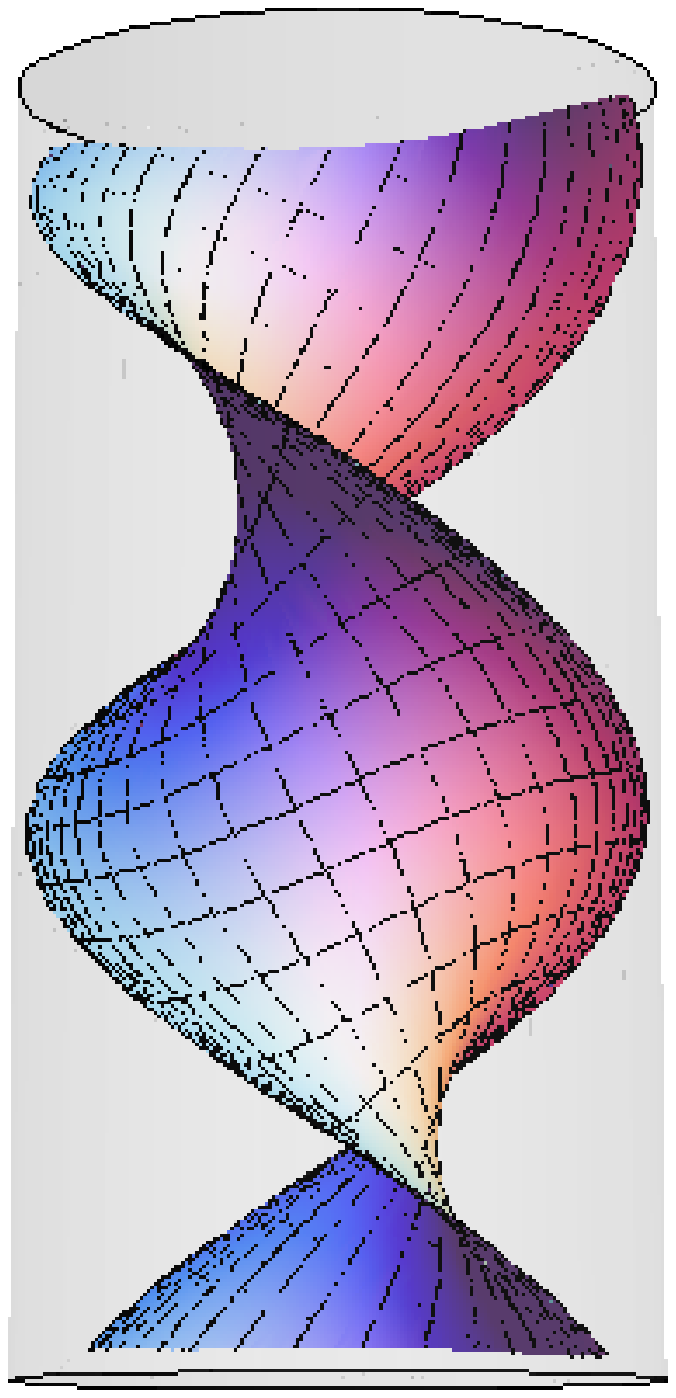} &&&\includegraphics[height=4cm,bb=0 0 266 432]{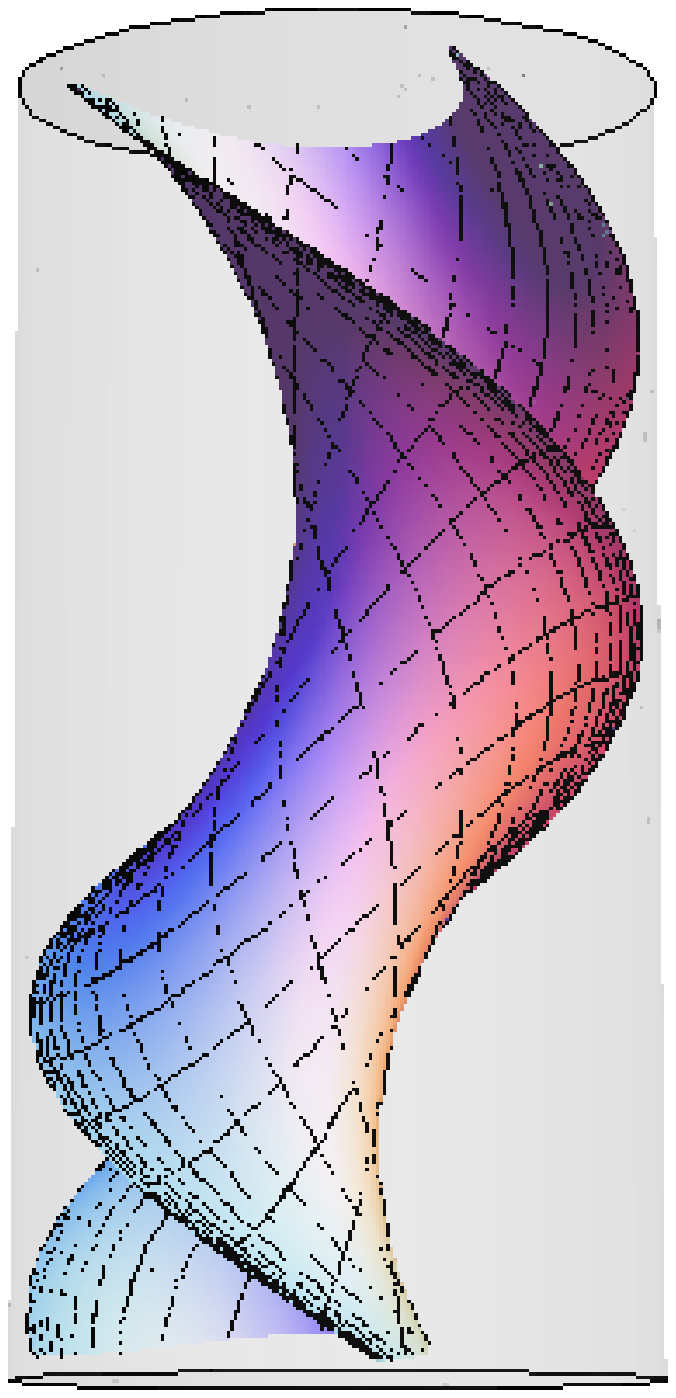} &&&\includegraphics[height=4cm,bb=0 0 266 432]{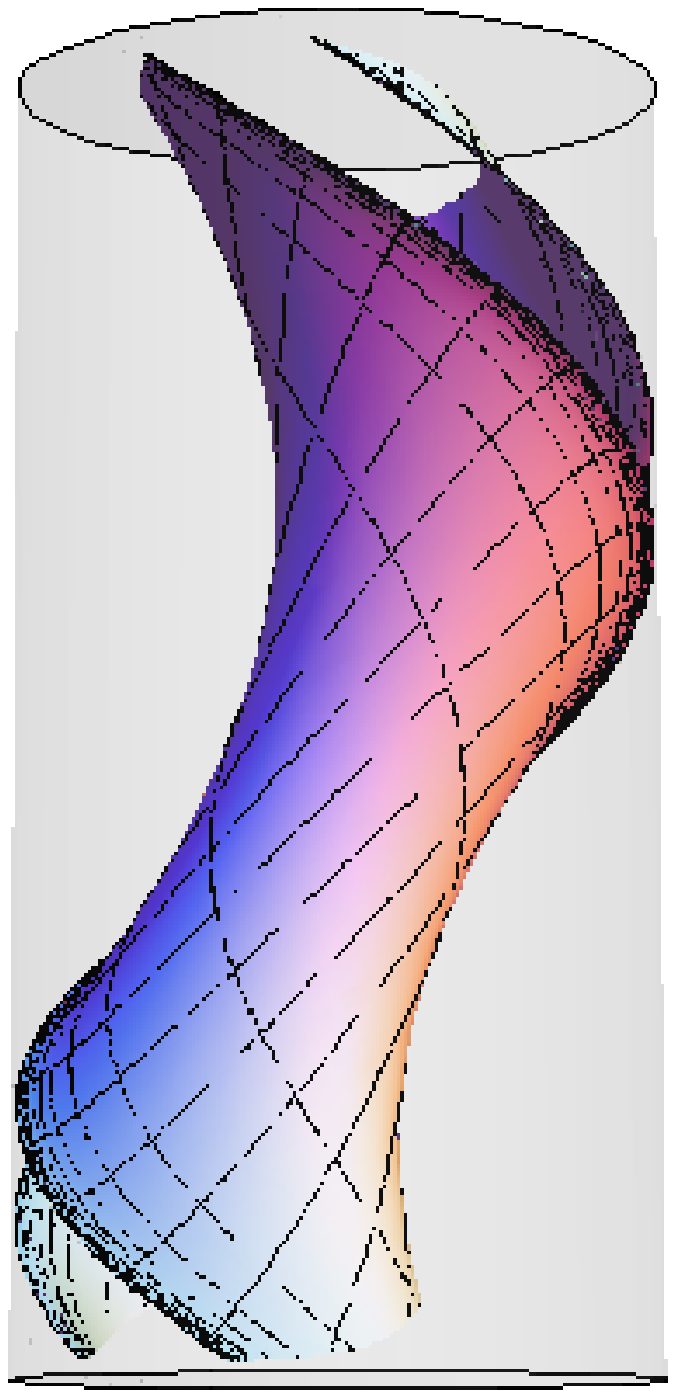} &&&\includegraphics[height=4cm,bb=0 0 266 432]{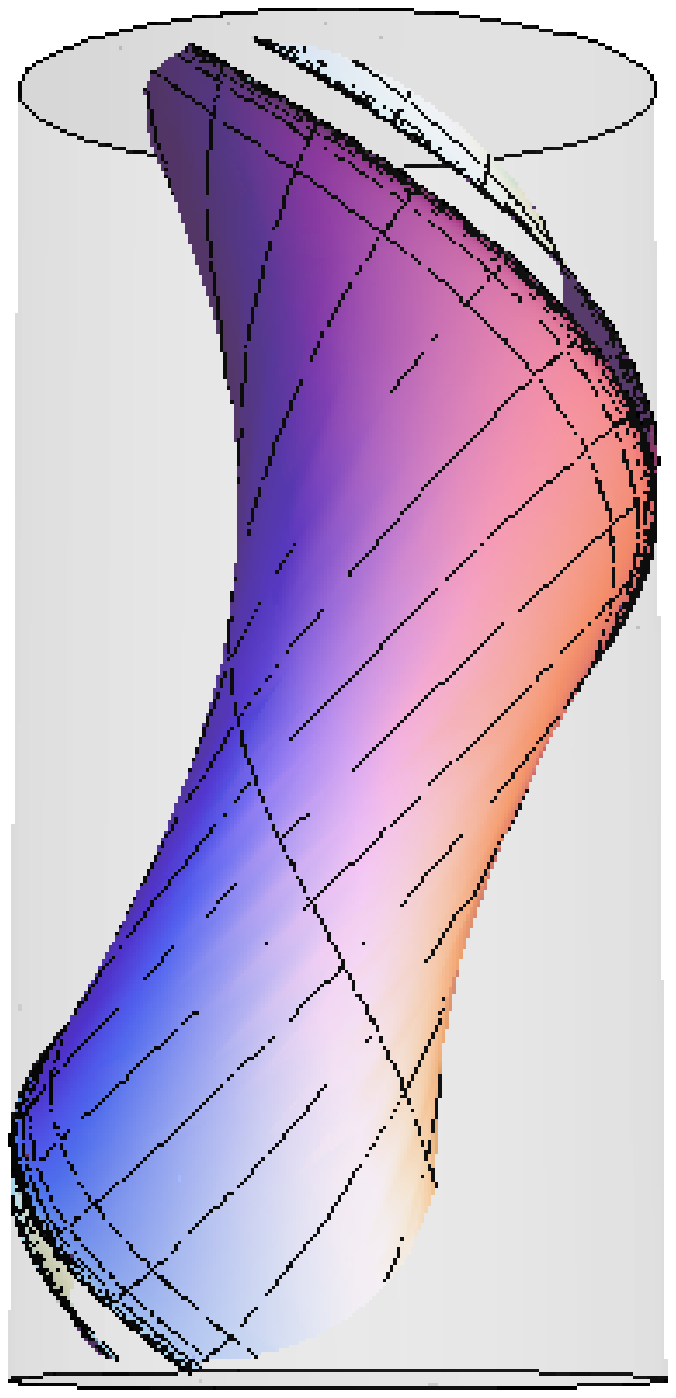}
\end{tabular}
\end{center}
\caption{Via boosts one can change the apparent shape of the time-like tube and the time-like two line solution to look similar to a time-like one line surface.}
\label{fig:boost}
\end{figure}

In fig.~\ref{fig:Bound}(a) and \ref{fig:Bound}(d)  we show the boundary configuration for one connected part of the intersections, i.e. for our previously constructed surfaces and depicted in fig.~\ref{plot}(a) and \ref{plot}(d). An additional shift in time and $\gamma$ has been implemented in the plot for graphical reasons in the time-like case. We also indicate in this and the following figures the boundary
of a suitable chosen Poincar\'e patch, i.e. a conformal representation of two-dimensional Minkowski space.
In a similar manner we can analyze the boundary behavior of all the other classes of surfaces shown in fig.~\ref{plot}. There are two light-like classes, fig.~\ref{plot}(b) and fig.~\ref{plot}(c), which we call the light-like two-gon and wedge, respectively. Their boundary is represented in fig.~\ref{fig:Bound}(b) and \ref{fig:Bound}(c). The remaining time-like cases are also visualized in fig.~\ref{fig:Bound}.
After this discussion let us summarize the properties of our classification. As indicated in fig.~\ref{Fig:RhoPhi}, to each point in the $(\rho,\phi)$-diagram corresponds a projection of the surface onto $\mbox{AdS}_3$ plus all information allowing to reconstruct, together with the analogous input from the $\mbox{S}^3$ projection, the full solutions in $\mbox{AdS}_3\times \mbox{S}^3$. This is true up to isometry transformations $\in \mbox{O}(2,2)$. If one further is interested only in the $\mbox{AdS}$ projection as a surface per se, points on the dashed lines in fig.~\ref{Fig:RhoPhi} have to be identified. Then using the $\mbox{O}(2,2)$ freedom we have generated for each class some simple explicit expressions for the embedding coordinates. Related to these representations are simple quadrics which define them as (parts of) intersections with the $\mbox{AdS}_3$ hyperboloid. In this manner we constructed within each class a canonical representative. While for parts of the classes the shape is fixed completely, for other classes there is a free parameter encoding the possible values for the constant mean curvature. The corresponding nine canonical $\mbox{AdS}$ projections are shown in fig.~\ref{plot}. All other projections can be generated out of them by applying $\mbox{O}(2,2)$ transformations (for illustration see fig.~\ref{fig:boost}).

To summarize, in the following, we indicate the name, the corresponding position in fig.~\ref{plot}, the intersecting quadrics, the range for $\rho$  and $\phi$ related to its positions in the diagram fig.~\ref{Fig:RhoPhi} and its mean curvature. We also add the topological characterization of the boundary as (parts) of left and/or right going null lines on $\rr\times \mbox{S}^1$.\\

\noindent{\bf Space-like tetragon, fig.~\ref{plot}(a):}\\[2mm]
$Y_0^2-Y_1^2=\sin ^2\theta~,~~~~~~~~$ region above line $AB~,~~~
H=\cot 2\theta~, ~~~~~$\\
two left, two right.\\[2mm]
{\bf Light-like two-gon, fig.~\ref{plot}(b):}\\[2mm]
$(Y^0-Y^1)(Y^{0'}+Y^2)=0~,~~~~~~~~$ point $A$ or $B$, $~~~H$ ill defined,\\
one left, one right.\\[2mm]
\vspace{3mm}

\noindent{\bf Light-like wedge, fig.~\ref{plot}(c):}\\[2mm]
$Y_0^2-Y_1^2=0~,~~~~~~~~$ open line segment $AB$, $~~~H$ ill defined,\\
one left, one right, part of boundary of the $\mbox{AdS}$ projection inside. \\[2mm]
\noindent{\bf Time-like tetragon, fig.~\ref{plot}(d):}\\[2mm]
$Y_0^2-Y_1^2=-\sinh ^2\theta~,~~~~~~~~$ interior of triangle $ABC~,~~~H=-\coth 2\theta~,$\\
one left, one right.\\[2mm]
\noindent{\bf Time-like special antisymmetric case, fig.~\ref{plot}(e):}\\[2mm]
$(Y^{0'}+Y^1)^2-(Y^0+Y^2)^2=1~,~~~~~~~~$ open line segments $AC$ or $BC~, ~~~~H=-1 ~,$\\
two left, one right $or$ one left two right.\\[2mm]
\noindent{\bf Time-like two-gon, fig.~\ref{plot}(f):}\\[2mm]
$(Y^{0'}+Y^1)^2=1~,~~~~~$ point $C~,~~~H=-1~,$\\
one left, one right.\\[2mm]
\noindent{\bf Time-like two line case, fig.~\ref{plot}(g):}\\[2mm]
$Y^{0'}Y^1-Y^0Y^2=\frac{1}{2}\sinh2\theta~,~~~~~~$ interior of triangles $ACD$ or $BCD~,~~~~H=-\tanh 2\theta~,$\\
two left $or$ two right.\\[2mm]
\noindent{\bf Time-like one line case, fig.~\ref{plot}(h):}\\[2mm]
$(Y^{0'}+Y^1)^2+(Y^0-Y^2)^2=1~,~~~~~~~~~$  open line segments $CD$ or $CE~,~~~~H=-1~,$
\\one left $or$ one right.\\[2mm]
\noindent{\bf Time-like tube, fig.~\ref{plot}(i):}\\[2mm]
$Y_1^2+Y_2^2=\sinh^2\theta~,~~~~~~~~~$ interior of triangle $DCE~,~~~~~H=-\coth 2 \theta~,$\\
does not touch boundary of $\mbox{AdS}$.\\

Isometry transformations on $\mbox{AdS}_3$ can be continued to the conformal boundary where they act as conformal transformations. By conformal transformation one can change the distance of two parallel light-like lines. Using this we can deform the apparent shape with respect to the chosen coordinate system of the surfaces,\footnote{Just in the same sense in which one can in any finite domain of Minkowski space or $\mbox{AdS}$ approximate a light ray by a ultrarelativistic massive particle trajectory.} but the geometric shape is of course untouched. For example, one can decrease the distance between the two lines in the two-line solution fig.~\ref{fig:Bound}(g) with a suitable isometry transformation. With an infinite boost, one can bring them exactly together and obtain the one-line solution fig.~\ref{fig:Bound}(h).

The boundary of the time-like tetragon fig.~\ref{fig:Bound}(d) solution consists of two left-moving and two right-moving lines. Using a boost one can decrease the distance between the right-moving lines and in the limit obtain the asymmetric solution fig.~\ref{fig:Bound}(f). There are further examples of how different classes of solutions are connected via infinite boosts.

\setcounter{equation}{0}
\section{Embedding in $\mbox{AdS}_5$ and a regularized area}
For potential use in the correspondence to 4-point gluon scattering amplitudes, with generic kinematics in the center of mass system, one has to consider boundary null tetragons in 3-dimensional Minkowski space conformal to parts of the boundary of $\mbox{AdS}_4\subset \mbox{AdS}_5$ . Here beyond the space-like tetragon solution, corresponding to a $u$-channel configuration and analyzed in \cite{Dorn:2009hs}, only the time-like
tetragon solution can be relevant. All other classes have either not enough cusps
or their boundary contains points which are antipodal in $\mbox{AdS}_3$. Antipodes in $\mbox{AdS}_3\subset \mbox{AdS}_5$ are automatically antipodes in $\mbox{AdS}_5$. The antipodal property is invariant under $\mbox{O}(2,4)$, hence there is no choice of Poincar\'{e} patches in $\mbox{AdS}_5$, such that the whole boundary of the surface fits in the conformal image of four-dimensional Minkowski space.

In the time-like tetragon one pair of opposite cusps is time-like and the other one space-like
separated. It corresponds to a $s$-channel scattering configuration. Seemingly there is no corresponding minimal surface in pure $\mbox{AdS}_5$. The area for the space-like tetragon case has been calculated in \cite{Dorn:2009hs}. Let us sketch the corresponding calculation for the time-like tetragon.

We start with (\ref{solution Y 3.1}) after some trivial renaming
\begin{equation}\label{solution Y 3.1'}
Y^{0'}= \sinh \th \, \sinh\eta~, ~ Y^0=\cosh\th \, \cosh \xi~, ~ Y^1=\cosh\th \, \sinh\xi~,~ Y^2=\sinh\th \, \cosh\eta~
\end{equation}
and $Y^4=Y^3=0$. By the application of isometry transformations in $\mbox{AdS}_4\subset \mbox{AdS}_5$ we can generate a boundary configuration
related to a generic scattering kinematics. $Y^4$ will be untouched during this procedure.

\begin{figure}
\centering
\includegraphics[scale=0.4]{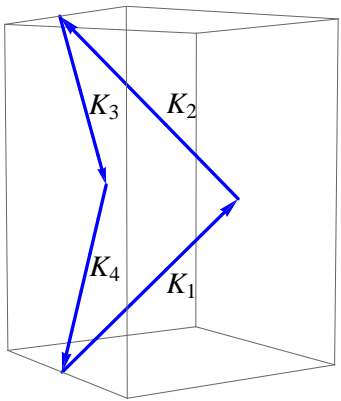}
\caption{The time-like tetragon configuration in Minkowski space corresponding to s-channel scattering.}
\label{cube}
\end{figure}

Let us consider the two $\mbox{SO}(2,3)$ matrices
\begin{equation}
A = \begin{pmatrix} 1&0&0&0&0\\0&\frac{1+a^2}{2a}&0&0&\frac{1-a^2}{2a}\\ 0&0&1&0&0\\0&0&0&1&0&\\0&\frac{1-a^2}{2a}&0&0&\frac{1+a^2}{2a} \end{pmatrix}~, \quad \quad \quad B = \begin{pmatrix} 1&0&0&0&0\\0&1&0&0&0\\0&0&1&0&0\\0&0&0&\frac{1}{\sqrt{1+b^2}}&\frac{-b}{\sqrt{1+b^2}}\\0&0&0&\frac{b}{\sqrt{1+b^2}}&\frac{1}{\sqrt{1+b^2}} \end{pmatrix}~.
\end{equation}
Calculating $A^I_{\,\,J} \cdot B^J_{\,\,K} \cdot  Y^K$ and then introducing Poincar\'{e} coordinates $(r,\,y^\mu),~ \mu \in \{0',1,2\}$ via
\begin{eqnarray}\label{Poincare coord}
Y^\mu = \frac{y^\mu}{r}, ~~~~~~Y^{0}+Y^3=\frac 1 r~,~~~~~
Y^{0}-Y^3=\frac{r^2 +y_\mu y^\mu}{r}~,
\end{eqnarray}
we find that
\be\ba\label{r=} r&=\frac{a\,\sqrt{1+b^2}}
{\sqrt{1+b^2}\cosh\theta\cosh\xi+b\sinh\theta\,\cosh\eta}~,&
y^{0'}&=\frac{a\sqrt{1+b^2}\,\sinh\theta\,\sinh\eta}
{\sqrt{1+b^2}\cosh\theta\cosh\xi+b\sinh\theta\,\cosh\eta}~,\\
y^1&=\frac{a\,\sqrt{1+b^2}\cosh\theta\,\sinh\xi}
{\sqrt{1+b^2}\cosh\theta\cosh\xi+b\sinh\theta\,\cosh\eta}~,&
y^2&=\frac{a\,\sinh\theta\,\cosh\eta}
{\sqrt{1+b^2}\cosh\theta\cosh\xi+b\sinh\theta\,\cosh\eta}
~. \ea\ee
Examining \eqref{solution Y 3.1'} we see that the cusps are approached by taking either $\xi$ or $\eta$ constant and sending the other variable to $\pm \infty$. These limits are $\theta$-independent. Thus we find the cusps located at
(see fig.~\ref{cube})
\begin{equation}
\vec c_1= \left(-\frac{a\sqrt{1+b^2}}{b}\,,0,\frac a b\right)~,~\vec c_2= \left(0,a,0\right)~,~\vec c_3= \left(\frac{a\sqrt{1+b^2}}{b}\,,0,\frac a b\right)~,~\vec c_4= \left(0,-a,0\right)~,
\end{equation}
with $\vec c = (y^{0'},y^1,y^2)$. It can be easily checked that the edges of the tetragon are light-like. The momenta associated with the edges of the tetragon are defined by $2\pi\,k^\mu=\Delta c^\mu$. In this way we get
\be\ba\label{momenta}
k_1&=\frac{1}{2}\left(\sqrt{s},\,\sqrt{-t},\,-\sqrt{-u}\right),&
k_2&=\frac{1}{2}\left(\sqrt{s},\,-\sqrt{-t},\,\sqrt{-u}\right),\\
k_3&=\frac{1}{2}\left(-\sqrt{s},\,-\sqrt{-t},\,-\sqrt{-u}\right),&
k_4&=\frac{1}{2}\left(-\sqrt{s},\,\sqrt{-t},\,\sqrt{-u}\right)~.
\ea\ee
Here $(s,\,t,\,u)$ are the Mandelstam variables,
which are related to the parameters $(a,\, b)$ by
\begin{equation}\label{Mandelstam ver}
s=-2k_1\cdot k_2=\frac{a^2(1+b^2)}{\pi^2b^2}\,~,~~~~~~-t=2k_1\cdot k_4=\frac{a^2}{\pi^2}
~, ~~~~~
-u=2k_1\cdot k_3=\frac{a^2}{\pi^2\,b^2}~.
\end{equation}
To calculate  the regularized action, we introduce a constant cutoff at $r=r_c$ in Poincar\'{e} coordinates.
 Terms that vanish for $r_c \rightarrow 0$ will be discarded. The metric tensor on the surface (including by both $\mbox{AdS}_3$ and $\mbox{S}^3$ parts) reads
\begin{equation}
g_{ab}(\sigma ,\tau )=f_{a\,b}+ (f_s)_{a\,b}=(\rho^2+\rho^2_s)\,\delta_{ab}~.
\end{equation}
Then the regularized action \eqref{action} is given by
\begin{equation}\label{S-reg}
S_{reg}=\frac{\sqrt{\lambda}}{2\pi}(\rho^2+\rho^2_s) \int_{r \geq r_c} \text{d} \tau \text{d} \sigma ~.
\end{equation}
The integral runs over that part of the $(\sigma,\,\tau$)- plane where
 $r\geq r_c$. In the $(\xi,\,\eta)$- plane this area is bounded by the contour
 \be\label{contour}
 \epsilon\,\cosh\eta+\epsilon'\,\sinh\xi=1,
 \ee
 with
 \be\label{epsilon, epsilon'}
 \epsilon=\frac{\sinh\theta}{\pi\,\sqrt s}~, \quad\quad \epsilon'=\frac{\cosh\theta}{\pi\,\sqrt{-t}}~,
 \ee
and we can rewrite \eqref{S-reg} as
\begin{equation}\label{S-reg 1}
S_{reg}= \frac{\sqrt{\lambda}}{2\pi}(\rho^2+\rho^2_s) \, 2 \, {\cal J}\, \cdot I(r_c)~.
\end{equation}
Here ${\cal J}$ is the Jacobian of the linear map between the $(\xi,\,\eta)$ and $(\s,\,\tau)$ coordinates and
\begin{equation}
I(r_c) = \frac 1 2 \int_{r \geq r_c} \text{d} \eta \text{d} \xi~.
\end{equation}
The dependence of $I(r_c)$ on $\epsilon$ and $\epsilon'$ can be taken from \cite{Dorn:2009hs}. Inserting the new definitions from \eqref{epsilon, epsilon'} we find
\begin{equation}
I(r_c)= \frac 1 4 \left(\log \frac{r_c^2 \sinh^2 \theta}{4\pi^2 \, s} \right)^2+\frac 1 4\left( \log\frac{r_c^2 \cosh^2 \theta}{-4\pi^2\, t} \right)^2 -\frac 1 4 \left(\log\frac{s\,\coth^2 \theta}{-t} \right)^2 - \frac{\pi^2}{3}~.
\end{equation}
With the Jacobian defined by \eqref{rescal} this yields
\begin{equation}\label{area}
S_{reg}=\frac{\sqrt{\lambda}}{2\pi}\frac{(\rho^2+\rho^2_s)\sinh 2 \theta}{\rho \sqrt{1-\rho^2}}  \, ~I(r_c)~.
\end{equation}

Formally this expression for $S_{reg}$ can also be obtained from the corresponding formula for the space-like tetragon in \cite{Dorn:2009hs}
by $\theta \rightarrow i\theta~$ and continuation of
$\rho ^2$ (remember footnote \ref{foot4})
from values above 1 to below 1. For the space-like case it was found, that the factor in front of $I(r_c)$ is always $\geq 1$ and approaches its lower bound for $\theta\rightarrow\frac{\pi}{4}~,~\rho^2\rightarrow\infty$ and
$\rho_s\rightarrow 0$.
In addition, for  $\theta =\frac{\pi}{4}$, $I(r_c)$ coincides with the pure $\mbox{AdS}$ case.

Now the situation is different. To analyze the prefactor in \eqref{area} we  use the inequality
\be\label{sinh>}
\sinh 2\theta (\rho,\phi  ) \geq \sinh 2\theta(\rho,\pi/4)=\frac{2\rho\sqrt{1-\rho^2}}{2\rho^2-1}~.
\ee
which follows from \eqref{thetadef}.
Setting $\rho_s=0$ and $\sinh2\theta$ to its minimal value  one finds  the prefactor $2\rho^2/(2\rho^2-1)$,
which is
$> 2$ in the interval $\rho\in(1/2,1)$. It approaches its lower bound $2$ for $\rho ^2\rightarrow 1$. This implies $\theta\rightarrow 0$,
which leads to a divergence of $I(r_c)$.
Therefore, altogether the meaning of (\ref{area}) for the correspondence to
scattering amplitudes remains unclear.

\section{Conclusions}

We have considered the subset of space-like minimal surfaces in $\mbox{AdS}_3 \times \mbox{S}^3$, defined by
the requirement that they
admit constant induced metric for both projections to $\mbox{AdS}_3$
and $\mbox{S}^3$. For almost all cases this leads also to constant second fundamental form, justifying the name vacuum solutions.  There is one exceptional case for light-like $\mbox{AdS}_3$ projection, in which the parameterization needs a free field with unit mass.  It would be interesting to further study this integrable subset of solutions.

For all the vacuum solutions a classification and explicit construction has been
achieved. Up to the freedom of isometries  in $\mbox{AdS}_3 \times \mbox{S}^3$,
they are parameterized by four parameters $\rho,~\phi,~\rho_s,~\phi_s$. In terms
of these parameters one can express the constant mean curvatures of the
projections of the surface to $\mbox{AdS}_3$ \footnote{For surfaces with constant mean curvature in pure AdS see \cite{Sakai:2010eh}.} and $\mbox{S}^3$, respectively, as well as the constant
ratios of the areas measured with the induced metrics of the projections to the area
measured with the metric induced from the total space $\mbox{AdS}_3 \times \mbox{S}^3$. For all values of $\rho,~\phi,~\rho_s,~\phi_s$ explicit
formul\ae{} for the embedding coordinates of the surfaces have been presented.

  The projection to $\mbox{S}^3$ is always a space-like torus. The projection to $\mbox{AdS}_3$
can be space-like, time-like or light-like. For the space-like case there is only one
class of projections, it has the same null tetragonal boundary behavior
as the pure $\mbox{AdS}_3$ solution of \cite{Kruczenski:2002fb, Alday:2007hr}.

In the time-like case one finds various different classes. Except for one class which stays completely inside, the
surfaces touch the conformal boundary of $\mbox{AdS}_3$ at null lines.  Concerning
their boundary behavior, the classes are distinguished by counting the
corresponding null lines in $\partial \mbox{AdS}_3 =\mbox{S}^1\times \mbox{S}^1$ (or $\rr\times \mbox{S}^1$).
The boundary is part of  either one null line, two with the same orientation, two
with opposite orientation, three, or four with pairwise opposite orientation.

In the case of light-like projections there are two different classes. One
reaches the $\mbox{AdS}_3$ boundary at a null two-gon built out of a left and a right going
null line. The other one reaches the $\mbox{AdS}_3$ boundary only at a wedge out of pieces
of two null lines, but the boundary of the surfaces $\mbox{AdS}_3$ projection is
completed by a line inside $\mbox{AdS}_3$.

Altogether nine different classes are related
to different pieces of the $(\rho~,\phi )$-diagram. Since $\rho$ and $\phi$
carry information both on the shape of the projection in pure $\mbox{AdS}_3$-sense
(mean curvature) as well as information on the relation to the surface in
the total product space, $\mbox{AdS}_3$ projections with the same shape are situated
on curves in the $(\rho~,\phi )$-diagram.

All $\mbox{AdS}_3$ projections can be described as (parts of) intersections of
quadrics with the $\mbox{AdS}$-hyperboloid.
In each class we have chosen a canonical form of the intersecting quadric
and plotted the resulting surfaces. All other  $\mbox{AdS}_3$-projections can be reached
from these pictures by varying the inhomogeneity parameters of these forms
and/or applying $\mbox{O}(2,2)$ isometry transformations.

While isometry transformations of course do not touch the geometric shape,
they can change the apparent shape relative to a chosen coordinate system.
We have discussed, how just in this sense one can approach some classes in the limit
of infinite boosts from other classes

We made the observation that our surfaces can be related to solutions
of the complex sin(h)-Gordon type equations, which for the light-like surfaces degenerate to a linear equation.

For potential use in the correspondence to gluon scattering amplitudes we have calculated  the regularized area of the surface with a generic  $s$-channel null tetragon
boundary, i.e. two non-consecutive cusps with space-like and two with time-like separation. This configuration is remarkable, because there is no corresponding solution in the pure $\mbox{AdS}$ case. The standard
tetragon of \cite{Kruczenski:2002fb, Alday:2007hr} is relevant for the $u$-channel case, where all four cusps have space-like momentum sum.

Finally, a corresponding classification of vacuum type time-like strings can be similarly carried out (as it was pointed out in \cite{Dorn:2009hs}).

\section*{\large Acknowledgments}

We thank Johannes Henn, Jan Plefka, Donovan Young and especially Nadav Drukker for useful discussions.

This work has been supported in part
by Deutsche Forschungsgemeinschaft via SFB 647 and by VolkswagenStiftung.
G.J. was also supported by GNSF.

\appendix

\setcounter{equation}{0}

\def\theequation{A.\arabic{equation}}

\section{Useful formul\ae{}}

\label{UF}

Here we present some useful formul\ae{} for the $\mbox{SU(2)}$ and $\mbox{SL}(2,\rr)$ groups.

\noindent
\subsection{$\mbox{SU}(2)$ group}

Using the $\mathfrak{su}(2)$ algebra \eqref{ss=} one gets ${\bf u}^2=-(u_nu_n)\,{\bf I}$, for any ${\bf u}=
u_n\,{\bf s}_n\in \mathfrak{su}(2)$,
and then
\begin{equation}\label{e^u}
e^{\bf u}=\cos u\,I+\sin u\,\hat{\bf u}~,\quad \mbox{with}\quad
u=\sqrt{u_nu_n}~, \quad \hat{\bf u}={\bf u}/u~.
\end{equation}
In particular,
\begin{equation}\label{e^sigma}
e^{i\theta\,\boldsymbol{\sigma}_2}=\left( \begin{array}{cr}~ \cos\theta&\sin\theta\\
-\sin\theta&\cos\theta
\end{array}\right)~, \quad\quad
e^{i\eta\,\boldsymbol{\sigma}_3}=\left( \begin{array}{cr}~ e^{i\eta}&0\\
0&e^{-i\eta}
\end{array}\right)~,
\end{equation}
and the $\mbox{SO}(2)$ adjoint orbit of $\bf{s}_3$ generated by  $\bf{s}_2$
is given by
\begin{equation}\label{s3 orbit}
e^{-i\beta\,\boldsymbol{\s}_2}\,
{\bf s}_3\,e^{i\beta\,\boldsymbol{\sigma}_2}=\cos2\b\,{\bf s}_3+
\sin2\b\,{\bf s}_1~.~~~~~~~~~~~~~~~~~~
\end{equation}
With these equations and the identity
\begin{equation}\label{e^ue^v}
e^{\left(e^{{\bf u}}\,{\bf v}\,e^{-{\bf u}}\right)}=e^{{\bf u}}\,e^{{\bf v}}\,e^{-{\bf u}}~,
\end{equation}
one easily obtains \eqref{Ls,Rs=}-\eqref{solution h 1}.

\subsection{$\mbox{SL}(2,\rr)$ group}

From (\ref{tt=}) follows $\mathfrak{a}^2=\langle \mathfrak{a}\,\mathfrak{a}\rangle\,{\bf I},$ for any
$\mathfrak{a}\in \mathfrak{sl}(2,\rr)$. In particular, if $\langle \mathfrak{a}\,\mathfrak{a}\rangle=0$, then
$\mathfrak{a}$ is nilpotent $\mathfrak{a}^2=0$.
A compact form of the exponent $\,e^{\mathfrak{a}}$ then reads
\begin{equation}\ba\label{e^a}
e^{\mathfrak{a}}&=\cosh\theta\,\,{\bf I}+\frac{\sinh\theta}{\theta}\,\,\mathfrak{a}~,~~
\mbox{with}~~~
\theta=\sqrt{\,\langle\,\,\mathfrak{a}\,\mathfrak{a}\,\rangle}~,~~~~
~~~~&\mbox{if}~~~\langle\,\mathfrak{a}\,\mathfrak{a}\,\rangle >0~;&\\
e^{\mathfrak{a}}&=\cos\theta\,\,\,I\,+\,\frac{\sin\theta}{\theta}\,\,
\mathfrak{a}~,~~~~\mbox{with}~~~ \theta=\sqrt{-\langle\,\mathfrak{a}\,\mathfrak{a}\,\rangle}~,
~~~~&\mbox{if}~~~ \langle\,\mathfrak{a}\,\mathfrak{a}\, \rangle <0~;&\\
e^{\mathfrak{a}}&={\bf I} +\mathfrak{a}~,
\quad &\mbox{if} \quad
\langle\,\mathfrak{a}\,\mathfrak{a}\, \rangle =0~.&
\ea\end{equation}
One gets the following one parameter subgroups
\begin{equation}\label{e^t}
e^{\theta\,{\bf t}_0}=\left( \begin{array}{cr}~ \cos\theta&\sin\theta\\
-\sin\theta&\cos\theta
\end{array}\right),\quad\quad
e^{\theta\,{\bf t}_1}=\left( \begin{array}{cr}~ \cosh\theta&\sinh\theta\\
\sinh\theta&\cosh\theta
\end{array}\right),\quad
e^{\theta\,{\bf t}_-}=\left( \begin{array}{cr}~ 1&0\\
\theta&1
\end{array}\right),
\end{equation}
with $ {\bf t}_-=\frac{1}{2}({\bf t}_1-{\bf t}_0)$, and the corresponding adjoint orbits
of ${\bf t}_2$ are
\begin{eqnarray}\label{e^t1 t2 e^t1}
e^{-\theta\,{\bf t}_0}\,{\bf t}_2\,e^{\theta\,{\bf t}_0}=
\cos2\theta\,{\bf t}_2 + \sin 2\theta\,{\bf t}_1~,\quad \quad
e^{-\theta\,{\bf t}_1}\,{\bf t}_2\,e^{\theta\,{\bf t}_1}=
\cosh2\theta\,{\bf t}_2 + \sinh 2\theta\,{\bf t}_0~,\\ \label{e^t t2 e^t}
e^{-\theta\,{\bf t}_+}\,{\bf t}_2\,e^{\theta\,{\bf t}_+}=
{\bf t}_2 + 2\theta\,{\bf t}_+~.~~~~~~~~~~~~~~~~~~~~~~~~~~~~~~~
\end{eqnarray}

The following summation rules become helpful in calculations
\begin{equation}\label{epsilon epsilon}
\epsilon_{\mu\nu\rho}\epsilon_{\mu'\nu'}\,^\rho=
\eta_{\mu\nu'}\,\eta_{\nu\mu'}-\eta_{\mu\mu'}\,\eta_{\nu\nu'}~,
\end{equation}
\begin{equation}\label{tmu-tmu}
({\bf t}_\mu)_{a\dot a}\,({\bf t}^\mu)_{\dot b b}=
2\delta_{ab}\delta_{\dot a\dot b}-\delta_{a\dot a}\delta_{b\dot b}~.
\end{equation}

\setcounter{equation}{0}
\def\theequation{B.\arabic{equation}}

\section{Left-right decomposition of $\mathfrak{so}(2,2)$}
\label{R(2,2)-gl(2)}

Let us consider the map $\mathcal{M}:\rr^{2,2}\mapsto \mathfrak{gl}(2,\rr)$ given by
\begin{equation}\label{R(2,2)}
S_{a\dot a}=Y^{0'}\,\delta_{a\dot a}+Y^\mu\,({\bf t}_\m)_{a\dot a}~.
\end{equation}
Here $S_{a\dot a}\in  \mathfrak{gl}(2,\rr)$ is a $2\times 2$ matrix
and ($Y^{0'}$, $Y^\mu$) $(\mu=0,1,2)$ are coordinates of $Y\in \rr^{2,2}$.
We also use $Y^J,$ $J=(0',0,1,2)$ and treat it as a row. A matrix of a linear
transformation acts then on $Y^J$ from the r.h.s.
The inverse map takes the form $Y^{0'}=\langle S\rangle,$ $Y_\mu=\langle {\bf t}_\mu\, S\rangle$.
Writing these linear maps
in the matrix form $S_{a\dot a}=Y^J\,M_{J,\,a\dot a}, $
$Y_J=S_{a\dot a}\,M^{-1}_{a\dot a,\,J}\,$, one gets the matrix components
\begin{equation}\label{M and M-inv} \ba
M_{0',\,a\dot a}=&\delta_{a\dot a}~,\quad\quad &M_{\mu,\,a\dot a}=({\bf t}_\mu)_{a\dot a}~,&\\
M^{-1}_{a\dot a,\,0'}=&-\frac{1}{2}\,\delta_{a\dot a}~,
\quad\quad
&M^{-1}_{a\dot a,\,\m}=\frac{1}{2}\,({\bf t}_\mu)_{\dot a a}~.&
\ea\end{equation}

With the help of the identity \eqref{epsilon epsilon} one can check that
\begin{equation}\label{S.T}
 Y\cdot Z=\langle S\,T\rangle -2\left\langle S\right\rangle \left\langle T\right\rangle~,
\end{equation}
where $Y\cdot Z=Y^J\, Z_J$ is the inner product in $\rr^{2,2}$,  $S=\mathcal{M}(Y)$ and $T=\mathcal{M}(Z)$.
The bilinear form \eqref{S.T} makes $\mathfrak{gl}(2,\rr)$ isometric to $\rr^{2,2}$.
This inner product\footnote{Note that this inner product differs from the standard Killing
form on $\mathfrak{gl}(2,\rr)$, which is degenerated.} is invariant under the infinitesimal left and right  multiplications
\begin{equation}\label{l-r in gl}
 S\mapsto S+\epsilon^\mu\, {\bf t}_\mu\, S~, \quad \quad S\mapsto S-\epsilon^\mu\,  S\,{\bf t}_\mu~.
\end{equation}
The linear transformations in $\mathfrak{gl}(2,\rr):$ $S\mapsto {\bf t}_\mu\, S$
and $S\mapsto -S\,{\bf t}_\mu$ are given by the matrices
\begin{equation}\label{L,Rmu}
(\hat L_\mu)_{b\dot b,\, a\dot a}=({\bf t}_\mu)_{ab}\,\delta_{\dot b\dot a}~,\quad\quad
(\hat R_\mu)_{b\dot b,\, a\dot a}=-\delta_{ba}\,({\bf t}_\mu)_{\dot b\dot a}~.
\end{equation}

By \eqref{M and M-inv} and \eqref{L,Rmu}, the corresponding matrices in $\rr^{2,2}$
\begin{equation}
(L_\mu)_{JK}=M_{J,\,b\dot b}\,(\hat L_\mu)_{b\dot b,\,a\dot a}\,M^{-1}_{a\dot a,\,K}~,\quad \quad
(R_\mu)_{JK}=M_{J,\,b\dot b}\,(\hat R_\mu)_{b\dot b,\,a\dot a}\,M^{-1}_{a\dot a,\,K}~.
\end{equation}
 become
\be \ba\label{Lmu}
 (L_\mu)_{0'0'}=&0~, \quad (L_\mu)_{0'\nu}=\eta_{\mu\nu}~,& \quad  (L_\mu)_{\nu 0'}=-\eta_{\mu\nu}~,& \quad
(L_\mu)_{\nu\nu'}=\epsilon_{\m\n\n'}~,&
\\
(R_\mu)_{0'0'}=&0~, \quad (R_\mu)_{0'\nu}=-\eta_{\mu\nu}~,& \quad (R_\mu)_{\nu 0'}=\eta_{\mu\nu}~,& \quad (R_\mu)_{\nu\nu'}=\epsilon_{\m\n\n'}~,&
\ea\ee	
and one obtains \eqref{L,R basis}.

Now we consider the matrix exponent
\be\label{exp 4}
\mathcal{E}_J\,^K=\left(e^{l^\mu L_\mu+r^\nu R_\nu}\right)_J\,^K~,
\ee
where $l^\m,\,$  $r^\n$ are 3d Minkowski vectors and $L_\m,\,$ $R_\n$ are the basis vectors \eqref{L,R basis}.
The exponent in \eqref{exp A_1,2} is a particular case of \eqref{exp 4}.
The map of the first row of \eqref{exp 4} to $\mbox{SL}(2,\rr),$ defined by \eqref{R(2,2)} can be written as
\be\label{exp 2}
g_{a\dot a}=\mathcal{E}_{0'}\,^K\,M_{K,\,a\dot a}=
\left(e^{l^\mu\,\hat L_\mu}\,e^{r^\nu\,\hat R_\nu}\right)_{b b,\,a\dot a}~,
\ee
with $\hat L_\mu,\,$ $\hat R_\mu,\,$ given by \eqref{L,Rmu}. Using \eqref{L,Rmu},  we obtain
\be\label{exp 2=}
g_{a\dot a}=\left(e^{l^\mu\,\bf{t}_\mu}\right)_{cb}\,\delta_{b\dot c}\,
\delta_{ac}\left(e^{-r^\nu\,\bf{t}_\nu}\right)_{\dot c\dot
a}=\left(e^{l^\mu\,\bf{t}_\mu}\right)_{ab}\,
\left(e^{-r^\nu\,\bf{t}_\nu}\right)_{b\dot a}~.
\ee
Taking $l_\m,$ $r_\n$ corresponding to \eqref{exp A_1,2}, from \eqref{exp
2=} we find the factorized form of the $g$-field  \eqref{solution g}.

\end{document}